\documentclass{jfm}
\usepackage{graphicx}

\usepackage{amssymb}
\usepackage[utf8]{inputenc}

\usepackage{amsmath}
\usepackage{tikz}
\usetikzlibrary{shapes}
\usepackage{comment}
\usepackage{bm}

\usepackage{adjustbox}
\usepackage{subfig}

\DeclareMathOperator*{\argmax}{argmax}

\shorttitle{Adjoint-based linear sensitivity of a hypersonic boundary layer}
\shortauthor{A. Poulain, C. Content, G. Rigas, E. Garnier and D. Sipp}

\title{Adjoint-based linear sensitivity of a hypersonic boundary layer to steady wall blowing-suction/heating-cooling}

\author{Arthur Poulain\aff{1}\corresp{\email{arthur.poulain@onera.fr}},
  Cédric Content\aff{1},
  Georgios Rigas\aff{2},
  Eric Garnier\aff{1},
 \and Denis Sipp\aff{1}}

\affiliation{\aff{1}DAAA, ONERA, Université Paris Saclay, F-92190 Meudon, France
\aff{2}Department of Aeronautics, Imperial College London, London SW7 2AZ, United Kingdom}

\begin{document}

\maketitle

\begin{abstract}
For a Mach $4.5$ flat-plate adiabatic boundary layer, we study the sensitivity of the first, second Mack modes and streaks to steady wall-normal blowing/suction and wall heat flux. The global instabilities are characterised in frequency space with resolvent gains and their gradients with respect to wall-boundary conditions are derived through a Lagrangian-based method. The implementation is performed in the open-source high-order finite-volume code BROADCAST and Algorithmic Differentiation is used to access the high-order state derivatives of the discretised governing equations.
For the second Mack mode, the resolvent optimal gain decreases when suction is applied upstream of Fedorov's mode S/mode F synchronisation point leading to stabilisation and conversely when applied downstream. The largest suction gradient is in the region of branch I of mode S neutral curve. For heat flux control, strong heating at the leading edge stabilises both the first and second Mack modes, the former being more sensitive to wall-temperature control. Streaks are less sensitive to any boundary control in comparison with the Mack modes. Eventually, we show that an optimal actuator consisting of a single steady heating strip located close to the leading edge manages to damp all the instabilities together.
\end{abstract}

\begin{keywords}
Compressible boundary layer, Boundary layer stability, Boundary layer control.
\end{keywords}

\section{Introduction}

The performance of all vehicles is significantly reduced when they are submitted to turbulent flows. At low and medium speeds, additional viscous drag is produced by the larger turbulent wall-shear stresses. At higher speed, they also generate high wall temperature requiring thicker thermal protection. Conversely, a turbulent flow may be desired to maintain a given flow topology or to avoid the detachment inherent to laminar flows on convex geometries. Controlling laminar-to-turbulent transition is a critical technology for design. Transition mechanisms depend strongly on the geometry of the configuration, the type and level of environmental disturbances. Small amplitude free-stream disturbances (vortical or acoustic waves) as well as wall-roughnesses may excite instabilities, through the receptivity process, which are subsequently amplified by various linear mechanisms, such as the Mack modes \citep{mack1963inviscid} or non-modal streaks  \citep{morkovin1994transition}. 

Linear stability theories predict the early stages of the amplification before the non-linear interactions play a leading role resulting in the breakdown towards laminar / turbulent transition. Hypersonic boundary layer flows can be studied by local stability analysis (LST) \citep{malik1989prediction} and Parabolised stability equations (PSE)  \citep{stuckert1995nonparallel} for weakly non-parallel flows. For more general configurations, resolvent analyses, which also take into account the non-modal phenomena arising from the non-normality of the Navier-Stokes operator \citep{sipp2013characterization}, have become computationally affordable in the recent years.

Stability analysis of supersonic flows displays an infinity of modes, called Mack modes \citep{mack1963inviscid} for high Mach numbers. At Mach 4.5, the first and second Mack modes are predominant among the others \citep{ma2003receptivity, bugeat20193d}. The oblique first Mack mode corresponds with a local unstable mode which consists in an inviscid wave located around the generalised inflection point. The two-dimensional second Mack mode is a trapped acoustic wave close to the wall resulting from the synchronisation of the phase speed of the local Fedorov's modes F and S \citep{fedorov2011high}, respectively a fast and a slow acoustic wave, that promotes the instability of the latter.

Laminar flow control includes two main families: wave-cancellation methods, which erase the instabilities with out-of-phase waves \citep{joslin1998overview,nibourel2023reactive} and steady mean-flow manipulations, which we will focus on in the present article, with a special interest in blowing/suction and heating/cooling control systems.

Firstly, suction control, when applied to incompressible flows, is optimal to damp the two-dimensional Tollmien-Schlichting waves when the actuator is located near branch I (similar location as the optimal forcing from the resolvent analysis) according to the asymptotic analysis \citep{reed1986numerical}, the experimental work \citep{reynolds1986experiments} and the local scattering approach \citep{huang2016effect}). For supersonic boundary layers, asymptotic analyses demonstrated that the second Mack mode is also highly receptive to unsteady blowing/suction located near Branch I of mode S \citep{fedorov2002receptivity}. \citet{wang2011response} completed the analysis by showing that unsteady blowing/suction strongly excites the mode S (slow) if located upstream of the point where mode F (fast) and S phase velocities synchronise (called synchronisation point in the rest of the article) while the effect is much lower when the actuator is downstream. The synchronisation point seems to be linked with the optimal location for local control as \citet{fong2014numerical} noticed an opposite effect on the growth of mode S if a roughness is located before or after the synchronization point in the DNS computation of a hypersonic boundary layer.

Secondly, cooling/heating control has been investigated. \citet{mack1993effect} showed that the growth of the Mack modes are sensitive to the wall temperature: a uniformly cooled wall damps the first Mack mode but destabilises the second Mack mode. Therefore, for Mach numbers below 4 \citep{mack1993effect}, cooling the wall in order to modify the base-flow represents a control technique to delay the laminar-to-turbulent transition. \citet{wang2009effect} found that the hypersonic boundary layer is less sensitive to unsteady temperature perturbation than unsteady blowing/suction. For subsonic boundary layers, a heating strip at the leading edge has a stabilising effect \citep{kazakov1995optimization} while this is the opposite if located further downstream \citep{masad1995transition2}. For supersonic boundary layers, \citet{masad1995transition} found by N-factor analysis that a heating strip upstream also stabilises by damping the first Mack mode. The experimental work \citep{sidorenko2015effect} on a cone at higher Mach number showed that the second Mack mode instability is damped by local wall cooling. To find the optimal location for a wall heating device, DNS simulations of hypersonic flows were carried out. \citet{fedorov2014numerical} pointed out the region upstream of the neutral point to be optimal to put a heating device in order to stabilise the second Mack mode but \citet{soudakov2015stability} underlined that this location depends on the receptivity region for a sharp cone at Mach 6. Recent studies \citep{zhao2018numerical, batista2020local} found that cooling upstream of the synchronisation point and heating downstream damp the second Mack mode. Furthermore, localised strips of wall cooling and heating combinations are nearly as effective as controlling the whole boundary \citep{batista2020local}. On a Mach 6 cone, \citet{oz2023local} showed by local stability analysis that complete wall cooling destabilises the boundary layer but a local wall cooling strip upstream of the synchronisation point damps the instabilities and conversely if located downstream.

The previous studies presented above were based on parametric analysis which cannot span the full range of optimal locations. General approaches using gradient-based optimisation were then offered. They rely on the adjoint-based linear sensitivity of the base-flow, i.e. the indicator of the regions where small modifications of the base-flow have the highest impact on the instabilities growth.

An adjoint method to find the optimum suction distribution on a Blasius boundary layer through the minimisation of N-factor was explored by \citet{balakumar1999optimum}. Parabolised stability equations and their adjoint equations were later solved to perform sensitivity analysis of compressible flows \citep{pralits2000sensitivity}. Sensitivity was later exploited to iteratively decrease the energy of the Tollmien-Schlichting (T-S) waves \citep{walther2001optimal} through wall transpiration or to damp the T-S waves, streaks and oblique waves \citep{pralits2002adjoint} via steady suction. At the same period, \citet{airiau2003methodology} developed a similar framework and extended the analysis to suction panels of finite length. While previous methods computed the optimal suction control to damp a fixed disturbance, \citet{zuccher2004algebraic} offered a "robust" control which damps the most disrupting instability for the controlled flow. The extension of the sensitivity analysis of steady blowing to the global analysis framework was later offered by \citet{brandt2011effect} on an incompressible boundary layer. Sensitivity of the global eigenvalue problem has then been computed for shape optimisation. Iterative methods \citep{wang2019enhanced, martinez-cava_sensitivity_2020} have been employed to optimise the geometry in order to gradually damp the growth rate of the most unstable mode. However, these techniques require to repeat the expensive computations of the base-flow, stability and sensitivity as they are valid only in the linear regime. \citet{boujo2021second} offered a second-order sensitivity method to enlarge the validity of the linear sensitivity and therefore reduce the total number of iterations to optimise a geometry.

The present work aims at finding the optimal location for small amplitude steady wall blowing/suction or heating/cooling actuators to damp the main instabilities in the Mach number 4.5 boundary layer by computing their linear sensitivity in the global stability framework. The sensitivity is computed here only once around the base-flow as the aim is not the optimisation of a finite-amplitude control, but the physical understanding of the local gradient for wall-based control around the base-flow.

\begin{figure}
\centering
\includegraphics[width=0.89\textwidth]{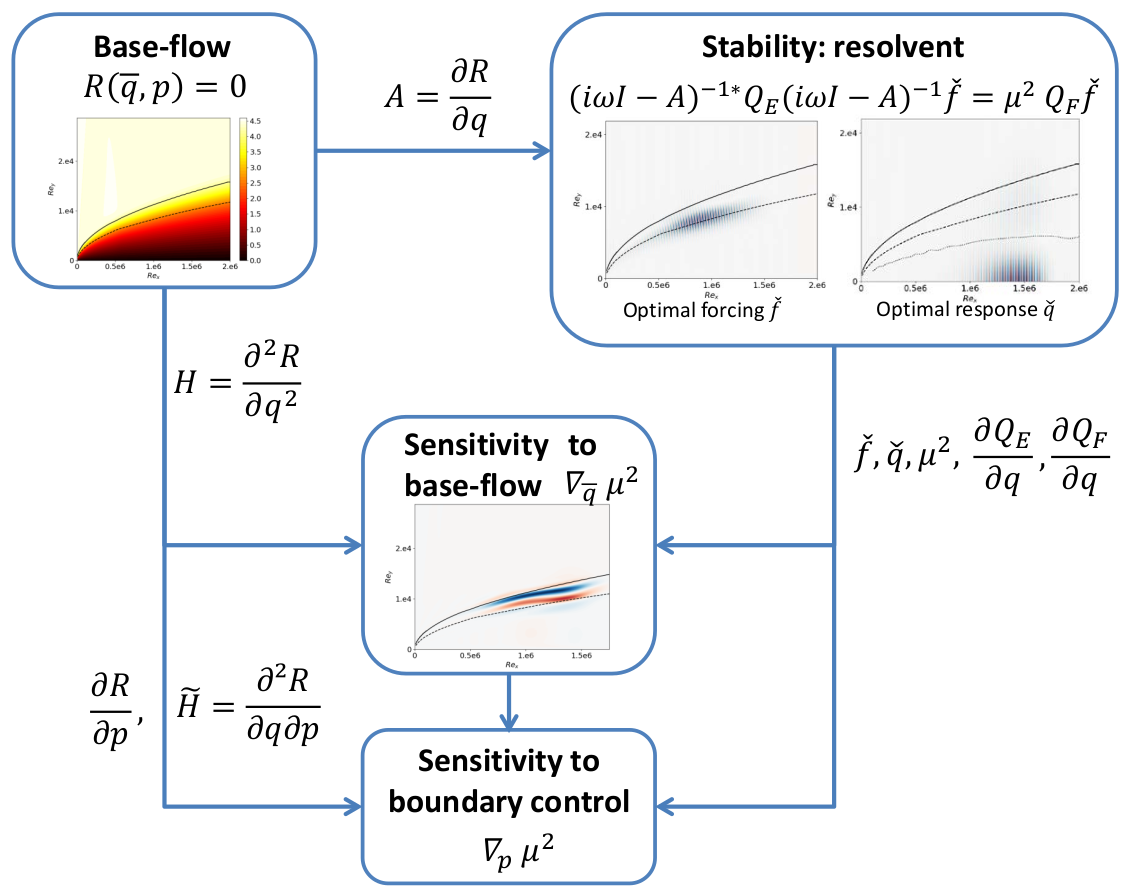}
\caption{Workflow for sensitivity computation. Application to the second Mack mode in the $M=4.5$ boundary layer. Mach number contours for base-flow $\mathbf{\overline{q}}$. Real part of the streamwise momentum forcing for the optimal forcing $\mathbf{\check{f}}$. Real part of the pressure disturbances for the optimal response $\mathbf{\check{q}}$. Sensitivity of the optimal gain to streamwise momentum base-flow modifications contours for the sensitivity to base-flow $\nabla_{\mathbf{\overline{q}}} \mu^2$.}
\label{diagbl}
\end{figure}

The outline of the paper grossly follows
the workflow for the computation of the sensitivity, shown in Figure \ref{diagbl} for boundary control of the second Mack mode. After presenting (\S \ref{sec:method}) the theoretical tools and implementation of stability and sensitivity analyses, we consider the hypersonic boundary layer case at $M=4.5$ in \S \ref{sec:hyper}. The base-flow is first computed (\S \ref{sec:bsf}) and resolvent analysis is performed to find the most predominant two-dimensional and three-dimensional modes (\S \ref{sec:stability}). Sensitivity to base-flow modifications is computed (\S \ref{sec:sensbf}) and projected to get the sensitivity of the three main instabilities to steady wall-blowing and to wall-heating (\S \ref{sec:sens3D}). Eventually, the initial design of an optimal wall-control actuator is attempted in \S \ref{sec:sens3Dapplied}.

\section{Methods} \label{sec:method}

After presenting the governing equations (\S \ref{sec:goveq}), we recall briefly the definitions of a base-flow and resolvent modes / optimal gains (\S \ref{sec:stab}). Then (\S \ref{sec:sensitivity}), we show the sensitivity of the optimal gains while the numerical details are given in \S \ref{sec:num}.

\subsection{Governing equations} \label{sec:goveq}

We consider the compressible Navier-Stokes equations written in conservative form 
\begin{equation}
 \frac{\partial \mathbf{q}}{\partial t} + \nabla \cdot \mathbf{F}(\mathbf{q}) =  0,
 \label{NS1}
\end{equation}
with $\mathbf{q} = (\rho, \rho \mathbf{v}, \rho E)$ designating respectively the density, momentum and total energy of the fluid and $ \mathbf{F}(\mathbf{q}) $ the associated fluxes. In developed form, the Navier-Stokes equations read
\begin{equation}
\frac{\partial \rho}{\partial t} + \nabla \cdot \left( \rho \mathbf{v} \right) = 0,
\label{NS0}
\end{equation}
\begin{equation}
\frac{\partial (\rho \mathbf{v})}{\partial t} + \nabla \cdot \left( \rho \mathbf{v} \mathbf{v} + p \mathbf{I} - \bm{\tau} \right) = 0,
\label{NS00}
\end{equation}
\begin{equation}
\frac{\partial (\rho E)}{\partial t} + \nabla \cdot \left( (\rho E +p) \mathbf{v} - \bm{\tau} \cdot \mathbf{v} - \lambda \nabla T \right) = 0, 
\label{NS000}
\end{equation}
with $E = p / (\rho (\gamma -1)) + \frac{1}{2}\mathbf{v} \cdot \mathbf{v} $,  $\bm{\tau} = \eta ( \nabla \mathbf{v} + (\nabla \mathbf{v})^T ) - \frac{2}{3} \eta (\nabla \cdot \mathbf{v}) \mathbf{I} $, $\mathbf{I}$ the identity matrix, $\lambda = \eta c_p / Pr$, $c_p$ the isobaric heat capacity and $P_r$ the Prandtl number ($P_r = 0.72$).
To close the system, two more equations are required. First, one assumes a homogeneous, thermally and calorically perfect gas. The perfect gas law is
\begin{equation}
p = \rho r T,
\label{NS2}
\end{equation}
with $r = 287.1$ $\textrm{J}.\textrm{kg}^{-1}.\textrm{K}^{-1}$ the specific gas constant.
Then, the Sutherland's law is selected to link the viscosity $\eta$ to the temperature \citep{sutherland1893lii},
\begin{equation}
\eta (T) = \eta_{ref} \left( \frac{T}{T_{ref}} \right)^{3/2} \frac{T_{ref} + S}{T + S},
\label{NS3}
\end{equation}
with $S = 110.4$ K the Sutherland's temperature, $\eta_{ref} = 1.716 \time 10^{-5}$ $\textrm{kg}.\textrm{m}^{-1}.\textrm{s}^{-1}$ and $T_{ref} = 273.15$ K.

After spatial discretisation (see \S \ref{sec:num}), the discrete residual is noted $\mathbf{R}(\mathbf{q}) = - \nabla \cdot \mathbf{F}(\mathbf{q})$.

\subsection{Base-flow, resolvent modes and optimal gains} \label{sec:stab}

The boundary layer base-flow $ \mathbf{\overline{q}} $ is a steady solution of the governing equations:
\begin{equation}
    \mathbf{R}(\mathbf{\overline{q}}) = 0.
\end{equation}
It is an amplifier flow \citep{huerre1990local}, for which all small-amplitude perturbations are exponentially stable in time. In such flows, it is more relevant to perform a resolvent analysis to study the linear dynamics of the flow and identify the pseudo-resonances of the flow. 
For this, we move to frequency-space and consider a small-amplitude forcing field $\mathbf{f}'(t)=e^{i\omega t} \mathbf{\check{f}}$ that is applied to the right-hand-side of equation \eqref{NS1} and which may be restricted, thanks to a prolongation matrix $\mathbf{P}$, to specific regions of the flow or specific components of the state. $\mathbf{P}$ is a rectangular matrix, containing only zeros and ones, and of size the dimension of the state $ \mathbf{q}$ times the dimension of the forcing $ \mathbf{f}$. The linear response of the flow $\mathbf{q}'(t)=e^{i\omega t} \mathbf{\check{q}}$ is then governed by
$
\mathbf{\check{q}} = \mathbf{\mathcal{R} P \check{f}}$,
with $\mathbf{\mathcal{R}} = \left( i\omega \mathbf{I} - \mathbf{A} \right)^{-1}$ denoting the resolvent operator and $\mathbf{I}$ the identity matrix. The resolvent corresponds to a transfer function between the input (forcing) and the response (perturbations). 
The optimal forcings / responses are then computed by optimising the input-output gain $\mu^2$ between the energy of the response and the energy of the forcing,
\begin{equation}
\mu^2 = \sup_{\mathbf{\check{f}}} \frac{\| \mathbf{\check{q}} \|^2_E}{\| \mathbf{\check{f}} \|^2_F},  
\label{gstab5}    
\end{equation}
with $\|\cdot\|_q$ and $\|\cdot\|_f$ the user-selected measures to evaluate the amplitude of the fluctuations and the forcing. These measures are defined with their associated discrete positive Hermitian matrices $\mathbf{Q}_q$ and $\mathbf{Q}_f$,
\begin{equation}
\| \mathbf{\check{q}} \|^2_q = \mathbf{\check{q}}^* \mathbf{Q}_q \mathbf{\check{q}} \;\; , \;\;
\| \mathbf{\check{f}} \|^2_f = \mathbf{\check{f}}^* \mathbf{Q}_f \mathbf{\check{f}},
\label{gstab6}    
\end{equation}
where only $\mathbf{Q}_f$ is required to be definite.
For compressible flows, a common choice for $\mathbf{Q}_q$ and $\mathbf{Q}_f$ consists in Chu's energy \citep{chu1965energy} in order to take into account the pressure ($\check{p}$) and entropy ($\check{s}$) disturbances, which writes for dimensionless fluctuations:
\begin{equation}
E_\text{Chu} = \mathbf{\check{q}}^* \mathbf{Q}_{\text{Chu}} \mathbf{\check{q}} = \frac{1}{2} \int_\Omega \left( \overline{\rho}\|\mathbf{\check{v}}\|^2 + \frac{1}{\gamma}\frac{\check{p}^2}{ \overline{p}} + \gamma (\gamma -1) M^4\, \overline{p}\, \check{s}^2 \right) \,d\Omega. 
\label{gstab7}    
\end{equation}
Chu's energy is the sum of the kinetic energy of the perturbation and a thermodynamic component (potential energy from compression and from heat exchange) with appropriate coefficients to exclude the conservative compression work \citep{hanifi1996transient} in order to obtain a norm which does not increase in time in the absence of sources of energy \citep{george2011chu}. 
Matrix $\mathbf{Q}_q = \mathbf{Q}_f = \mathbf{Q}_{\text{Chu}}$ for Chu's energy norm is block-diagonal and may be written with conservative variables, as detailed for instance in \citet{bugeat20193d}. 

Solving for $ \mu_i^2 $ over a range of frequencies $ \omega $ provides the most receptive frequency
(where $ \mu_i(\omega)^2 $ is the largest) and the associated optimal forcing mode $\mathbf{\check{f}}_i$. 

From an algorithmic point of view, we solve for the optimal gain in eq. \eqref{gstab5} by rewriting
\begin{equation}
\mu^2 = \sup_{ \mathbf{\check{f}}} \frac{\mathbf{\check{q}}^* {\mathbf{Q}_q} \mathbf{\check{q}} }{\mathbf{\check{f}}^* \mathbf{Q}_f \mathbf{\check{f}}} = \sup_{ \mathbf{\check{f}}}  \frac{(\mathbf{\mathcal{R}P \check{f}})^* {\mathbf{Q}_q} (\mathbf{\mathcal{R}P \check{f}})}{(\mathbf{P \check{f}})^* \mathbf{Q}_f (\mathbf{P \check{f}})} =  \sup_{ \mathbf{\check{f}}}  \frac{ \mathbf{\check{f}}^* \mathbf{P}^* \mathbf{\mathcal{R}}^* {\mathbf{Q}_q} \mathbf{\mathcal{R}P \check{f}}}{ \mathbf{\check{f}}^* {\mathbf{Q}_f} \mathbf{\check{f}}}.
\label{gstab9}    
\end{equation}
The optimisation problem defined by equation \eqref{gstab9} is the Rayleigh quotient. It is equivalent to the generalised Hermitian eigenvalue problem
\begin{equation}
\mathbf{P}^* \mathbf{\mathcal{R}}^* {\mathbf{Q}_q} \mathbf{\mathcal{R}P \check{f}}_i = \mu_i^2 {\mathbf{Q}_f} \mathbf{\check{f}}_i.
\label{gstab11}    
\end{equation}
Its eigenvalues are ranked such that $\mu_i^2 \geq \mu_{i+1}^2$ and the associated eigenvectors are $\mathbf{\check{f}}_i$, which we normalise to $\mathbf{\check{f}}_i^* \mathbf{Q}_f \mathbf{\check{f}}_i=1$.
The normalized responses ($\mathbf{\check{q}}_i^* \mathbf{Q}_q \mathbf{\check{q}}_i=1$) are then obtained through:
\begin{equation}
\mathbf{\check{q}}_i = \mu_i^{-1}\mathbf{\mathcal{R} P \check{f}}_i.
\label{gstab8}    
\end{equation}
The bases $ \mathbf{\check{f}}_i$ and $\mathbf{\check{q}}_i$ are orthonormal bases of the input and output spaces.

\subsection{Sensitivity of optimal gains to base-flow variations, steady forcing and parameter variations} \label{sec:sensitivity}

Linear sensitivity of eigenvalues to a general flow parameter has been addressed by \citet{martinez2019direct}.
Here we focus on optimal gains, as in \citet{brandt2011effect}.
Following the discrete framework introduced in \citet{mettot2014computation} for eigenvalues, we extend the work of \citet{martinez2019direct} to optimal gains and stress the link with the concepts 
of linear sensitivity of the optimal gain to base-flow modifications $\delta \mathbf{\overline{q}}$ and steady forcing $\delta \mathbf{\overline{f}}$, that were initially described in details in \citet{marquet2008sensitivity}. We consider the general case of the sensitivity of the optimal gain $\mu_i^2$ to any flow parameter written $\mathbf{p}$, such that:
\begin{equation} \label{eq:grad}
    \delta\mu_i^2=(\nabla_{\mathbf{p}} \mu_i^2)^*\mathbf{Q}_p\delta \mathbf{p},
\end{equation}
where $ \mathbf{Q}_p$ is a given scalar-product.
$ \mathbf{p} $ can be either a scalar number as $Re$, $M$, $T_\infty$,... or a large-dimensional vector as the prescribed inlet profile or control vectors (wall-normal velocity or wall-temperature profiles). The latter will be considered for application in the next section. The only restriction in the following is that the control parameter $\mathbf{p}$ has to be invariant in time and in the $z$-direction, as the base-flow is assumed to remain steady and two-dimensional when varying the control parameter. We consider that the residual depends on the parameter $\mathbf{p}$, $\mathbf{R}(\mathbf{q})=\mathbf{R}(\mathbf{q},\mathbf{p})$, and consequently also the Jacobian $\mathbf{A}(\mathbf{q},\mathbf{p}) = \partial \mathbf{R}(\mathbf{q},\mathbf{p}) / \partial \mathbf{q}$.

The objective of the optimisation is the optimal gain $\mu_i^2$ and the constraint is given by the eigenvalue problem written in Eq. \eqref{gstab11} and that the base-flow must remain solution of $\mathbf{R}(\mathbf{\overline{q}},\mathbf{p}) = 0$.
Because of the space or component restriction ($\mathbf{P} $ may be degenerate),
 it is necessary to split the generalised eigenvalue problem \eqref{gstab11} into three equations so as to handle only matrices without inverses:
\begin{eqnarray}
\mathbf{P\check{f}}_i = \mu_i (i\omega\mathbf{I}-\mathbf{A})\mathbf{\check{q}}_i, \;\;\;\; & \mu_i \mathbf{Q}_q \mathbf{\check{q}}_i = (-i\omega\mathbf{I}-\mathbf{A}^*) \mathbf{\check{a}}, \;\;\;\; & \mathbf{P}^*\mathbf{\check{a}} = \mu_i^2 \mathbf{Q}_f\mathbf{\check{f}}_i.
\label{sensapp}
\end{eqnarray}
This system involves an additional component  $\mathbf{\check{a}}$ within the eigenproblem.
We therefore define the Lagrangian function $\mathcal{L}$ as a function of the state (the optimal gain $\mu_i^2$, the optimal forcing $\mathbf{\check{f}}_i$ and response $\mathbf{\check{q}}_i$, the additional variable $\mathbf{\check{a}}$, the base-flow $\mathbf{\overline{q}}$), the four Lagrangian multipliers $\bm{\lambda}_1$, $\bm{\lambda}_2$, $\bm{\lambda}_3$ and $\bm{\lambda}_4$, and the control vector $ \mathbf{p}$:
\begin{eqnarray}
\mathcal{L}([\mu_i^2,\mathbf{\check{f}}_i,\mathbf{\check{q}}_i,\mathbf{\check{a}},\mathbf{\overline{q}}],\bm{\lambda}_{1\cdots4},\mathbf{p}) &=& \mu_i^2 + \langle \bm{\lambda}_1, \mathbf{P\check{f}}_i-  \mu_i (i\omega\mathbf{I}-\mathbf{A}(\mathbf{\overline{q}},\mathbf{p}))\mathbf{\check{q}}_i \rangle \nonumber \\ &+& \langle \bm{\lambda}_2 ,  \mu_i\mathbf{Q}_q(\mathbf{\overline{q}}) \mathbf{\check{q}}_i + (i\omega\mathbf{I}+\mathbf{A}(\mathbf{\overline{q}},\mathbf{p})^*) \mathbf{\check{a}} \rangle \nonumber  \\
 &+& \langle \bm{\lambda}_3 , \mathbf{P}^*\mathbf{\check{a}} - \mu_i^2 \mathbf{Q}_f(\mathbf{\overline{q}})\mathbf{\check{f}}_i \rangle + \langle \bm{\lambda}_4, \mathbf{R}(\mathbf{\overline{q}},\mathbf{p}) \rangle.
\label{sensp1}
\end{eqnarray}
Here $\langle \mathbf{a},\mathbf{b} \rangle = \mathbf{a}^*\mathbf{b} $ is the Hermitian scalar product.
By zeroing the variation of $\mathcal{L}$ with the state, taking into account that $  \langle \mathbf{\check{f}}_i, \mathbf{Q}_f \mathbf{\check{f}}_i \rangle=1$, we obtain that
\begin{equation}
\bm{\lambda}_1= \mathbf{\check{a}},  \;\;\;\;
\bm{\lambda}_2 =\mu_i \mathbf{\check{q}}_i,  \;\;\;\; \bm{\lambda}_3 = \mathbf{\check{f}}_i,  \;\;\;\;
\bm{\lambda}_4 = \mathbf{Q}_f\nabla_{\mathbf{\overline{f}}} \mu_i^2,
\end{equation}
where
\begin{eqnarray}
 \frac{\nabla_{\mathbf{\overline{f}}} \mu_i^2}{\mu_i^2}&=&   - \mathbf{Q}_f^{-1}\mathbf{A}^{-1*}\mathbf{Q}_q \frac{\nabla_{\mathbf{\overline{q}}} \mu_i^2}{\mu_i^2}, \\
\frac{\nabla_{\mathbf{\overline{q}}} \mu_i ^2}{\mu_i^2} &=& \mathbf{Q}_q^{-1} \left[2 \operatorname{Re}\left(  \mathbf{H}'^* \mathbf{\mathcal{R}}^{*}\mathbf{Q}_q \mathbf{\check{q}}_i \right) +  \left( \frac{\partial (\mathbf{Q}_q \mathbf{\check{q}}_i)}{\partial \mathbf{q}} \right)^* \mathbf{\check{q}}_i -  \left( \frac{\partial (\mathbf{Q}_f \mathbf{\check{f}}_i)}{\partial \mathbf{q}} \right)^* \mathbf{\check{f}}_i\right], \label{sensapp3}
\end{eqnarray}
are two vectors called sensitivity to steady volume forcing and sensitivity to base-flow variations \citep{marquet2008sensitivity,brandt2011effect,mettot2014computation} that verify:
\begin{equation} \label{eq:grad2}
\delta \mu_i^2 = (\nabla_{\mathbf{\overline{f}}} \mu_i^2)^* \mathbf{Q}_f \delta\mathbf{\overline{f}}  = (\nabla_{\mathbf{\overline{q}}} \mu_i^{2})^* \mathbf{Q}_q \delta \mathbf{\overline{q}}.
\end{equation}
Matrix $\mathbf{H}'$ is defined as $\mathbf{H}' \delta \mathbf{\overline{q}}= \mathbf{H}(\mathbf{\check{q}}_i, \delta \mathbf{\overline{q}})$ for all $\delta \mathbf{\overline{q}}$, where $\mathbf{H}={\partial \mathbf{A}}/{\partial \mathbf{q}}={\partial^2 \mathbf{R}}/{\partial \mathbf{q}^2}$ is the Hessian rank-3 tensor. Details of the computation of the sensitivity to base-flow modifications is given in appendix \ref{sec:bfvar}.

In expression \eqref{sensapp3}, the components ${\partial \mathbf{Q}_{q,f}}/{\partial \mathbf{q}}$ are non-zero in the case of Chu's energy norm. As an aside, \citet{brandt2011effect} has introduced an additional physical constraint (the time-invariant continuity equation \eqref{NS0}) to study the sensitivity to physically relevant base-flow modifications. Details of the computation of the sensitivity to divergence-free base-flow modifications is given in appendix \ref{sec:appdivfree}.
 
Finally, the variation of $\mathcal{L}$ with the control vector $\mathbf{p}$ provides the gradient we are looking for:
\begin{equation}
\frac{\nabla_{\mathbf{p}} \mu_i^2}{\mu_i^2} = \mathbf{Q}_p^{-1} \left[ \left(\frac{\partial \mathbf{R}}{\partial \mathbf{p}}\right)^* \mathbf{Q}_f \frac{\nabla_{\mathbf{\overline{f}}} \mu_i^2}{\mu_i^2} + 2 \operatorname{Re}\left(  \mathbf{\tilde{H}}'^* \mathbf{\mathcal{R}}^{*}\mathbf{Q}_q \mathbf{\check{q}} \right) \right],
\label{sensp3}
\end{equation}
with $\mathbf{\tilde{H}}'$ defined as $\mathbf{\tilde{H}}' \delta \mathbf{p} = \mathbf{\tilde{H}}  (\mathbf{\check{q}}_i, \delta \mathbf{p}) $ for all $\delta \mathbf{p}$, $\mathbf{\tilde{H}}={\partial \mathbf{A}}/{\partial \mathbf{p}}={\partial^2 \mathbf{R}}/{\partial \mathbf{q}\partial\mathbf{p}}$ being a rank-3 tensor. 

In expression \eqref{sensp3}, the first term is interpreted as the variation of the optimal gain induced by the modification of the Jacobian due to the change of the base-flow $ \mathbf{A}(\mathbf{q}) $ while the second term is the variation of the Jacobian due to the variation of the control parameter $ \mathbf{A}(\mathbf{p}) $, keeping the base-flow constant. It has been described in other words in \citet{guo2021sensitivity} for eigenvalue sensitivity: Route I is the distortion of the base-flow which induces a modification of the eigenvalue problem, Route II is the direct distortion of the linear operator. Depending on the parameter $\mathbf{p}$ chosen to compute the sensitivity, one route or another is favoured.

\subsection{Interpretation of $ \nabla_{\mathbf{p}}\mu_i^2 $ and $\nabla_{\overline{\mathbf{q}}} \mu_i^2$ } \label{sec:interp}

We consider the following optimisation problem:
\begin{equation}
\delta\mathbf{{p}}^{m}=\argmax_{\|  \delta\mathbf{{p}} \|_{p} =1} \delta \mu_i^2(\delta\mathbf{{p}}), 
\end{equation}
where $ \delta\mu_i^2$ and $ \delta{\mathbf{p}}$ are related through eq. \eqref{eq:grad}.
It is straightforward to show that:
\begin{equation} \label{eq:gradp} 
     \delta\mathbf{{p}}^{m}=\frac{\nabla_{\mathbf{p}}\mu_i^2}{\| \nabla_{\mathbf{p}}\mu_i^2\|_p}.
\end{equation}
The gradient is therefore the best profile (of given small amplitude measured with $ \| \cdot \|_p$) that optimally increases the gain and therefore optimally strengthens the instability. Conversely, because of linearity, $-\delta \mathbf{{p}}^{m}$ is the optimal open-loop control to damp the optimal gain.

For interpretation, $ \left. \delta \mu_i^2\right|^{m}:=\delta \mu_i^2 ( \delta\mathbf{p}^{m})$ may be rewritten as
\begin{equation}
      \frac{\left. \delta \mu_i^2\right|^{m}}{\mu_i^2} =\underbrace{\left[2 \mathbf{Q}_p^{-1}\operatorname{Re}\left(  \mathbf{\tilde{H}}'^* \mathbf{\mathcal{R}}^{*}\mathbf{Q}_q \mathbf{\check{q}} \right)\right]^* \mathbf{Q}_p \delta \mathbf{p}^{m}}_{\delta E^{m}_{\mathbf{A}}} +\underbrace{\left(\frac{\nabla_{\mathbf{\overline{q}}} \mu_i^{2}}{\mu_i^2}\right)^* \mathbf{Q}_q \delta \mathbf{\overline{q}}^{m}}_{\delta E^{m}_{\mathbf{\overline{q}}}},
\label{deltamum}
\end{equation}
where the first term on the rhs, $\delta E^{m}_{\mathbf{A}}$, is the variation of the gain due to the induced modification of the Jacobian, and   
the second part, $\delta E^{m}_{\mathbf{q}}$, is the variation due to the induced change in base-flow,
\begin{equation}
\delta\overline{\mathbf{q}}^{m}=-\mathbf{A}^{-1}\left(\frac{\partial \mathbf{R}}{\partial \mathbf{p}}\right)\delta\mathbf{p}^{m}.
\label{indbsfvar}
\end{equation}
Finally, the optimal gain variation due to the induced base-flow modification may be broken down into three pieces:
\begin{equation}
      \delta E^{m}_{\mathbf{\overline{q}}}=\delta E^{m}_c+\delta E^{m}_p+\delta E^{m}_s,
\label{chusm}
\end{equation}
where 
\begin{equation}
      \delta E^{m}_c=\int_\Omega \frac{\overline{\rho}\delta{\mathbf{\overline{v}}}^{b} \cdot\delta{\mathbf{\overline{v}}}^{m}}{2} d\Omega, \;
      \delta E^{m}_p=\int_\Omega
      \frac{\delta\overline{p}^{b}\delta\overline{p}^{m}}{2\gamma\overline{p}} d\Omega, \;
      \delta E^{m}_s=\frac{\gamma(\gamma-1)M^4}{2}\int_\Omega
      \overline{p}\delta{{\overline{s}}}^{b}\delta{{\overline{s}}}^{m} d\Omega.
\end{equation}
Here, the notation
$ (\delta{\mathbf{\overline{v}}}^{b},\delta{\overline{p}^{b}},\delta{{\overline{s}}^{b}})$ corresponds to the velocity-pressure-entropy variations associated to $\delta{\mathbf{\overline{q}}}^{b}:=\frac{\nabla_{\mathbf{\overline{q}}} \mu_i^{2}}{\mu_i^2}$ and 
$ (\delta{\mathbf{\overline{v}}}^{m},\delta{{\overline{p}}}^{m},\delta{{\overline{s}}}^{m})$ to 
$ \delta\overline{\mathbf{q}}^{m}$.
In the following, we will represent the contributions $\delta E^{m}_{\mathbf{A}}$, $\delta E^{m}_c$, $\delta E^{m}_p$ and $\delta E^{m}_s$  to assess the importance of Jacobian, base-flow kinetic energy, pressure and entropy modifications in the gain variation associated to the optimal control $ \delta\mathbf{p}^{m}$.  

Finally, it is also straightforward to show that the above $\delta\mathbf{\overline{{q}}}^{b}$ is also the solution to the following optimisation problem:
\begin{equation}
\delta\mathbf{\overline{{q}}}^{b}=\argmax_{\|  \delta\mathbf{\overline{{q}}} \|_{q} =\| \mu_i^{-2}\nabla_{\overline{q}}\mu_i^2\|_q} \delta \mu_i^2(\delta\mathbf{\overline{{q}}}), 
\end{equation}
where $ \delta\mu_i^2$ and $ \delta\mathbf{\overline{{q}}}$ are related through eq. \eqref{eq:grad2}. Hence, the maximum
$ \left. \delta \mu_i^2\right|^{b}:=\delta \mu_i^2 ( \delta \mathbf{\overline{q}}^{b})$, can be rewritten and decomposed as:
\begin{equation}
   \frac{\left. \delta \mu_i^2\right|^{b}}{\mu_i^2}=\left\| \frac{\nabla_{\overline{q}}\mu_i^2}{\mu_i^{2}}\right\|_q^2={ \delta E^{b}_c + \delta E^{b}_p + \delta E^{b}_s   }
\label{chusp}
\end{equation}
where
\begin{equation}
      \delta E^{b}_c=\int_\Omega \frac{\overline{\rho}\|\delta{\mathbf{\overline{v}}}^{b}\|^2}{2} d\Omega, \;
      \delta E^{b}_p=\int_\Omega
      \frac{\left(\delta\overline{p}^{b}\right)^2}{2\gamma\overline{p}} d\Omega, \;
      \delta E^{b}_s=\frac{\gamma(\gamma-1)M^4}{2}\int_\Omega
      \overline{p}\left(\delta{{\overline{s}}}^{b}\right)^2 d\Omega.
\end{equation}
In the following, we will represent the contributions $\delta E^{b}_c$, $\delta E^{b}_p$ and $\delta E^{b}_s$ to assess the importance of base-flow kinetic energy, pressure and entropy modifications in the gain variation associated to the base-flow modification $ \delta{\overline{\mathbf{q}}}^{b}$.  
 
\subsection{Numerical methods} \label{sec:num}

\subsubsection{Numerical discretisation and algorithms}

The BROADCAST code includes all the tools required to compute the base-flow, the global stability analysis and the linear sensitivity analysis. A thorough description of the various numerical methods and their validation for stability and sensitivity in BROADCAST can be found in \citet{poulain2023broadcast}.

The two-dimensional space discretisation for the inviscid flux follows the 7th order FE-MUSCL (Flux-Extrapolated-MUSCL) scheme \citep{cinnella2016high} which had been assessed in hypersonic flow simulations by \citet{sciacovelli2021assessment} showing excellent results in accuracy and shock capturing features. The viscous fluxes are computed on a five-point compact stencil which is fourth-order accurate \citep{shen2009high}.

The Jacobian as well as all the other operators derived to compute the gradients are constructed by Algorithmic Differentiation (AD) through the software \textit{TAPENADE} \citep{hascoet2013tapenade}. The matrix operators are explicitly built by matrix-vector products \citep{mettot2013linear}. They correspond to linearised discrete residuals given by AD and stored in a sparse format. For sensitivity computation, for instance in the case of the wall velocity ($\mathbf{p}=\mathbf{v}_w$), the numerical method to compute the sparse residual and Jacobian operators $\partial \mathbf{R} / \partial \mathbf{v}_w$ and $\partial \mathbf{A} / \partial \mathbf{v}_w$ which appear in $\nabla_{\mathbf{v}_w} \mu^2$ consists in writing the wall boundary condition which appears inside the residual $\mathbf{R}$ as a function of the input $\mathbf{v}_w$ (taken equal to $0$ for base-flow) and linearise through AD. A description of the implementation of the wall boundary condition is given in Appendix \ref{sec:bcwall}.

All linear systems involving sparse matrices are then solved using the PETSC software interface \citep{balay2019petsc} which includes the direct sparse LU solver from MUMPS \citep{amestoy2001fully}. BROADCAST code being written in Python language, the petsc4py version is used \citep{dalcin2011parallel}. 
The base-flow solution is solved with a Newton method which consists of an iterative method, where from a state $ \mathbf{q}^n $, we build $ \mathbf{q}^{n+1} = \mathbf{q}^n + \delta \mathbf{q}^n $, with 
$ \mathbf{A}(\mathbf{q}^n)\, \delta \mathbf{q}^n = - \mathbf{R}(\mathbf{q}^n) $ and $\mathbf{A}(\mathbf{q}^n) =\left.  \frac{\partial \mathbf{R}}{\partial \mathbf{q}} \right|_{\mathbf{q}^n} $ is the Jacobian operator evaluated at $ \mathbf{q}=\mathbf{q}^n$.
To ease convergence, a pseudo-transient continuation method (or relaxation method) is used following \citet{crivellini2011implicit}.
To solve the generalised eigenvalue problem from resolvent analysis, we use the SLEPc library \citep{roman2015slepc}, which implements various Krylov-Schur methods \citep{hernandez2007krylov}.


The gradients have been validated by comparing the results with a finite difference method:
\begin{equation}
\lim_{\epsilon \rightarrow 0}\frac{\mu_i^2(\mathbf{p} + \epsilon \nabla_\mathbf{p} \mu_i^2) - \mu_i^2(\mathbf{p})}{\epsilon} \rightarrow \| \nabla_\mathbf{p} \mu_i^2\|^2.
\label{validgain}
\end{equation}
Furthermore, a comparison of the gradient of the optimal gain to wall blowing between the discrete and continuous frameworks is performed on a low Mach boundary layer in Appendix \ref{sec:valid}.

\subsubsection{Building the Jacobian and Hessian for 3D perturbations}

The extension of global stability analysis to linear 3D perturbations follows \citet{bugeat20193d} and has been adapted to BROADCAST in \citet{poulain2023broadcast}.
The base-flow being assumed homogeneous in the $z$-direction, the perturbation field can be searched under the form
\begin{equation}
\mathbf{q}'(x,y,z,t) = \mathbf{\check{q}}(x,y) e^{i(\omega t + \beta z)},
\label{gstab3}
\end{equation}
where $ \beta $ is the real wavenumber in the $ z$-direction.
A similar form is assumed for the optimal forcing.
These perturbations can therefore be studied on the same 2D mesh without discretisation of the $ z$-direction. The $z$-dependency of the forcing and response are taken into account analytically. One can split the 3D residual $\mathbf{R}_\text{3D}$ as the sum of the 2D discretised residual $\mathbf{R}$ and its z-derivative components $\mathbf{R}_z$,
\begin{equation}
\mathbf{R}_\text{3D} (\mathbf{q}) = \mathbf{R} (\mathbf{q}) + \mathbf{R}_z(\mathbf{q}).
\label{3d}
\end{equation}
For the compressible Navier-Stokes equations, the $\mathbf{R}_z$ residual in conservative form can be written as the sum of four functions whose expressions can be found in \citet{poulain2023broadcast},
\begin{equation}
\mathbf{R}_z (\mathbf{q}) = \mathbf{B}(\mathbf{q}) \frac{\partial \mathbf{q}}{\partial z} + \mathbf{C}_1(\mathbf{q}) \frac{\partial^2 \mathbf{q}}{\partial z^2} + \frac{\partial \mathbf{C}_2}{\partial \mathbf{q}} \frac{\partial \mathbf{q}}{\partial z} \frac{\partial \mathbf{q}}{\partial z} + \mathbf{D}_1(\mathbf{q}) \frac{\partial \mathbf{q}}{\partial z} \odot \mathbf{D}_2(\mathbf{q}) \frac{\partial \mathbf{q}}{\partial z}.
\label{3d2}
\end{equation}
Notation $\odot$ refers to the element-wise product of two matrices or vectors (Hadamard product). Given that the base-flow is homogeneous in the $z$-direction and keeping only the first-order terms for small fluctuations $\mathbf{q}'$, the linearisation of equation \eqref{3d} yields
\begin{equation}
\mathbf{A}_\text{3D}\, (\mathbf{\overline{q}}) \mathbf{q}' = (\mathbf{A}(\mathbf{\overline{q}}) + \mathbf{A}_z(\mathbf{\overline{q}}) )\, \mathbf{q}' = \left(\mathbf{A}(\mathbf{\overline{q}}) + i\beta \mathbf{B}(\mathbf{\overline{q}}) - \beta^2 \mathbf{C}_1(\mathbf{\overline{q}}) \right) \mathbf{q}'.
\label{3d5}
\end{equation}

Linear sensitivity described in \S \ref{sec:sensitivity} may also be extended to 3D perturbations. However, the following developments are correct only for 2D sensitivity (homogeneous gradient in $z$-direction) of eigenvalue/optimal gain of 3D modes.
Similarly to the equation \eqref{3d5}, the 3D Hessian operator can be written as $\mathbf{H}_\text{3D}(\mathbf{\check{q}},\mathbf{\overline{q}}) = \mathbf{H}(\mathbf{\check{q}},\mathbf{\overline{q}}) + \mathbf{H}_z(\mathbf{\check{q}},\mathbf{\overline{q}})$. One should notice that the base-flow $\mathbf{\overline{q}}$ remains 2D and only the response $\mathbf{\check{q}}$ brings a new 3D contribution. The three dimensional Hessian is defined as
\begin{equation}
\mathbf{H}_\text{3D}(\mathbf{\check{q}},\mathbf{\overline{q}}) = \left.\frac{\partial \left( \mathbf{A}_\text{3D} \mathbf{\check{q}} \right)}{\partial \mathbf{q}}\right|_{\mathbf{\overline{q}}}.
\label{hess3D}
\end{equation}
From the equation \eqref{3d5}, the following expression may be derived,
\begin{equation}
\mathbf{H}_\text{3D}' = \mathbf{H}' + i\beta \frac{\partial \mathbf{B}}{\partial \mathbf{q}} \mathbf{\check{q}} - \beta^2 \frac{\partial \mathbf{C}_1}{\partial \mathbf{q}} \mathbf{\check{q}}_i.
\label{hess3D2}
\end{equation}

Therefore, the 2D sensitivity of a 3D mode is given by the same equations as the one of a 2D mode but by replacing the 2D Hessian by the 3D Hessian written in equation \eqref{hess3D2}.

\section{Hypersonic boundary layer} \label{sec:hyper}

\subsection{Configuration and base-flow} \label{sec:bsf}

We consider an adiabatic flat plate close to the configuration described in \citet{bugeat20193d}.
All quantities are made non-dimensional with the following density, velocity, length and temperature scales: $\rho_\infty,\: U_\infty,\: \nu_\infty/U_\infty, \; T_\infty $. The free-stream Mach number and free-stream temperature are respectively $M=4.5$ and $T_\infty = 288$ K.
The spatial coordinates $(x,y,z)$ then correspond to $ (Re_x, Re_y, Re_z) $, where for example $Re_x=U_\infty x /\nu_\infty $.

The flat plate geometry is studied in a rectangular computational domain. The domain starts with a thin boundary layer profile at $Re_{x,\text{in}} = 8160$ and ends at $Re_{x,\text{out}} = 2 \times 10^6$. The height of the domain is high enough in order not to affect the development of the boundary-layer or the stability analysis. In practice, the domain height is about $9\,\delta^*_\text{out}$ with $\delta^*_\text{out}$ the compressible displacement thickness at the outlet. This gives $Re_{y,\text{top}} = 119000$. The Cartesian mesh is equi-spaced in the x-direction and stretched in the wall-normal direction (y-direction). The stretching has the following properties: maximum $y^+ \leq 1$ at the wall, cell height geometric growth rate of $2 \%$ from $y=0$ till $y=3\delta_{99}$ with $\delta_{99}$ the boundary layer thickness, where $ \Delta y^+ \approx 10$ is reached and then a growth rate increase from $2 \%$ till $10 \%$ from $y=3\delta_{99}$ till $y=L_y$ where $\Delta y^+ \approx 130$. The mesh has the size $(N_x, N_y) = (1000, 150)$ which gives $N = 150 000$ grid points. 

Four different boundary conditions are applied around the rectangular domain. At the inlet, a Dirichlet boundary condition is applied since the flow is supersonic. The imposed flow profile corresponds to a compressible self-similar solution for $u$, $v$, $\rho$ and $T$. At the outlet, an extrapolation boundary condition is imposed (the flow is overall assumed supersonic). An adiabatic no-slip wall is prescribed at the bottom while a non-reflecting condition \citep{poinsot1992boundary} is employed at the top boundary.

\begin{figure}
    \centering
    \includegraphics[width=0.5\textwidth]{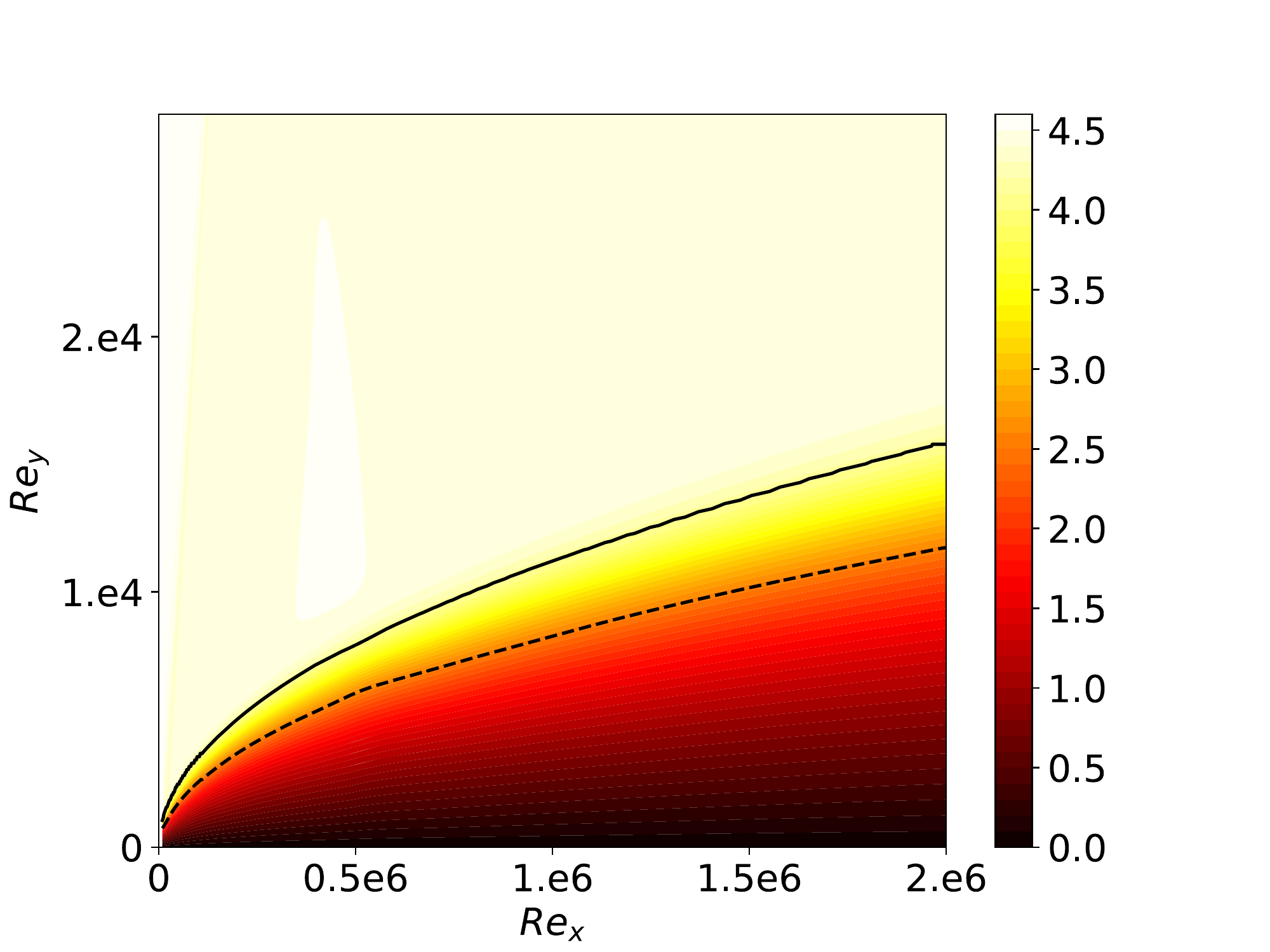}
    \caption{Mach number. Solid line indicates the boundary layer thickness $ \delta_{99}$. Dashed line indicates the displacement thickness $ \delta^*$.}
    \label{mach}
\end{figure}

The two-dimensional steady base-flow (Figure \ref{mach}) is computed by a pseudo-transient continuation method, the compressible self-similar solution being taken as initial state. The algorithm converges in 7 iterations (high initial CFL as the self-similar solution is close to the base-flow solution) and decreases the residual $L^2$ norms by 12 orders of magnitude. 

\subsection{Stability} \label{sec:stability}

The global stability analysis of the $M=4.5$ boundary layer over an adiabatic flat plate has been thoroughly studied by \citet{bugeat20193d} and validated with the present tools in \citet{poulain2023broadcast}. Therefore, more details can be found in the first cited paper, only the most important stability results are recalled here because they represent necessary steps to perform the sensitivity analysis.

\begin{figure}
\centering
\subfloat[\label{gain3d}]{\includegraphics[width=0.5\textwidth]{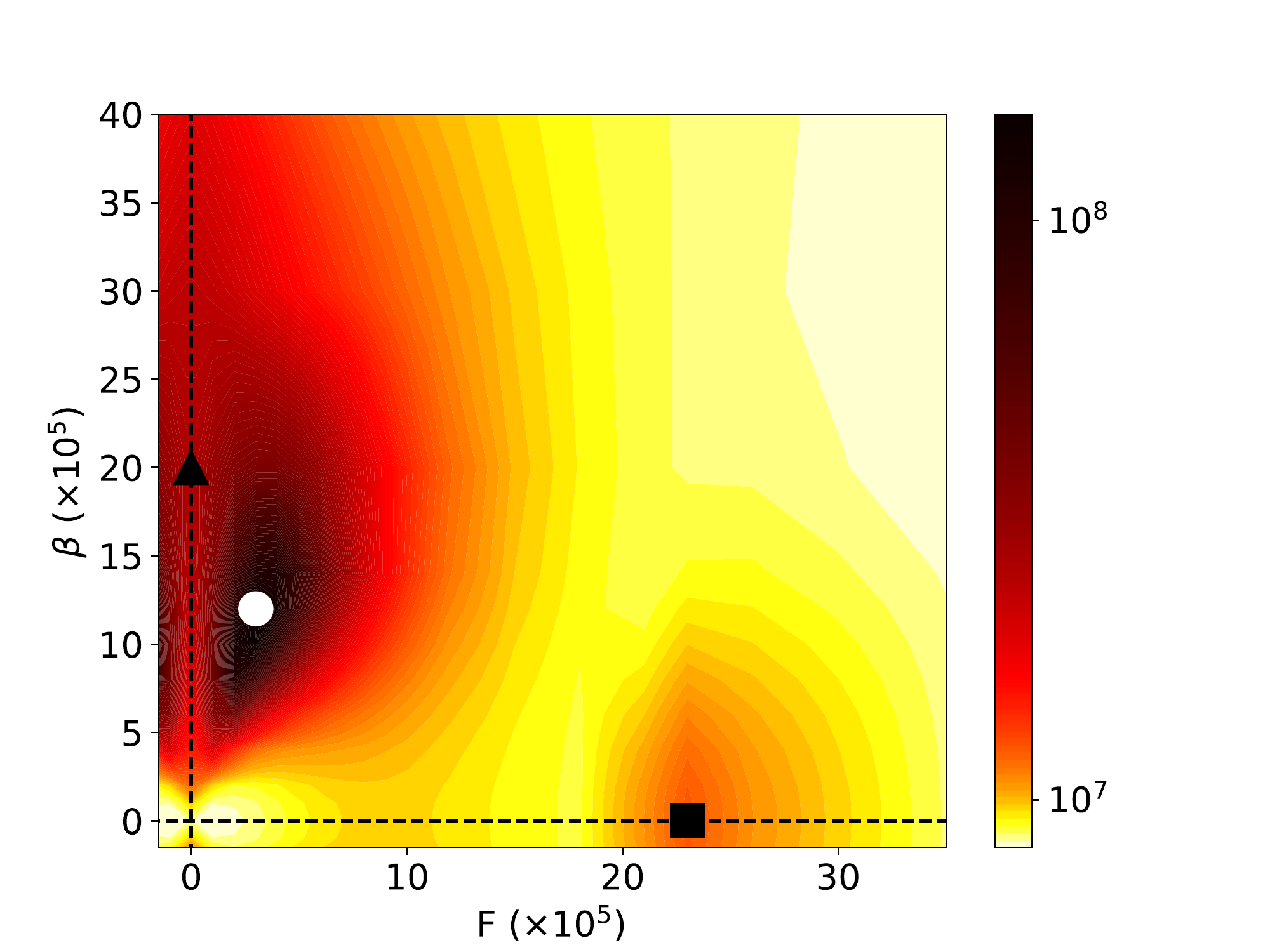}}
{\subfloat[\label{dfdchu}]{\includegraphics[width=0.5\linewidth]{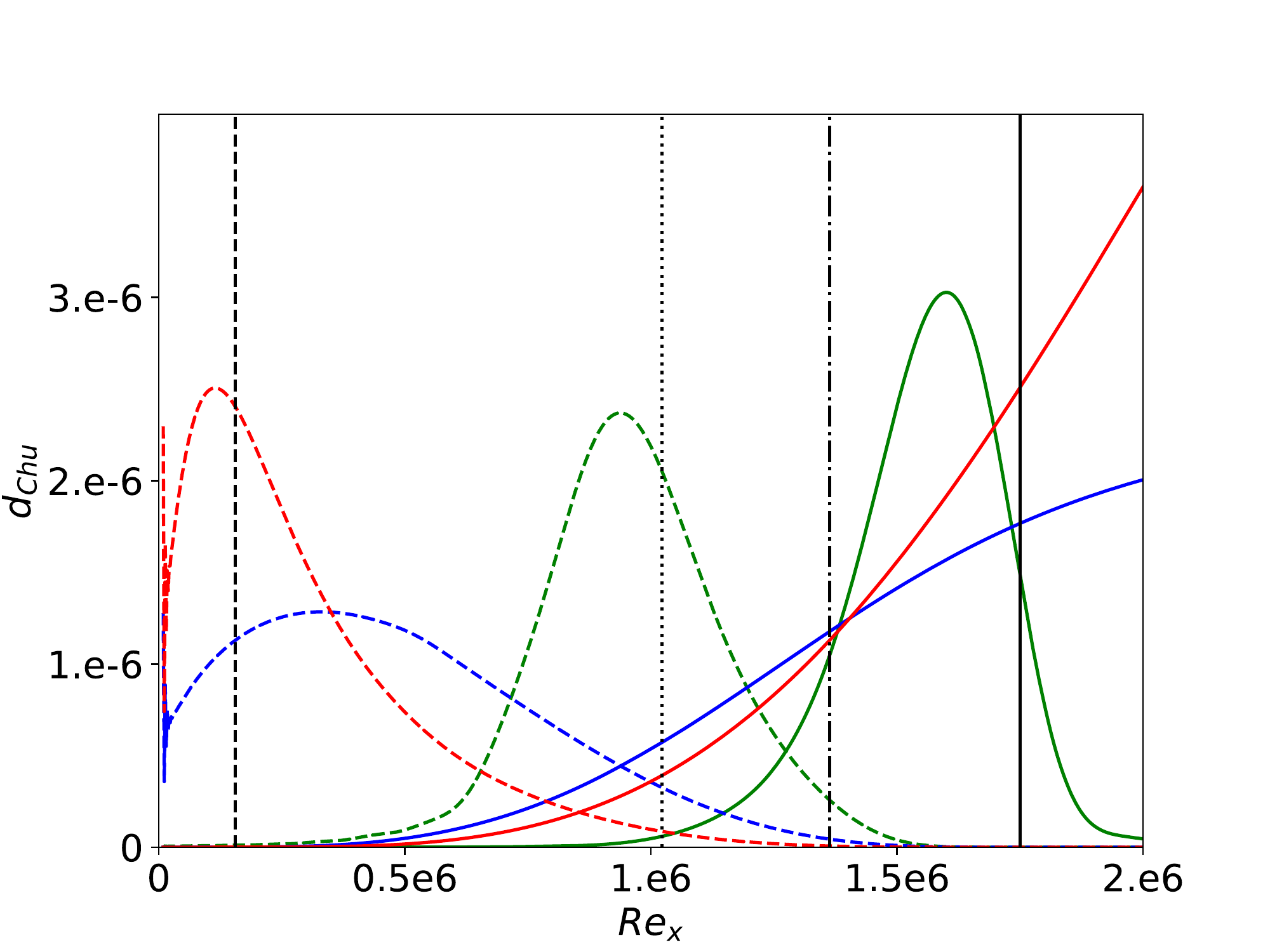}}}
\caption{Resolvent analysis. (a) Optimal gain $\mu_0$ with respect to the frequency $F$ and the spanwise wavenumber $\beta$. White circle denotes the first Mack mode, black triangle the streaks and black square the second Mack mode. (b) Energy density $d_\text{Chu}$ of the optimal forcing (dashed lines) and response (solid lines) of the streaks (blue), the first (red) and second (green) Mack modes. Black vertical lines indicate branch I of first Mack mode (dashed line), branch I of mode S (dotted line), branch II of mode S (dashed-dotted line) and the end of the optimisation domain for resolvent analysis (solid line).}
\label{gain2ds}
\end{figure}

We recall that the frequency is normalised as $F = \omega \nu_\infty/ U_\infty^2$ and the spanwise wavenumber $\beta:=\beta \nu_\infty/U_\infty $.
The measures $ \mathbf{Q}_q$ and $\mathbf{Q}_f$ correspond to Chu's energy, both being restricted to $Re_{x} \leq 1.75 \times 10^6$ and $ Re_y \leq 59500 $ to remove the top and outlet boundary conditions from the optimisation domain. The forcing $ \check{\mathbf{f}}$ is therefore only defined in this region and is applied on all five equations. At $M=4.5$, the second Mack mode exhibits a large gain, $\mu_0 = 1.80 \times 10^{7}$ which is obtained for $ \beta=0$ and $F=2.3 \times 10^{-4}$. The optimal gains are also computed with $\beta \neq 0$ (Figure \ref{gain3d}). The three-dimensional gains highlight streaks ($\mu_0 = 4.66 \times 10^{7}$) around $\beta = 2 \times 10^{-4}$ at zero frequency and the first oblique Mack mode is the strongest instability ($\mu_0 = 1.16 \times 10^{8}$) for around $\beta = 1.2 \times 10^{-4}$ and $F=3 \times 10^{-5}$.

Then the streamwise evolution of the resolvent modes is illustrated through their streamwise energy growth. An energy density is defined as the integral of Chu's energy in the wall-normal direction. For example,  $d_\text{Chu} (x) = \int_0^{y_\text{max}} \check{\mathbf{q}}(x,y)^* \mathbf{Q}_\text{Chu}(x,y)\check{\mathbf{q}}(x,y)\,dy $ for the optimal response. Chu's energy densities of the forcing and response for the streaks, first and second Mack modes are plotted in Figure \ref{dfdchu}. A spatial local stability analysis (description of the method in Appendix \ref{sec:lst}) has been performed at the frequency ($ \omega $) and wavenumber ($ \beta $) of the first and second Mack modes (see symbols in Figure \ref{gain3d}). The streamwise locations of branch I (resp. II) which is the beginning (resp. end) of the unstable region in the $x$-direction of the local modes are also plotted in Figure \ref{dfdchu}. The maximal optimal forcing location for the first and second Mack modes agree well with branch I location of their respective local modes. Branch II of the first Mack mode is downstream of the domain end while branch II of the local mode S (linked to the second Mack mode) is slightly upstream of the maximal optimal response location of the second Mack mode \citep{sippmarquet}.

\subsection{Sensitivity to base-flow modifications and steady forcing} \label{sec:sensbf}

The sensitivities of the optimal gain of the first and second Mack modes to base-flow modifications $\nabla_{\mathbf{\overline{q}}} \mu_ 0^2$ are computed. To highlight the most predominant contributions in the sensitivity, the total optimal gain variation has been split (Eq. \eqref{chusp}) into base-flow variation components of streamwise kinetic energy $\delta E_{c_x}^b$, wall-normal kinetic energy $\delta E_{c_y}^b$, pressure  $\delta E_{p}^b$ and entropy $\delta E_{s}^b$. These quantities have been reported in Table \ref{tab:chu}. The largest component for both Mack modes seems to be the wall-normal kinetic energy. However, as noticed in previous studies \citep{brandt2011effect, park2019sensitivity}, a strong wall-normal base-flow velocity modification is not physical since base-flow momentum fields are divergence-free (${\partial_x \delta \bm{\overline{\rho u}}} + {\partial_y \delta \bm{\overline{\rho v}}}= 0$). The addition of this constraint  strongly damps the wall-normal velocity component of the sensitivity. Therefore, the largest component of the sensitivity becomes the streamwise velocity for both Mack modes followed by the entropy $\delta E_s^b$ and pressure $\delta E_p^b$ components, which are one order of magnitude lower for both Mack modes. Overall, the second Mack mode is an order of magnitude more sensitive than the first Mack mode. The same conclusions are drawn for the streaks with a sensitivity an order of magnitude below the first Mack mode.

\begin{table}
    \begin{center}
\def~{\hphantom{0}}
    \begin{tabular}{c|c|cccc|c}
&  & $ \delta E_{c_x}^b$ & $ \delta E_{c_y}^b$ &  $\delta E_p^b $ &  $ \delta E_s^b$ & $ \delta \mu_i^2/\mu_i^2$  \\ 
     \hline
1st & $\nabla_{\overline{\mathbf{q}}} \mu_0^2 /\mu_0^2$ & $2.6 \times 10^{-4}$ & $1.7 \times 10^{-2}$ & $3.1 \times 10^{-6}$ & $1.1 \times 10^{-5}$ & $1.7 \times 10^{-2}$  \\
Mack & $\nabla_{\overline{\mathbf{q}}}^{df} \mu_0^2 /\mu_0^2$ & $3.0 \times 10^{-4}$ & $4.8 \times 10^{-6}$ & $9.4 \times 10^{-6}$ & $3.4 \times 10^{-5}$ & $3.5 \times 10^{-4}$  \\
mode & $\nabla_{\overline{\mathbf{f}}} \mu_0^2 /\mu_0^2$ & $2.0 \times 10^{+7}$ & $2.5 \times 10^{+5}$ & $7.4 \times 10^{+6}$ & $1.8 \times 10^{+6}$ & $2.9 \times 10^{+7}$ \\
    \hline
2nd & $\nabla_{\overline{\mathbf{q}}} \mu_0^2 /\mu_0^2$ & $1.3 \times 10^{-3}$ & $1.4 \times 10^{-2}$ & $4.2 \times 10^{-5}$ & $1.5 \times 10^{-5}$ & $1.5 \times 10^{-2}$  \\
Mack & $\nabla_{\overline{\mathbf{q}}}^{df} \mu_0^2 /\mu_0^2$ & $1.1 \times 10^{-3}$ & $7.6 \times 10^{-8}$ & $7.1 \times 10^{-5}$ & $1.2 \times 10^{-4}$ & $1.3 \times 10^{-3}$   \\
mode & $\nabla_{\overline{\mathbf{f}}} \mu_0^2 /\mu_0^2$ & $4.5 \times 10^{+7}$ & $7.1 \times 10^{+5}$ & $9.1 \times 10^{+6}$ & $4.2 \times 10^{+6}$ & $5.9 \times 10^{+7}$ \\
    \hline
 & $\nabla_{\overline{\mathbf{q}}} \mu_0^2 /\mu_0^2$ & $4.6 \times 10^{-8}$ & $1.9 \times 10^{-4}$ & $4.4 \times 10^{-8}$ & $4.9 \times 10^{-8}$ & $1.9 \times 10^{-4}$  \\
Streaks & $\nabla_{\overline{\mathbf{q}}}^{df} \mu_0^2 /\mu_0^2$ & $1.3 \times 10^{-5}$ & $1.1 \times 10^{-6}$ & $1.3 \times 10^{-6}$ & $3.0 \times 10^{-6}$ & $1.8 \times 10^{-5}$   \\
 & $\nabla_{\overline{\mathbf{f}}} \mu_0^2 /\mu_0^2$ & $6.4 \times 10^{+6}$ & $4.3 \times 10^{+5}$ & $6.7 \times 10^{+6}$ & $3.2 \times 10^{+4}$ & $1.4 \times 10^{+7}$ \\
     \hline
    \end{tabular}
    \caption{Chu energy norm components of the sensitivity to base-flow modifications, to divergence-free base-flow modifications and to steady forcing. The last column $ \delta \mu_i^2/\mu_i^2$ represents the sum of the Chu energy norm components (Eq. \eqref{chusp}).}
    \label{tab:chu}
    \end{center}
\end{table}

\begin{figure}
\centering
\subfloat[\label{sensbfum1}]{\includegraphics[width=0.49\textwidth]{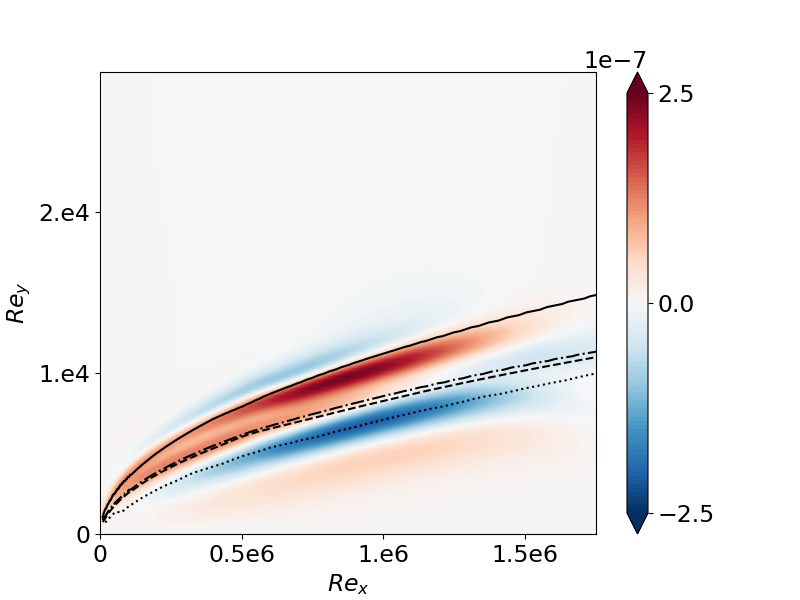}}
{\subfloat[\label{sensbftm1}]{\includegraphics[width=0.49\linewidth]{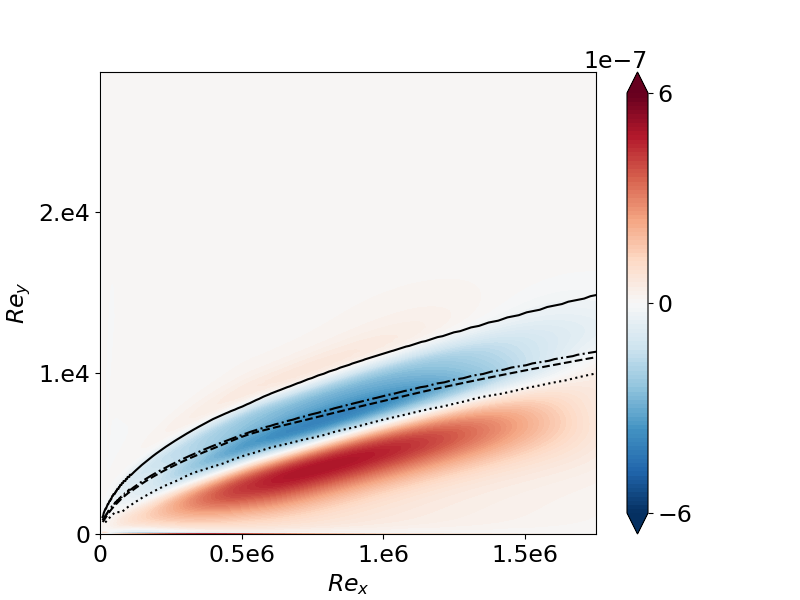}}}
\subfloat[\label{sensbfu}]{\includegraphics[width=0.49\textwidth]{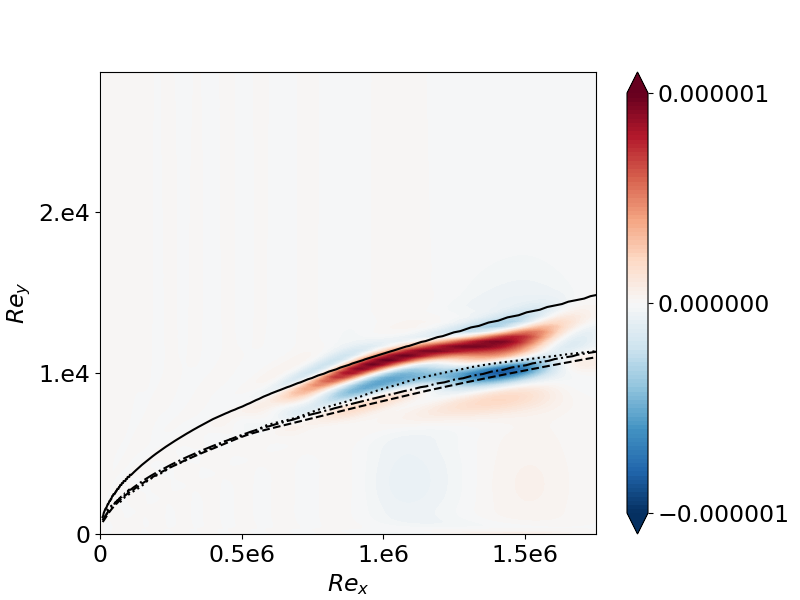}}
{\subfloat[\label{sensbft}]{\includegraphics[width=0.49\linewidth]{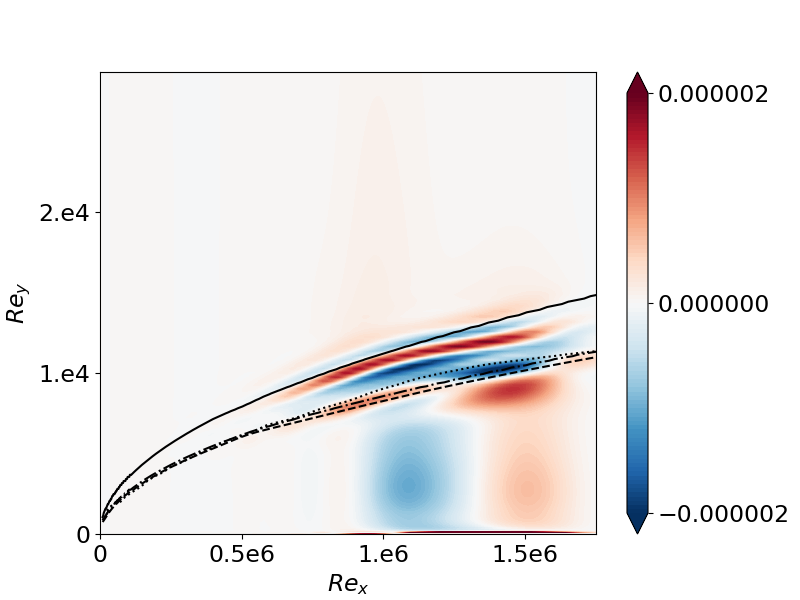}}}
\caption{Linear sensitivity of the optimal gain to base-flow modifications $\nabla_{\overline{\mathbf{q}}} \mu_0^2  /\mu_0^2$. Solid line represents the boundary layer thickness, dash-dotted line indicates the generalised inflection point, dashed line is the displacement thickness and dotted line is the critical layer. (a-b) Sensitivity of the first Mack mode. (c-d) Sensitivity of the second Mack mode. (a,c) Streamwise component of the momentum $\nabla_{\overline{\bm{{\rho u}}}} \mu_0^2  /\mu_0^2$ (similar to the divergence-free component). (b,d) Temperature component $\nabla_{\overline{\mathbf{{T}}}} \mu_0^2  /\mu_0^2$.} 
\label{sensbf}
\end{figure}

In Figure \ref{sensbf}(a,c), the sensitivity to local modifications of the streamwise base-flow momentum $\nabla_{\overline{\bm{{\rho u}}}} \mu_0^2$ is plotted for both Mack modes. The sensitivity of the first Mack mode extends longer in the streamwise direction than the second Mack mode as expected from the wavemaker support (overlap of optimal forcing and response from the resolvent analysis). These sensitivities are strongest in the region between the displacement thickness $\delta^*$ and the boundary layer thickness $\delta_{99}$ ($u(y=\delta_{99}) = 0.99 U_\infty$). This reminds the local results found for Fedorov's mode S \citep{park2019sensitivity} or for the incompressible Tollmien-Schlichting waves \citep{brandt2011effect}, the sensitivities $\nabla_{\overline{\bm{{\rho u}}}} \mu_0^2$ of the first and second Mack modes being negative at the critical layer $y_c$ ($u(y_c)=c_r$ with $c_r=Re(\omega/\alpha)$ the phase velocity of the mode) and positive in the vicinity above and below. \citet{park2019sensitivity} and \citet{guo2021sensitivity} showed that the mean shear modification contributes the most to the sensitivity resulting in a receptive region around the critical layer. Besides, from Figure \ref{sensbf}(c), it is seen that, for the second Mack mode, the generalised inflection line $y_{GIP}$ ($\partial [\rho \partial u/\partial y]/\partial y(y_{GIP}) = 0$) of the base-flow, where the optimal forcing and optimal entropy response are maximal, is close to the critical layer $y_c$ of the optimal response.

In Figure \ref{sensbf}(b,d), the sensitivity to local modifications of the base-flow temperature $\nabla_{\overline{\mathbf{{T}}}} \mu_0^2$ is plotted for both Mack modes. The gradient $\nabla_{\overline{\mathbf{{T}}}} \mu_0^2$ is computed from the components of $\nabla_{\overline{\mathbf{{q}}}} \mu_0^2$ through the linearisation of the definition of the total energy $E$ (chain rule):
\begin{equation}
\nabla_{\overline{\mathbf{{T}}}} \mu_0^2 = \frac{(\gamma -1)\gamma M^2}{\overline{\rho}} \left( \left( \frac{1}{2} (\overline{u}^2 + \overline{v}^2) - \overline{e} \right) \nabla_{\overline{\bm{{\rho}}}} \mu_0^2 - \overline{u} \nabla_{\overline{\bm{{\rho u}}}} \mu_0^2 - \overline{v} \nabla_{\overline{\bm{{\rho v}}}} \mu_0^2 + \nabla_{\overline{\bm{\rho E}}} \mu_0^2 \right).
\label{defgradt}
\end{equation}
The gradient $\nabla_{\overline{\mathbf{{T}}}} \mu_0^2$ is not only located between $y_c$ and $\delta_{99}$ anymore. A strong region of the gradient also exists closer to the wall for both Mack modes. In the case of the first Mack mode, the critical layer $y_{c}$ line separates a positive ($y<y_{c}$) and a negative ($y>y_{c}$)) region of the gradient, likely resulting once again from the mean shear distortion. For the second Mack mode, although the sensitivity $\nabla_{\overline{\mathbf{{T}}}} \mu_0^2$ is similar to $\nabla_{\overline{\bm{{\rho u}}}} \mu_0^2$, there is an additional region of large sensitivity to changes of the base-flow temperature close to the wall from $Re_x=10^6$ until the end of the domain. 

From the sensitivity to base-flow modifications $\nabla_{\overline{\mathbf{q}}} \mu_0^2$, the sensitivity of the optimal gain to a steady volume forcing term $\nabla_{\overline{\mathbf{f}}} \mu_0^2$ is deduced. The Chu energy norm of the sensitivity $\nabla_{\overline{\mathbf{f}}} \mu_0^2$, as well as its decomposition into different components, is reported in Table \ref{tab:chu}. The different contributions reflect those of the divergence-free base-flow gradient $\nabla_{\overline{\mathbf{q}}}^{df} \mu_0^2$ for both Mack modes, however the components induced by base-flow pressure variations is larger than those induced by entropy variations and their sum $\delta E_p^b + \delta E_s^b$ exhibit a magnitude similar to the streamwise kinetic energy component $\delta E_{c_x}^b$. For the streaks, the pressure component $\delta E_{p}^b$ is equivalent to the streamwise kinetic energy component $\delta E_{c_x}^b$ while its entropy component $\delta E_{s}^b$ is much smaller. The three instabilities display a sensitivity of similar magnitude.

\begin{figure}
\centering
\subfloat[\label{sensfum1}]{\includegraphics[width=0.49\textwidth]{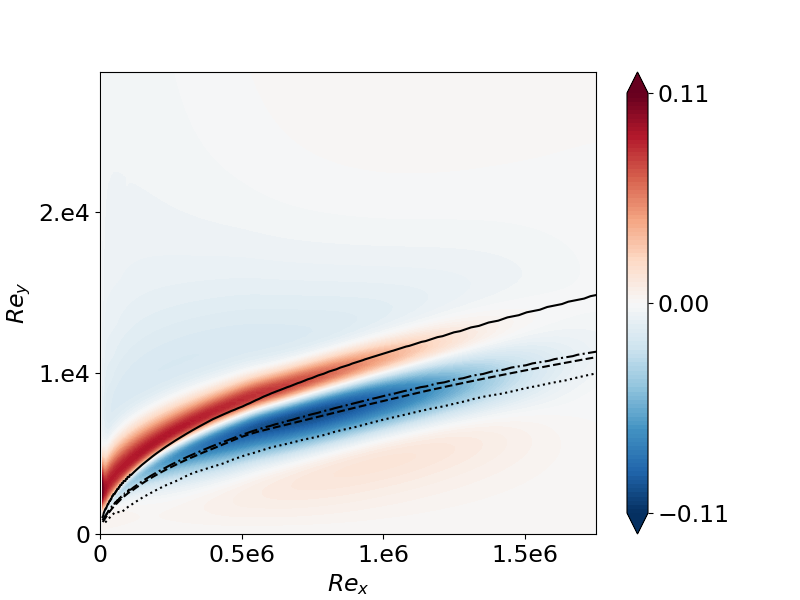}}
{\subfloat[\label{sensftm1}]{\includegraphics[width=0.49\linewidth]{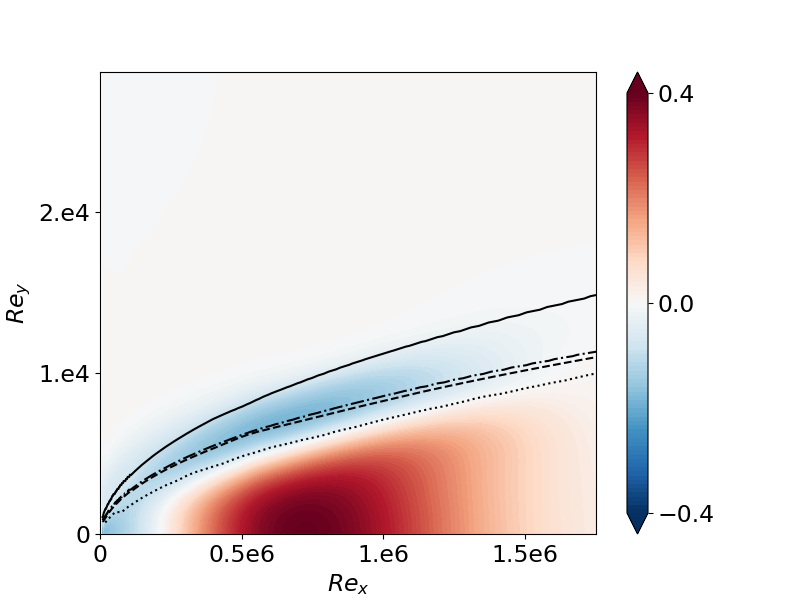}}}
{\subfloat[\label{sensfu}]{\includegraphics[width=0.49\textwidth]{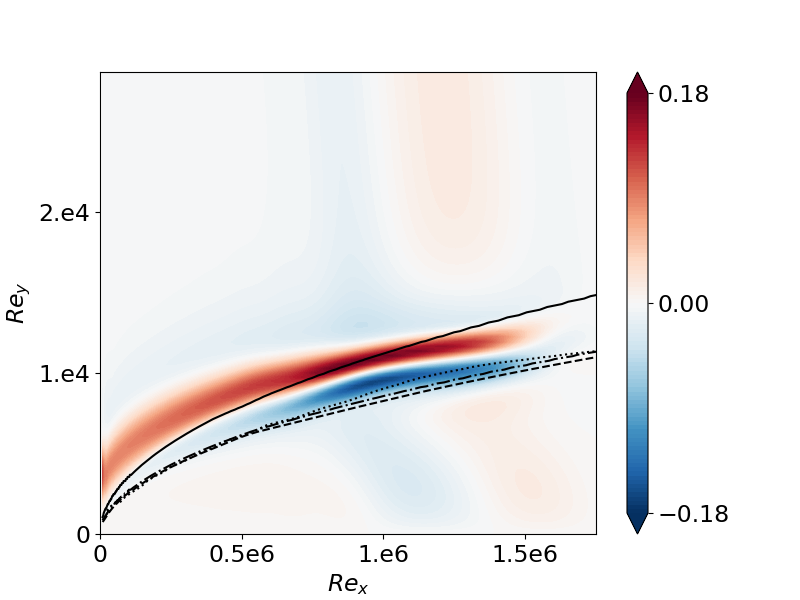}}}
{\subfloat[\label{sensft}]{\includegraphics[width=0.49\linewidth]{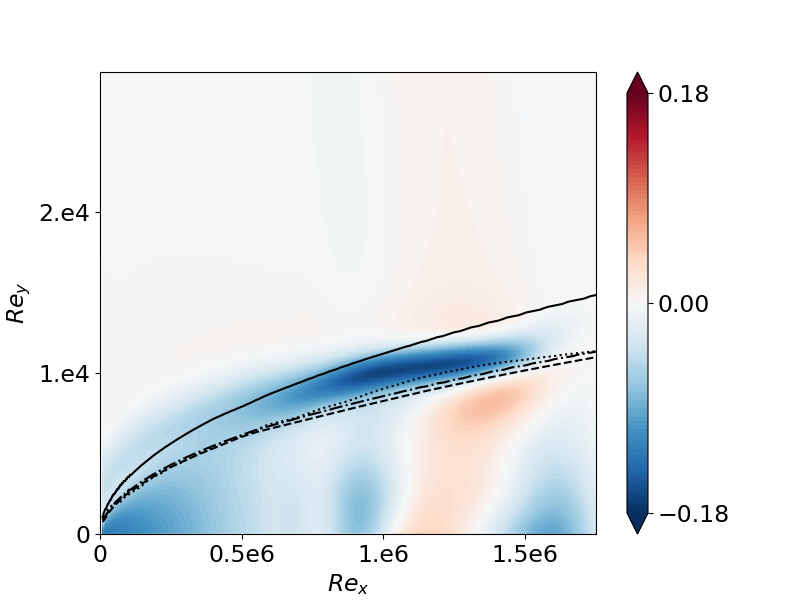}}}
\caption{Linear sensitivity of the optimal gain to steady forcing $\nabla_{\overline{\mathbf{f}}} \mu_0^2  /\mu_0^2$. Solid line represents the boundary layer thickness, dash-dotted line indicates the generalised inflection point, dashed line is the displacement thickness and dotted line is the critical layer. (a-b) Sensitivity of the first Mack mode. (c-d) Sensitivity of the second Mack mode. (a,c) Streamwise component of the momentum $\nabla_{\overline{\bm{f_{\rho u}}}} \mu_0^2  /\mu_0^2$. (b,d) Temperature component $\nabla_{\overline{\mathbf{f_{T}}}} \mu_0^2  /\mu_0^2$.}
\label{sensf}
\end{figure}

In Figure \ref{sensf}(a,c), the sensitivity to a steady streamwise momentum source term $\nabla_{\overline{\bm{f_{\rho u}}}} \mu_0^2$ is plotted for both Mack modes. The gradients for both Mack modes are very similar with a positive region around the boundary layer thickness $\delta_{99}$ and a negative region around the displacement thickness $\delta^*$. The mechanism below this gradient might be the local thickening of the boundary layer by the increase of the streamwise mean velocity above the edge and its decrease below.

Then, the sensitivity to a steady heating source term $\nabla_{\overline{\mathbf{f_{T}}}} \mu_0^2$, computed with Eq. \eqref{defgradt} by replacing the components of $\nabla_{\overline{{\mathbf{q}}}} \mu_0^2$ by $\nabla_{\overline{\mathbf{f}}} \mu_0^2$, is plotted in Figure \ref{sensf}(b,d) for both Mack modes. As for $\nabla_{\overline{\mathbf{{T}}}} \mu_0^2$, the sensitivity $\nabla_{\overline{\mathbf{f_{T}}}} \mu_0^2$ has strong distortions inside the boundary layer. For the first Mack mode, the gradient is mainly positive from the wall up to the critical layer ($y<y_c$) except in the region close to the leading edge. However, in the case of the second Mack mode, the largest region of sensitivity (negative gradient) is between the displacement thickness and the boundary layer thickness. Inside the boundary layer, closer to the wall, the gradient varies a lot in the streamwise direction for the second Mack mode. From an overall perspective, a steady heating source term has an opposite effect inside the boundary layer between the first and the second Mack modes. This recalls the well-known effect of stabilisation by cooling for the first Mack mode and by heating for the second Mack mode \citep{mack1993effect}.

\subsection{Sensitivity to wall boundary control} \label{sec:sens3D}

After the computation of the sensitivity to base-flow modifications and steady forcing in \S \ref{sec:sensbf}, the sensitivity of the streaks, first and second Mack modes to two types of modifications of the wall boundary condition are analysed:
\begin{itemize}
    \item Small amplitude wall-normal blowing/suction $\delta v_w(x)$ at the surface of the flat plate (\S \ref{sec:sens2Dblow}),
    \item Small amplitude heat flux $\delta \phi_w(x)$ at the surface of the flat plate (\S \ref{sec:sens2Dheat}).
\end{itemize}
In both cases, the identity $\mathbf{Q}_p = \mathbf{I}$ is chosen for the norm associated to the parameter spaces $ \mathbf{p}=v_w$ and $ \mathbf{p}=\phi_w $ as the discretisation along the $x$-direction is uniform.

\begin{figure}
\centering
\subfloat[\label{sens3d2vw}]{\includegraphics[width=0.5\textwidth]{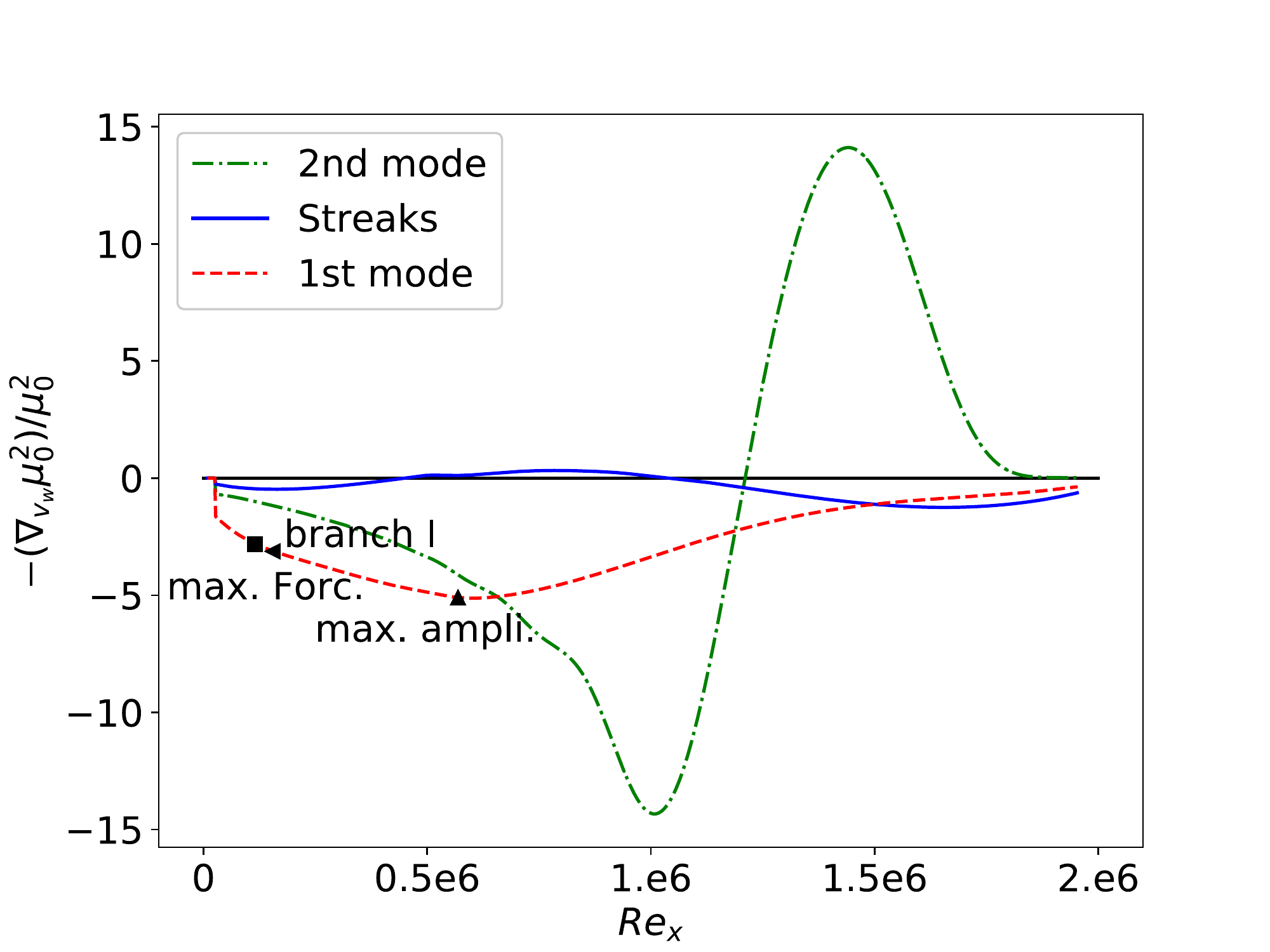}}
{\subfloat[\label{sens3d2flux}]{\includegraphics[width=0.5\linewidth]{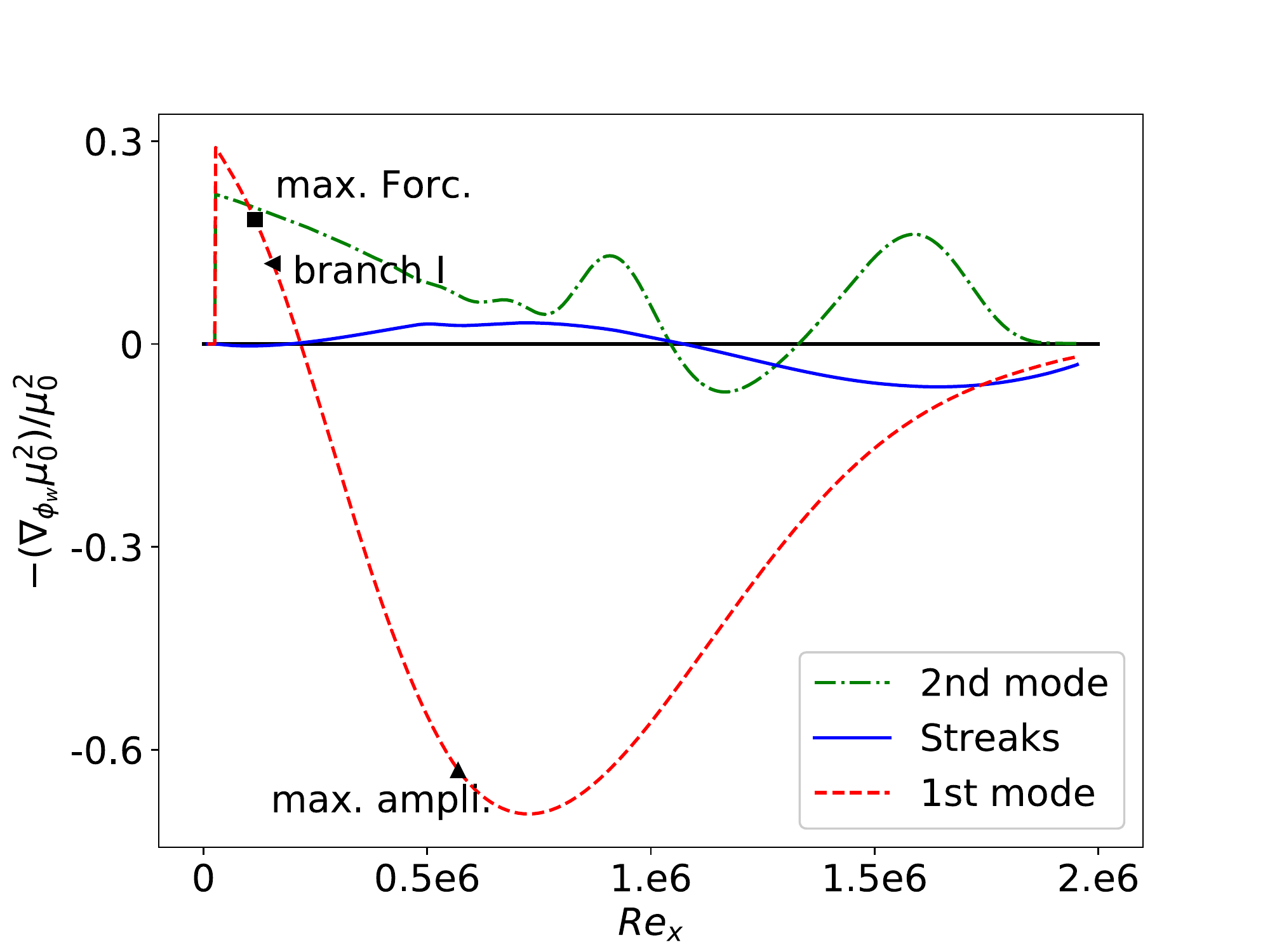}}}
\caption{Optimal wall control profiles to damp the different instabilities i.e. opposite of the sensitivity of the optimal gain for each instability: $-\delta \mathbf{p}^m$. (a) Optimal wall-velocity profile $-\delta \mathbf{v}_w $. (b) Optimal wall heat flux profile $-\delta \bm{\phi}_w $.}
\label{sens3d3}
\end{figure}

In the following, we discuss the opposite of the sensitivity $-\delta \mathbf{p}^m $ (see \eqref{eq:gradp}), which corresponds to the optimal wall profile to be prescribed at the surface in order to mitigate the instability. This profile is plotted for blowing/suction control in Figure \ref{sens3d2vw} and for heat flux control in Figure \ref{sens3d2flux}. 
In Figure \ref{sens3d3}, the marker "max. Forc." refers to the maximal optimal forcing location of the first Mack mode, "branch I" to the streamwise location of branch I of the local mode linked to the first Mack mode and "max. ampli." to the location of the maximal amplification rate ($\max(-\alpha_i)$) of the local mode.

Both wall velocity and heat flux control gradients highlight that the streaks are not sensitive to boundary control in comparison with the Mack modes. Furthermore, the profile to damp the streaks has a similar behaviour for both types of wall-control i.e negative in the downstream part of the domain where the optimal response of the streaks lies. In Figure \ref{sens3d2vw}, the second Mack mode is confirmed to be the most sensitive instability to wall velocity control. Furthermore, it is the only instability among the three which is stabilised by steady wall-blowing (downstream $Re_x = 1.2 \times 10^6$). The gradient for the second Mack mode is studied in detail in \S \ref{sec:sens2Dblow}. Optimal suction actuator locations to stabilise each instability are different: $Re_x = 0.6 \times 10^6$ for the first Mack mode (close to the location of the maximal amplification rate of the local mode), $Re_x = 1 \times 10^6$ for the second Mack mode and $Re_x = 1.7 \times 10^6$ for the streaks. However, a suction actuator located anywhere upstream $Re_x = 1.2 \times 10^6$ would damp both Mack modes without affecting much the streaks. The effect of applying the second Mack mode gradient on the stability of the other instabilities is investigated in \S \ref{sec:sens3Dapplied}, where the design of an optimal wall-control actuator is attempted.

In figure \ref{sens3d2flux}, it is seen that the first Mack mode is the most sensitive to wall heat flux control. The profile to damp the optimal gain of the first Mack mode has one heating zone upstream until $Re_x = 0.21 \times 10^6$ (located slightly downstream of branch I) and one cooling region downstream with the largest sensitivity close to the location of the maximal amplification rate of the local mode. The variations of the optimal wall heat flux profile to damp the second Mack mode are more complex and are studied in detail in \S \ref{sec:sens2Dheat}. By comparing both Mack modes' wall heat flux profiles, they overall display an opposite behaviour with respect to wall heat flux changes \citep{mack1993effect}, however, in some streamwise regions, they follow similar trends. First, at the leading edge, upstream $Re_x = 0.21 \times 10^6$, heating the wall would damp both Mack modes and secondly, downstream, between $Re_x = 1.045 \times 10^6$ and $Re_x = 1.33 \times 10^6$, cooling would produce the same effect. The effect of the application of the first Mack mode gradient on the stability of the other instabilities is investigated in \S \ref{sec:sens3Dapplied}. Furthermore, based on the zones where the gradients' sign matches between the first and second Mack modes, a control that damps both instabilities will also be discussed in this section.

One may remark that the optimal wall control profiles given by Figure \ref{sens3d3} may have been partially predicted from the sensitivity $\nabla_{\mathbf{f}} \mu_0^2/\mu_0^2$ in Figure \ref{sensf}. Indeed, by looking at the sign of the gradient close to the wall, one may see that the gradient $\nabla_{\bm{f_{\rho u}}} \mu_0^2/\mu_0^2$ stays positive all along $Re_x$ for the first Mack mode while it becomes negative downstream $Re_x=1 \times 10^6$ for the second Mack mode, inducing similar trends for the wall gradient $\nabla_{\mathbf{v}_w} \mu_0^2$. Comparing the gradients $\nabla_{\bm{f_{T}}} \mu_0^2/\mu_0^2$ and $\nabla_{\bm{\phi}_w} \mu_0^2$ leads to similar observations.

From the optimal wall control profile $\delta \mathbf{p}^m$, one may compute the linear base-flow variations $\delta \overline{\mathbf{q}}^m$ induced by $\delta \mathbf{p}^m$ using Eq. \eqref{indbsfvar}. This informs us on the base-flow variations induced by the wall-control, which in turn then change the optimal gain of the instability.

\begin{figure}
\centering
\subfloat[\label{ubsfvwm1}]{\includegraphics[width=0.49\textwidth]{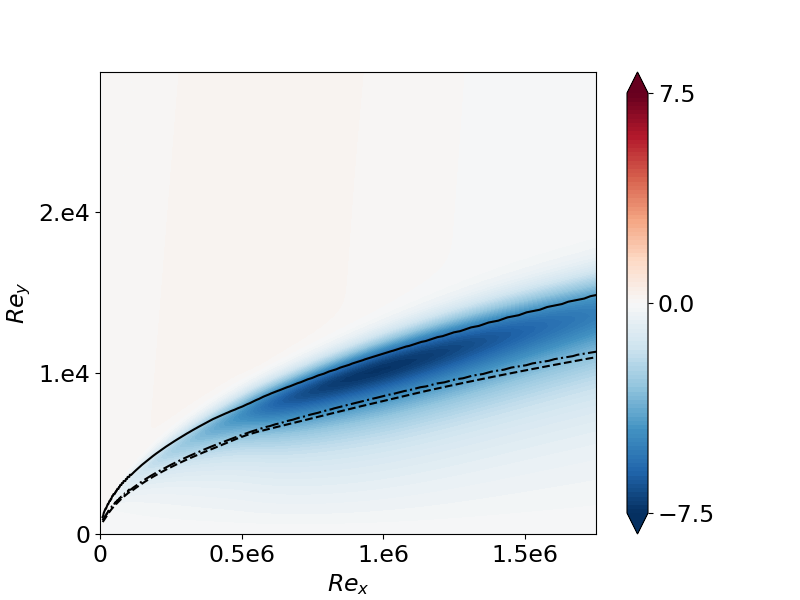}}
{\subfloat[\label{Tbsfvwm1}]{\includegraphics[width=0.49\linewidth]{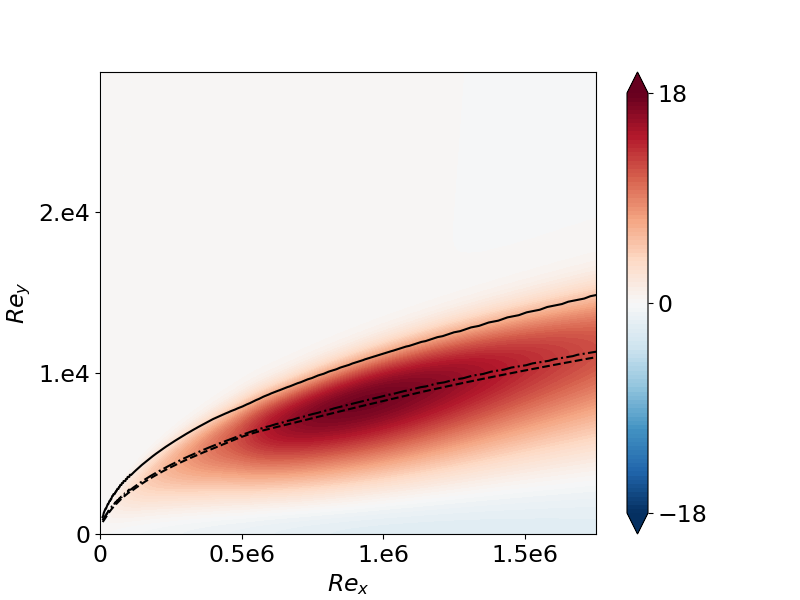}}}
{\subfloat[\label{ubsfvw}]{\includegraphics[width=0.49\textwidth]{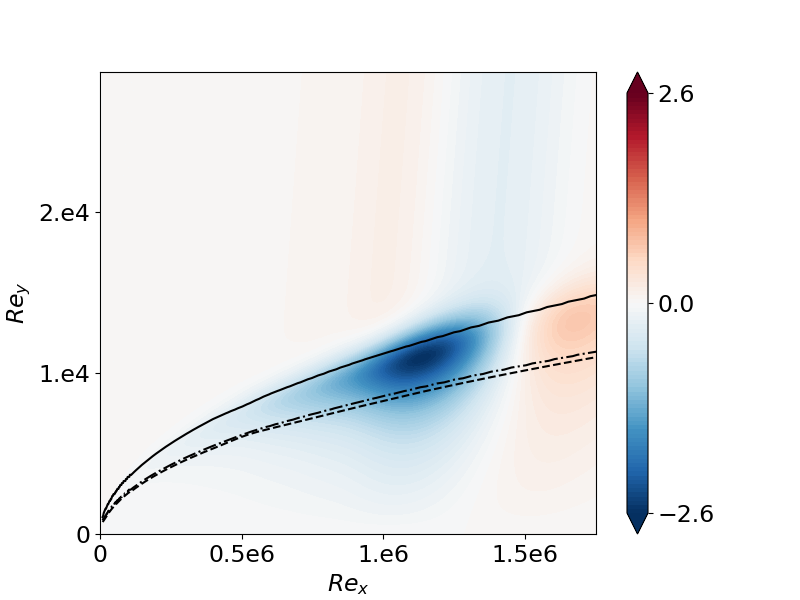}}}
{\subfloat[\label{Tbsfvw}]{\includegraphics[width=0.49\linewidth]{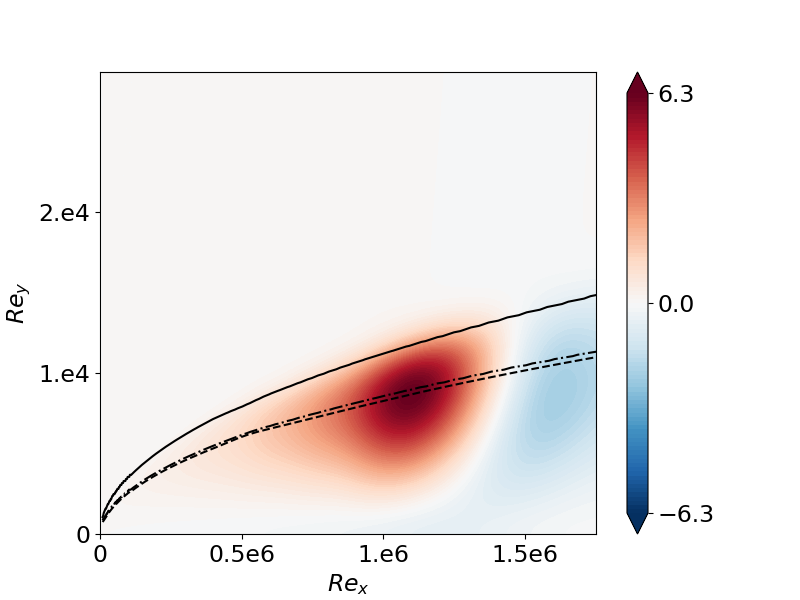}}}
\caption{Linear base-flow variations $\delta \mathbf{\overline{q}}^m$ induced by the optimal wall-velocity control $\delta \mathbf{p}^m=\delta \mathbf{v}_w$. Solid line represents the boundary layer thickness, dash-dotted line indicates the generalised inflection point and dashed line is the displacement thickness. (a-b) First Mack mode ($\delta \mathbf{v}_w$ corresponds to blowing B). (c-d) Second Mack mode ($\delta \mathbf{v}_w$ corresponds to blowing/suction BS). (a,c) Streamwise component of the momentum $\delta \bm{\overline{\rho u}}^m$. (b,d) Temperature component $\delta \mathbf{\overline{T}}^m$. }
\label{bsfvw}
\end{figure}

First, the optimal wall-normal velocity control $\delta \mathbf{v}_w$ (which consists in blowing) of the first Mack mode induces a decrease of streamwise momentum (Figure \ref{ubsfvwm1}) between the boundary layer thickness $\delta_{99}$ and the displacement thickness $\delta^*$ and a temperature increase (Figure \ref{Tbsfvwm1}) close to the displacement thickness leading to the thickening of the boundary layer. The patterns are quite constant in the streamwise direction, which reflects the quasi-uniform blowing profile $\delta \mathbf{v}_w$ of the first Mack mode. The variations of temperature (positive here) are opposite to those of streamwise momentum, showing that a deceleration of the flow in the boundary layer goes with an increase of its temperature. Second, as the wall-normal velocity control $\delta \mathbf{v}_w$ of the second Mack mode involves blowing and suction along $Re_x$, the base-flow variations in streamwise momentum (Figure \ref{ubsfvw}) and temperature (Figure \ref{Tbsfvw}) exhibit similar streamwise wavy patterns, since these fields are affected both by the induced advection linked to the blowing/suction profile. By comparing the variations $\delta \mathbf{\overline{q}}^m$ induced by wall-normal velocity control $\delta \mathbf{v}_w$ (Figure \ref{bsfvw}) with the gradient $\nabla_{\overline{\mathbf{q}}} \mu_0^2  /\mu_0^2$ (Figure \ref{sensbf}), which is the optimal base-flow variation $\delta \mathbf{\overline{q}}^b$, one may notice how wall control efficiently acts in the sensitive regions. The wall-normal velocity control $\delta \mathbf{v}_w$ indeed manages to act in the most sensitive region of the momentum component of the base-flow, it however fails to induce a shear, which is optimal to control the Mack modes.

\begin{figure}
\centering
\subfloat[\label{ubsffluxm1}]{\includegraphics[width=0.49\textwidth]{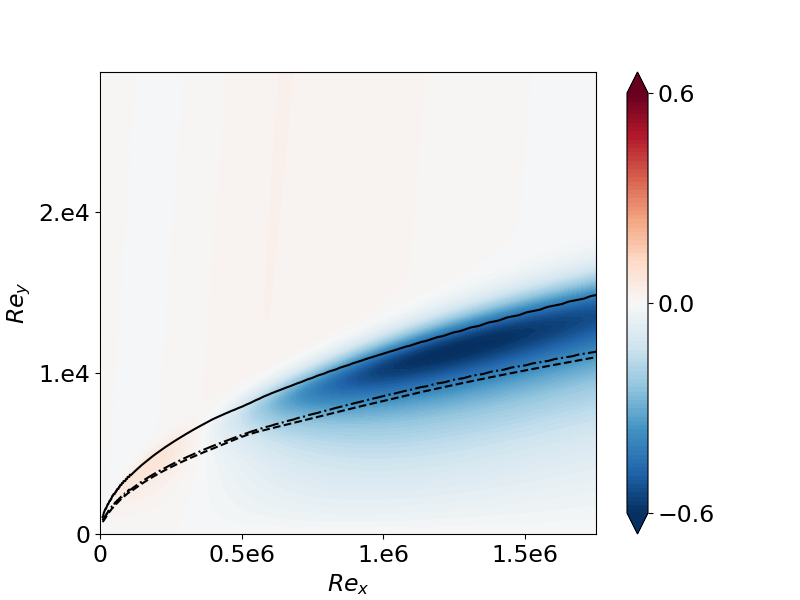}}
{\subfloat[\label{Tbsffluxm1}]{\includegraphics[width=0.49\linewidth]{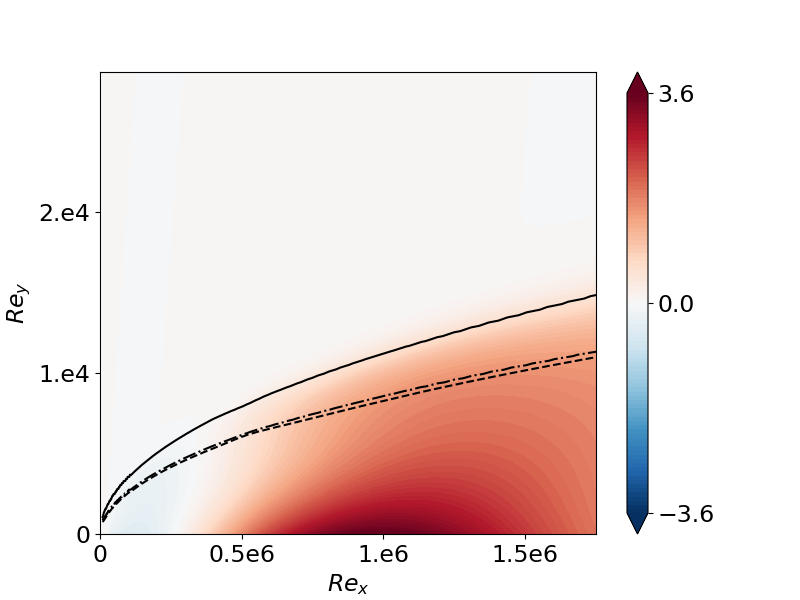}}}
{\subfloat[\label{ubsfflux}]{\includegraphics[width=0.49\textwidth]{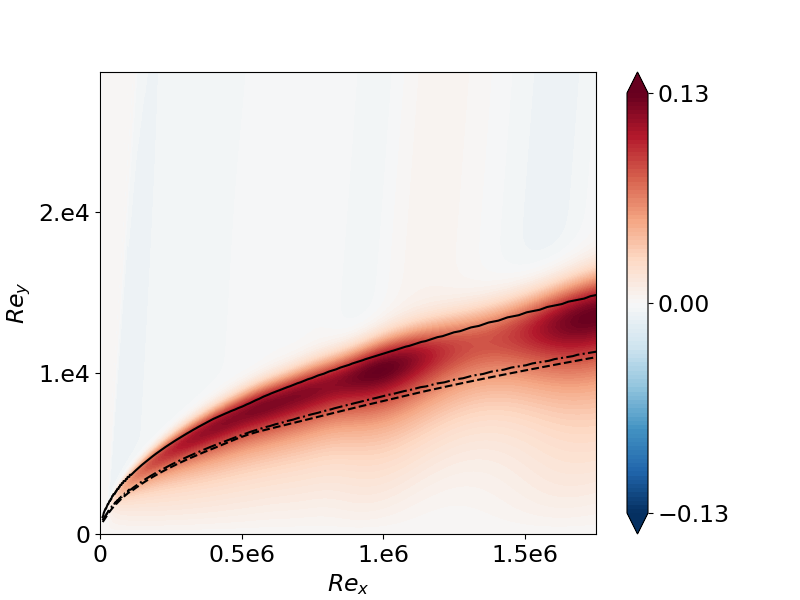}}}
{\subfloat[\label{Tbsfflux}]{\includegraphics[width=0.49\linewidth]{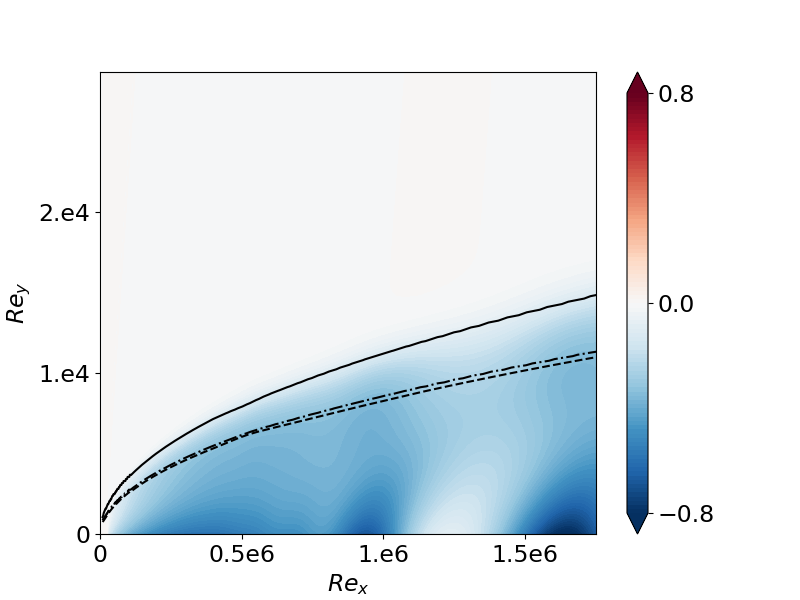}}}
\caption{Linear base-flow variations $\delta \mathbf{\overline{q}}^m$ induced by the optimal wall heat flux control $\delta \mathbf{p}^m=\delta \bm{\phi}_w$. Solid line represents the boundary layer thickness, dash-dotted line indicates the generalised inflection point and dashed line is the displacement thickness. (a-b) First Mack mode ($\delta \mathbf{\phi}_w$ corresponds to cooling/heating cH). (c-d) Second Mack mode ($\delta \mathbf{\phi}_w$ corresponds to cooling/heating/cooling ChC). (a,c) Streamwise component of the momentum $\delta \bm{\overline{\rho u}}^m$. (b,d) Temperature component $\delta \mathbf{\overline{T}}^m$.}
\label{bsfflux}
\end{figure}

Secondly, the wall heat flux control $\delta \bm{\phi}_w$ of both Mack modes induces a variation of streamwise momentum (Figures \ref{ubsffluxm1} and \ref{ubsfflux}) between the boundary layer thickness $\delta_{99}$ and the displacement thickness $\delta^*$, acting in the same region as the wall-normal velocity control $\delta \mathbf{v}_w$. The variations of temperature induced by the wall heat flux control $\delta \bm{\phi}_w$ of both Mack modes reproduce the  optimal base-flow variations $\delta \mathbf{\overline{q}}^b$ within the boundary layer (Figures \ref{Tbsffluxm1} and \ref{Tbsfflux} compared with \ref{sensftm1} and \ref{sensft}). Because of its application at the wall, the control is maximal there, which favours the first Mack mode where the largest sensitivity region is at the wall, while the largest sensitivity region of the second Mack mode is above the critical layer. This may explain why the wall heat flux control has a larger effect on the first Mack mode (a factor 3 in terms of magnitude with respect to the second Mack mode).

Finally, the components of the linear gain variation induced by wall boundary control are computed for both Mack modes and reported in Table \ref{tab:chuvw}. As it will be confirmed in the next section for the second Mack mode, the variation of gain induced by the modification of the Jacobian is small in comparison with the one induced by the base-flow modification for both Mack modes. By analysing the contributions driven by the base-flow modification, the pressure component is always the smallest. The streamwise momentum and entropy components represent most of the energy, except for the gain variation of the first Mack mode induced by the optimal heat flux profile and the streaks where the wall-normal component contributes as much as the entropy in the energy. The streamwise momentum energy change always exhibits an effect opposite to the one of the entropy for both Mack modes. 
Finally, in nearly all cases, the linear gain variation is driven by the streamwise momentum base-flow change, except for the second Mack mode with heat flux control where it is the entropy modification of the base-flow. This raises the possibility of simultaneous control of both Mack modes if one manages to design a heat flux actuator which triggers variations of same sign in the streamwise momentum and entropy components (the optimal profiles discussed here trigger opposite effects in these two components).

\begin{table}
    \begin{center}
\def~{\hphantom{0}}
    \begin{tabular}{c|c|ccccc|c}
& & $ \delta E_{\mathbf{A}}^m$ & $ \delta E_{c_x}^m$ & $ \delta E_{c_y}^m$ &  $\delta E_p^m $ &  $ \delta E_s^m$ &  $ \delta \mu_i^2/\mu_i^2$ \\ 
     \hline
1st Mack & $\delta \mathbf{v}_w$ (B) & $0.058$ & $476.1$ & $-34.4$ & $1.8$ & $-342.2$ & $101.4$\\
mode & $\delta \bm{\phi}_w$ (cH) &$-0.006$ & $34.1$ & $8.3$ & $-0.11$ & $-8.9$ & $33.4$ \\
     \hline
2nd Mack &$\delta \mathbf{v}_w$ (BS) & $-1.9$ & $133.6$ & $7.6$ & $-1.2$ & $-43.2$ & $94.9$ \\
mode & $\delta \bm{\phi}_w $ (ChC) &$-0.006$ & $-2.0$ & $0.15$ & $-0.069$ & $3.0$ & $1.1$ \\ 
     \hline
Streaks &$\delta \mathbf{v}_w$ (bsB) & $0.01$ & $38.1$ & $-31.7$ & $-0.2$ & $35.3$ & $41.5$ \\
 & $\delta \bm{\phi}_w $ (hCH) &$-0.001$ & $1.1$ & $-1.3$ & $-0.034$ & $2.4$ & $2.2$ \\ 
     \hline
    \end{tabular}
    \caption{Chu energy norm components of the linear gain variation induced by the wall boundary control. BS means Blowing (large region)-Suction (large region), cH means cooling (small region)-Heating (large region). The last column $ \delta \mu_i^2/\mu_i^2$ represents the sum of the Chu energy norm components (Eq. \eqref{deltamum},\eqref{chusm}).}
    \label{tab:chuvw}
    \end{center}
\end{table}

\subsubsection{Sensitivity of the second Mack mode to steady wall blowing} \label{sec:sens2Dblow}

The variations of the optimal wall-normal velocity profile to damp the second Mack mode is analysed in further details. Then, linear predictions are compared to nonlinear computations for increasing blowing amplitudes to assess the predicting capabilities of linear gradients.

\begin{figure}
\centering
\includegraphics[width=0.5\textwidth]{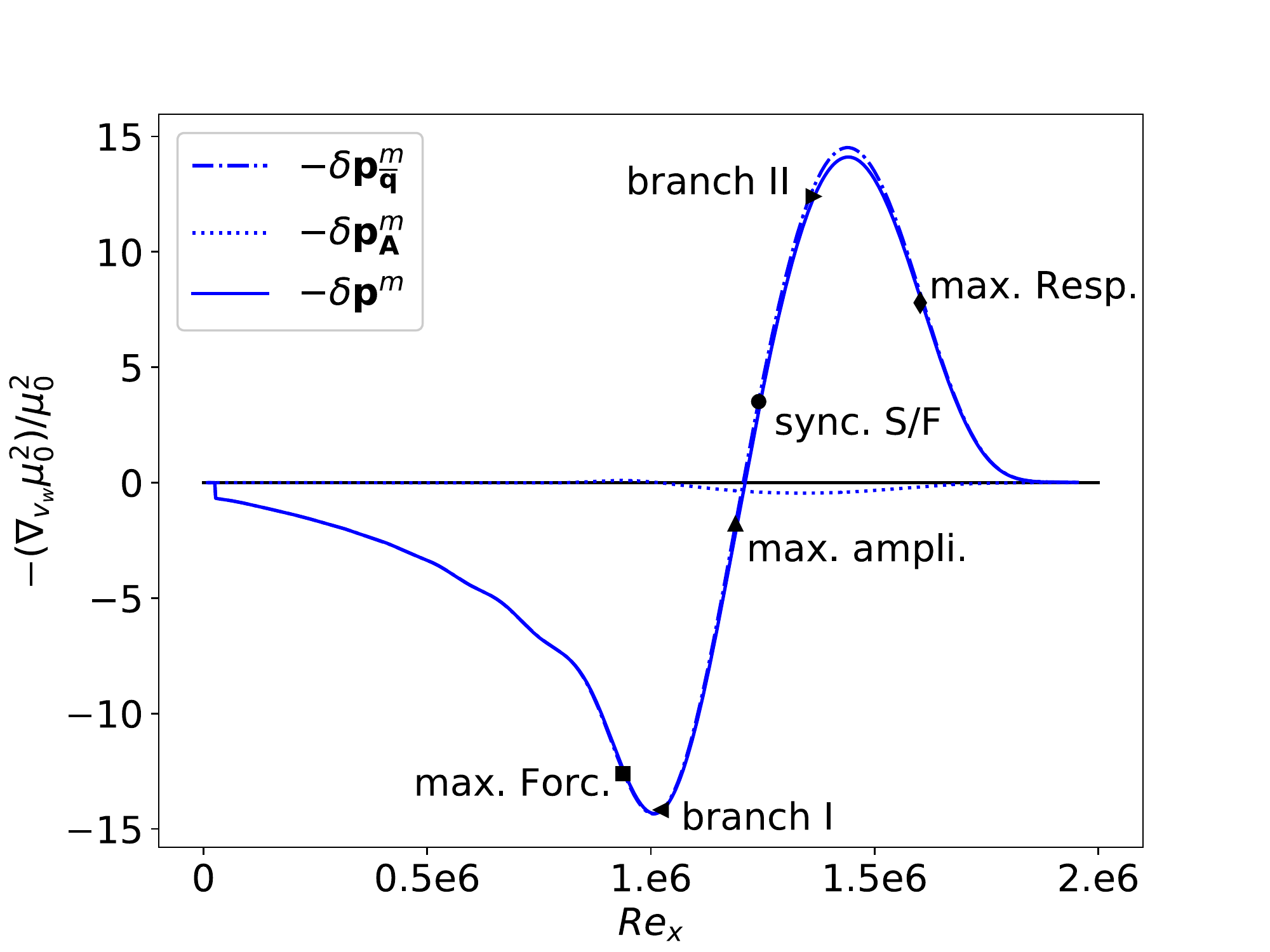}
\caption{Optimal wall-velocity profile to damp the second Mack mode i.e. opposite of the sensitivity of the optimal gain of the second Mack mode to wall blowing $ -\delta \mathbf{p}^{m} = - \delta \mathbf{v}_w$. Base-flow component $- \delta \mathbf{p}_{\overline{\mathbf{q}}}^m $ and Jacobian component $ -\delta \mathbf{p}_{\mathbf{A}}^m $ are also indicated, the latter being almost zero.}
\label{sens2dvwbis}
\end{figure}

First, the optimal steady wall-normal velocity profile to damp the second Mack mode and its decomposition into various components is plotted in Figure \ref{sens2dvwbis}. As seen from the integral energy in Table \ref{tab:chuvw} the gradient $\nabla_{\mathbf{v}_w}\mu_0^2$ is mainly produced from the sensitivity term due to the base-flow variation as the term due to the Jacobian variation is of smaller amplitude. The low impact of the Jacobian modifications on the gradient is due to the fact that the wall-normal velocity prescribed at the wall only appears in the energy equation in the linearised wall boundary condition.

The total gradient may be split into two main parts: suction in the upstream region up to $Re_x \approx 1.2 \times 10^6$ and blowing downstream. In Figure \ref{sens2dvwbis}, the markers "max. Forc" and "max. Resp" indicate respectively the locations of maximal optimal forcing and response for the second Mack modes (peak values from Figure \ref{dfdchu}). The markers "branch I" and "branch II" refer respectively to the beginning and end in the downstream direction of the unstable region of the local mode S. The marker "max. ampli." denotes the location of the largest negative amplification rate of the local mode S. The marker "sync. S/F" indicates the synchronisation point which is the streamwise location where the phase velocities of mode S and F are equal. 
Furthermore, we give evidence in appendix \ref{sec:appsize} that the location of the above points within the gradients (for both types of wall-control) remain relatively constant for different frequencies. 

In the following, to fix the control amplitude, we introduce the momentum coefficient $C_\theta$ as the ratio between the momentum injected at the wall and the free-stream momentum deficit,
\begin{equation}
C_\theta = \frac{ \int_{y=0} |(\rho v^2)'| \,dx}{\int_{x=x_{out}} (\rho_\infty U_\infty^2 - \rho u^2)\,dy },
\label{cmu}
\end{equation}
with $(\rho v^2)'$ the difference of wall-normal momentum $\rho v^2$ between controlled and uncontrolled base-flow.
To understand the effect of the control profile shown in Figure \ref{sens2dvwbis}, we compare local stability (spatial LST) analysis results applied on the uncontrolled base-flow and on the controlled base-flow with the full stabilising gradient at $C_\theta = 3.2 \times 10^{-6}$. \citet{fong2014numerical} indeed showed that the location of a roughness element upstream or downstream the synchronisation point results in opposite stabilisation effects. We find similar trends as the intersection of the phase velocity of Fedorov's mode F and S is close to the point where the gradient is null in Figure \ref{sens2dvwbis}. We also highlight that the locations of branch I and branch II of mode S are respectively close to the maxima for suction and for blowing.

\begin{figure}
\centering
\subfloat[\label{cr2}]{\includegraphics[width=0.5\textwidth]{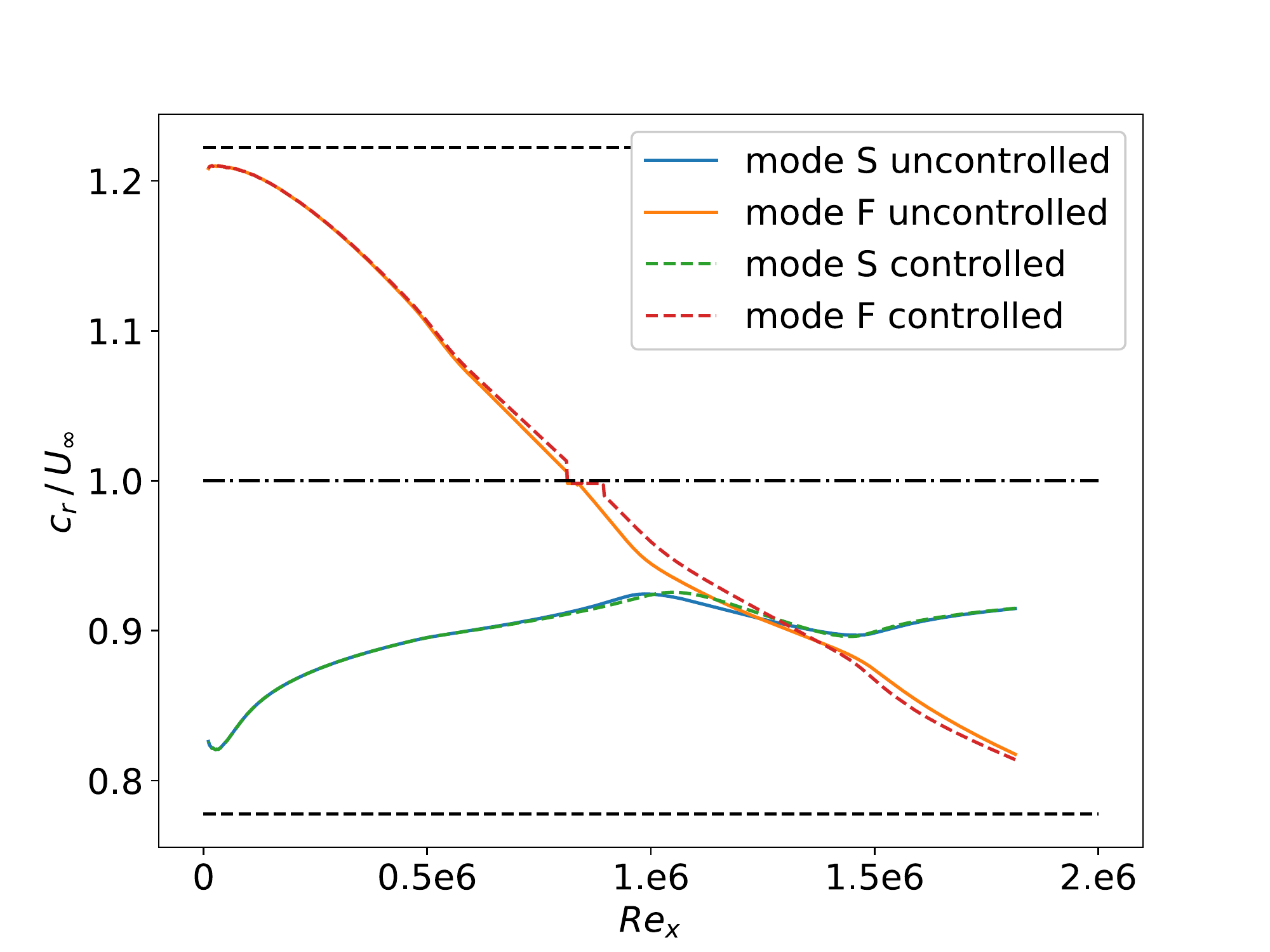}}
{\subfloat[\label{alphai2}]{\includegraphics[width=0.5\linewidth]{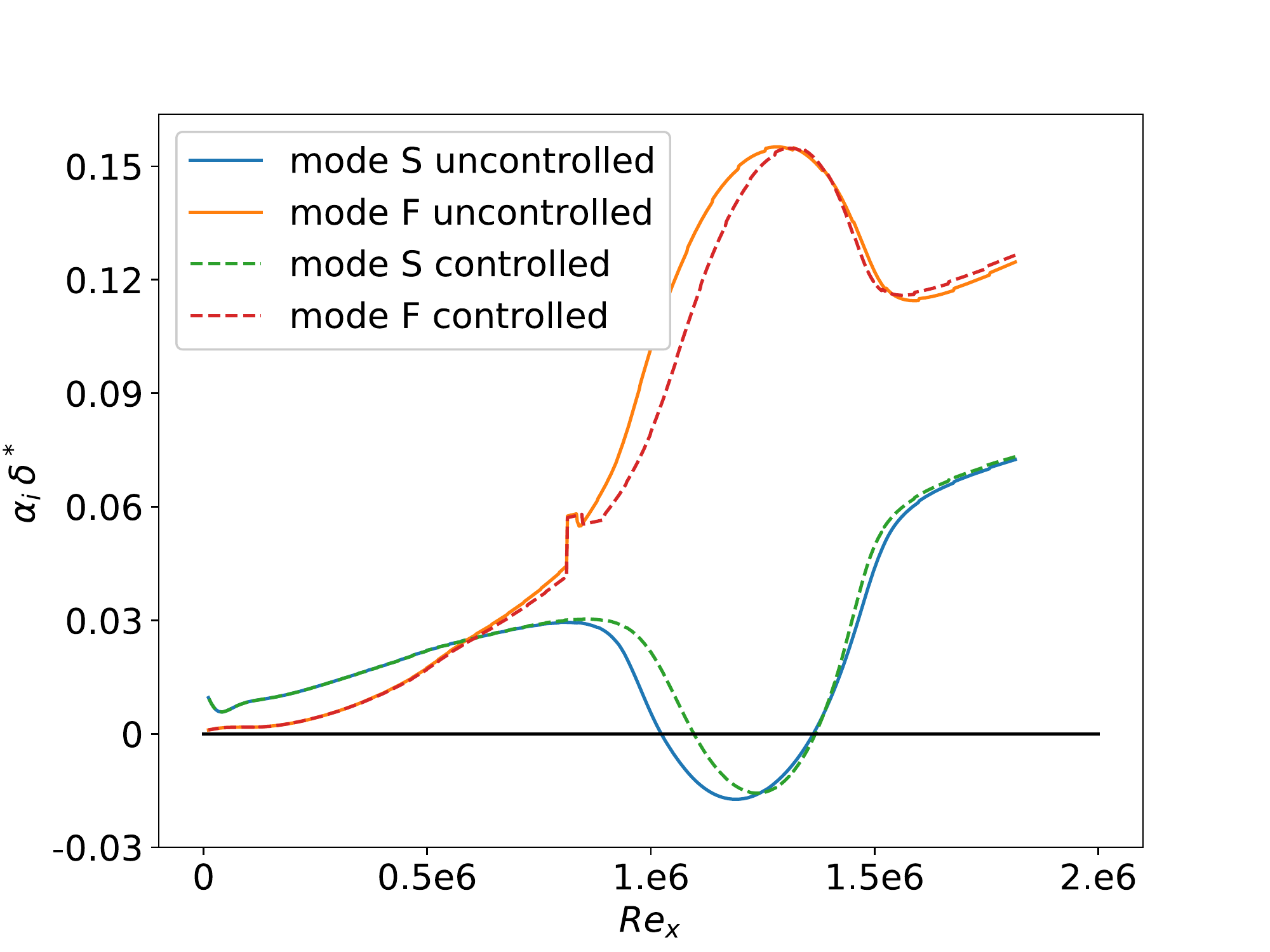}}}
\caption{Local stability analysis of an uncontrolled base-flow (solid lines) and a controlled base-flow (dashed lines) with the optimal wall-velocity profile $-\nabla_{\mathbf{v}_w} \mu_0^2  /\mu_0^2$ at $C_\theta = 3.2 \times 10^{-6}$. (a) Phase velocity $c_r/U_\infty$. Dashed lines denote respectively from top to down the phase velocities $1 + 1/M_\infty$, $1$ and $1-1/M_\infty$. (b) Amplification rate $\alpha_i  \delta^*$.}
\label{LSTblow2}
\end{figure}

Phase velocities and amplification rates of modes S and F are shown in Figure \ref{LSTblow2} with and without the application of the stabilising wall profile. The phase velocity of mode S remains quite similar in both cases while the phase velocity of mode F of the controlled base-flow slightly deviates from the original in the control region. This results in a shorter synchronisation region between mode F and S leading to a shorter unstable region for the amplification rate of mode S induced by the modification of branch I location. \citet{zhao2018numerical} noticed the same behaviour for the phase velocity in the case of heating and cooling strips control.

In the following, we compare the prediction obtained by the gradient approach and the exact results obtained by computing the modified base-flow and the associated resolvent gains.

\begin{figure}
\centering
\subfloat[\label{gain2dvw}]{\includegraphics[width=0.5\textwidth]{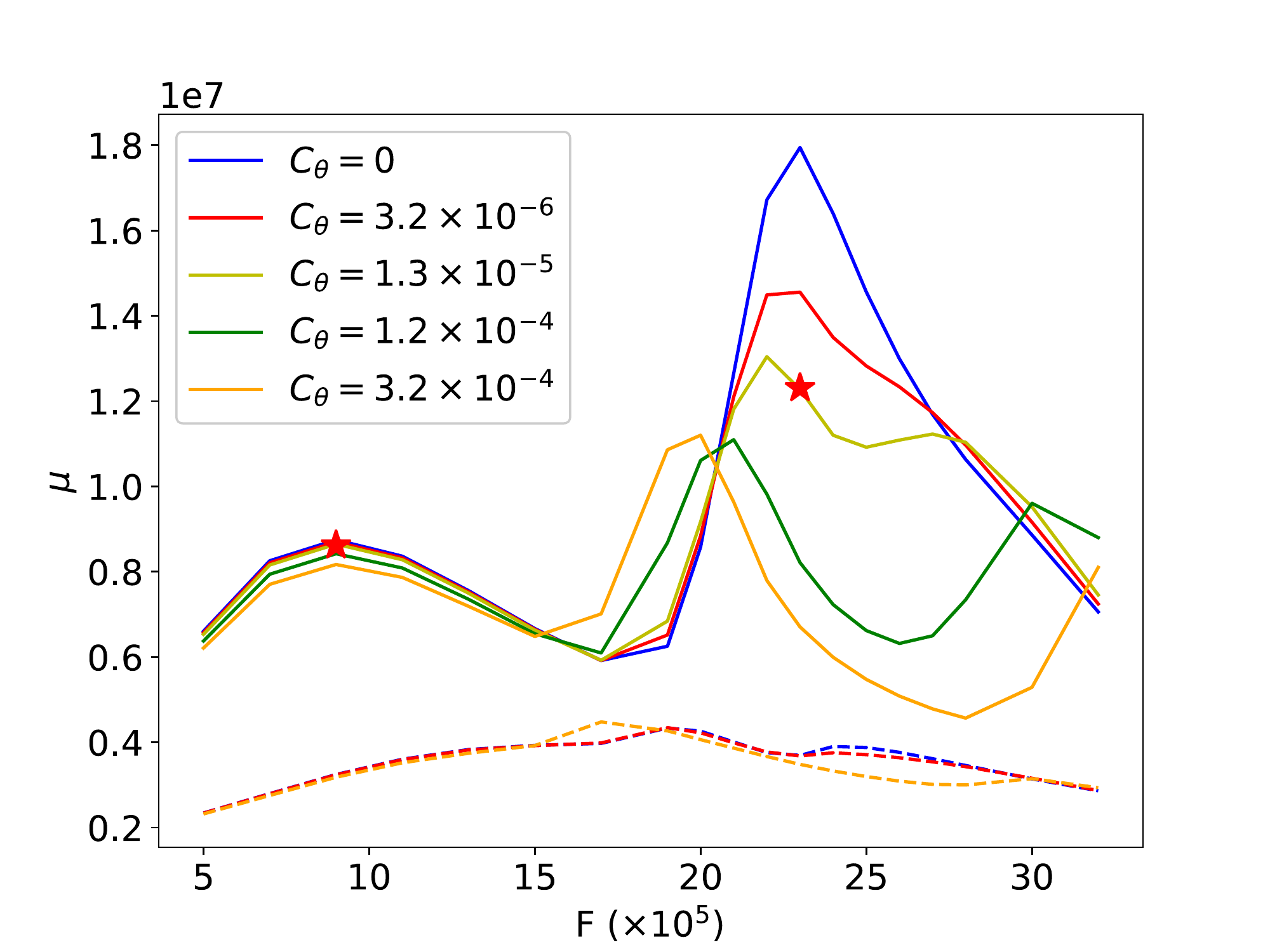}}
{\subfloat[\label{dfdchuvw}]{\includegraphics[width=0.5\linewidth]{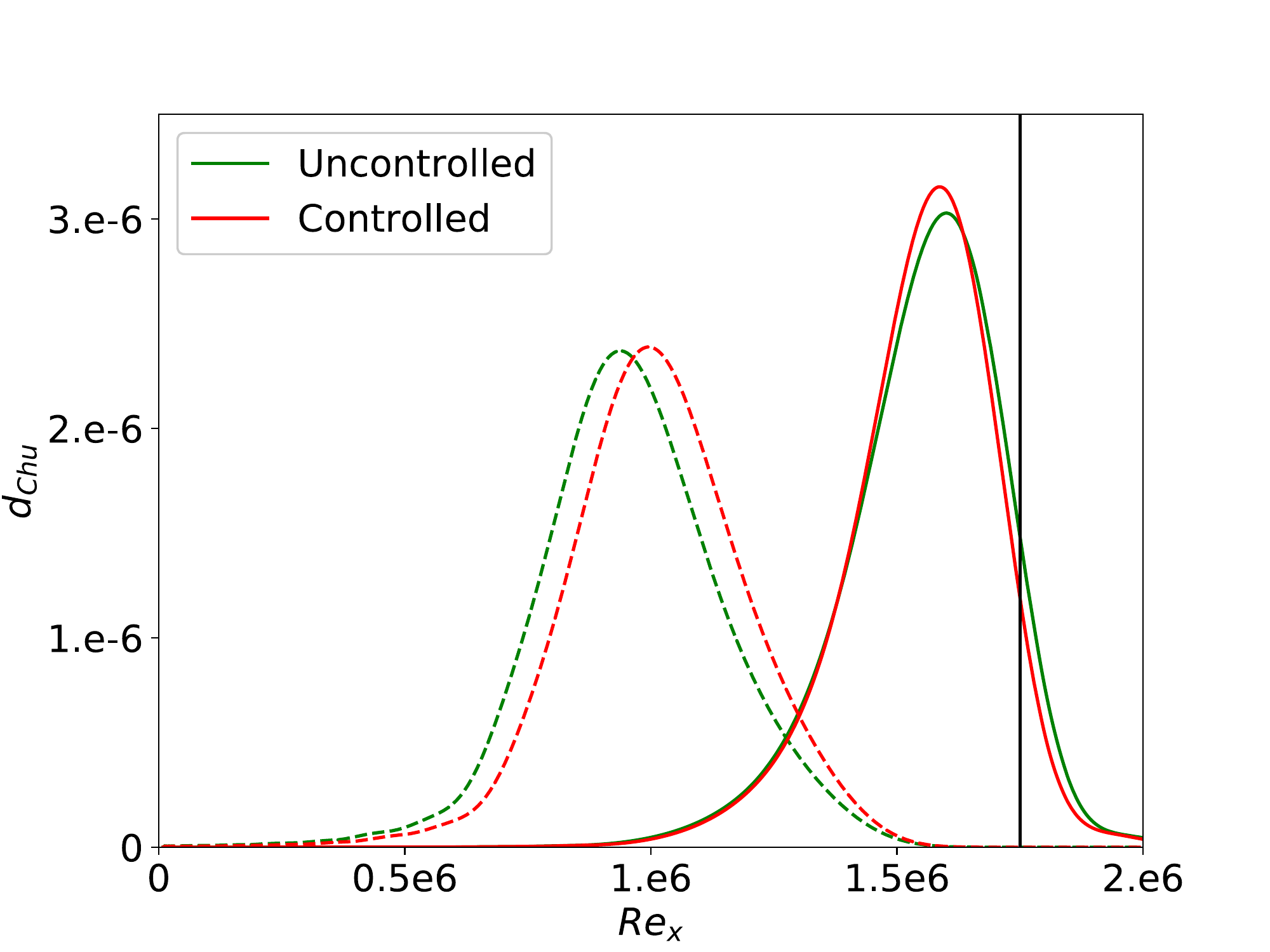}}}
\caption{Resolvent analysis for different base-flows controlled with the optimal wall-velocity profile $-\nabla_{\mathbf{v}_w} \mu_0^2 /\mu_0^2$. (a) Optimal gain $\mu$ for $ \beta=0 $ with respect to the frequency $F$ for $C_\theta=0$ (blue),  $C_\theta=3.2 \times 10^{-6}$ (red),  $C_\theta=1.3 \times 10^{-5}$ (yellow),  $C_\theta=1.2 \times 10^{-4}$ (green) and $C_\theta=3.2 \times 10^{-4}$ (orange). Solid lines indicate the optimal gain $\mu_0$ while dashed lines indicate the first suboptimal gain $\mu_1$. The stars indicate the linear gain predicted from the gradient. (b) Energy density $d_\text{Chu}$ of the optimal forcing (dashed lines) and response (solid lines) of the second Mack mode without control (green) and for the controlled base-flow at $C_\theta=3.2 \times 10^{-6}$ (red). Black vertical line indicates the end of the optimisation domain for resolvent analysis.}
\label{gain2vws}
\end{figure}

We plot in Figure \ref{gain2dvw} the optimal gain variations for $ \beta=0 $ as a function of frequency $ F $ (around the first and second Mack modes' frequencies). The exact "nonlinear" optimal gain variations are given with lines and the linear predictions with stars of same colour. The linear predictions are evaluated by writing Eq. \eqref{validgain} as $\mu_0^2(\mathbf{p} - \epsilon \nabla_\mathbf{p} \mu_0^2) = \mu_0^2(\mathbf{p}) - \epsilon \| \nabla_\mathbf{p} \mu_0^2\|^2$ with $\epsilon$ a function of $C_\theta $. The maximal optimal gain reduction for different values of $C_\theta$ is reported in Table \ref{tab:blow}. At $C_\theta = 3.2 \times 10^{-8}$, the linear prediction remains accurate while from $C_\theta = 3.2 \times 10^{-6}$, the nonlinear ($0.811$) and linear ($0.686$) predicted optimal gain ratios of the second Mack mode deviate from each others (the two-dimensional first Mack mode being well predicted by the linear sensitivity according to Figure \ref{gain2dvw}). Increasing further the blowing momentum coefficient $C_\theta$ allows reaching an optimal gain ratio of $0.62$ at $C_\theta = 1.2 \times 10^{-4}$. However, we observe from the curve $\mu(F)$ in Figure \ref{gain2dvw} that the control damps efficiently the optimal gain at the frequency where it was computed but not for lower or higher frequencies resulting in a split of the second Mack mode into two peak regions: a first peak at $F=2 \times 10^{-4}$ and a second for higher frequencies above $F=3 \times 10^{-4}$. This effect was also noticed by \citet{miro2018effect} on the growth rate. The optimisation procedure could in principle be pursued by computing a new descend direction to eventually reach a local/global minimum. 

Figure \ref{gain2dvw} overall shows that the control applied on the base-flow strongly reduces the optimal gain $\mu_0$ for the initial frequency of the second Mack mode without modifying the gains of the sub-optimal gain $\mu_1$. While the optimal gain is effectively damped with an increased wall-normal velocity control, the sub-optimal gain remains of similar magnitude compared to the no-control case and does not increase. This results in a low-rankness loss of the system (ratio of approximately 4 between optimal and suboptimal gains which decreases to 2 for $C_\theta = 1.2 \times 10^{-4}$). Therefore, for larger control intensity, the optimal gain alone is not sufficient anymore to characterise the linear dynamics of the boundary layer.

The Chu's energy densities of the optimal forcing and response with blowing/suction control at $C_\theta = 3.2 \times 10^{-6}$ are plotted in Figure \ref{dfdchuvw} and compared to the results without control. The optimal forcing has been shifted slightly downstream while the optimal response exhibits a slightly smaller support (we recall that the response energy is normalised by $ \langle \mathbf{\check{q}}, \mathbf{Q}_q \mathbf{\check{q}} \rangle = 1$ so that the integral under the curve is $1$). The  downstream shift of the optimal forcing is expected as the boundary layer thickness has been locally reduced because of the suction at the wall until $Re_x = 1.2 \times 10^6$. The response is not shifted downstream as the wall-blowing control increases again the boundary layer thickness to almost recover its value without control (an example of boundary layer thicknesses with and without control is shown in Figure \ref{machbl}). This can be understood from spatial LST results, where Branch I is shifted downstream while Branch II does not move.

\begin{table}
    \begin{center}
\def~{\hphantom{0}}
    \begin{tabular}{c|c|c|c}
 $C_\theta$ & Gain $\mu_0$ & Gain ratio & Linear gain ratio  \\ 
     \hline
    $3.2 \times 10^{-8} $  & $1.755 \times 10^7$  & $0.978$ & $0.973$ \\
    $3.2 \times 10^{-6} $  & $1.456 \times 10^7$  & $0.811$ & $0.686$ \\
    $1.3 \times 10^{-5} $  & $1.304 \times 10^7$  & $0.727$ & X \\
    $1.2 \times 10^{-4} $  & $1.110 \times 10^7$  & $0.618$ & X \\
    $3.2 \times 10^{-4} $  & $1.120 \times 10^7$  & $0.624$ & X 
    \end{tabular}
    \caption{Optimal gain evolution of the second Mack mode with respect to the optimal wall-velocity profile $-\nabla_{\mathbf{v}_w} \mu_0^2  /\mu_0^2$ at various momentum coefficient $C_\theta$ intensity. Gain ratio is computed by $\mu_0(C_\theta \neq 0)/\mu_0(C_\theta = 0)$. The expected linear gain ratio is computed by $\sqrt{\mu_0^2- \epsilon \|\nabla_{\mathbf{p}} \mu_0^2 \|^2}/\mu_0$ with $\epsilon$ being a function of  $C_\theta$. The cross X indicates that the linear gain ratio would predict a negative value.}
    \label{tab:blow}
    \end{center}
\end{table}

\subsubsection{Sensitivity of the second Mack mode to steady wall heating} \label{sec:sens2Dheat}

\begin{figure}
\centering
\includegraphics[width=0.5\textwidth]{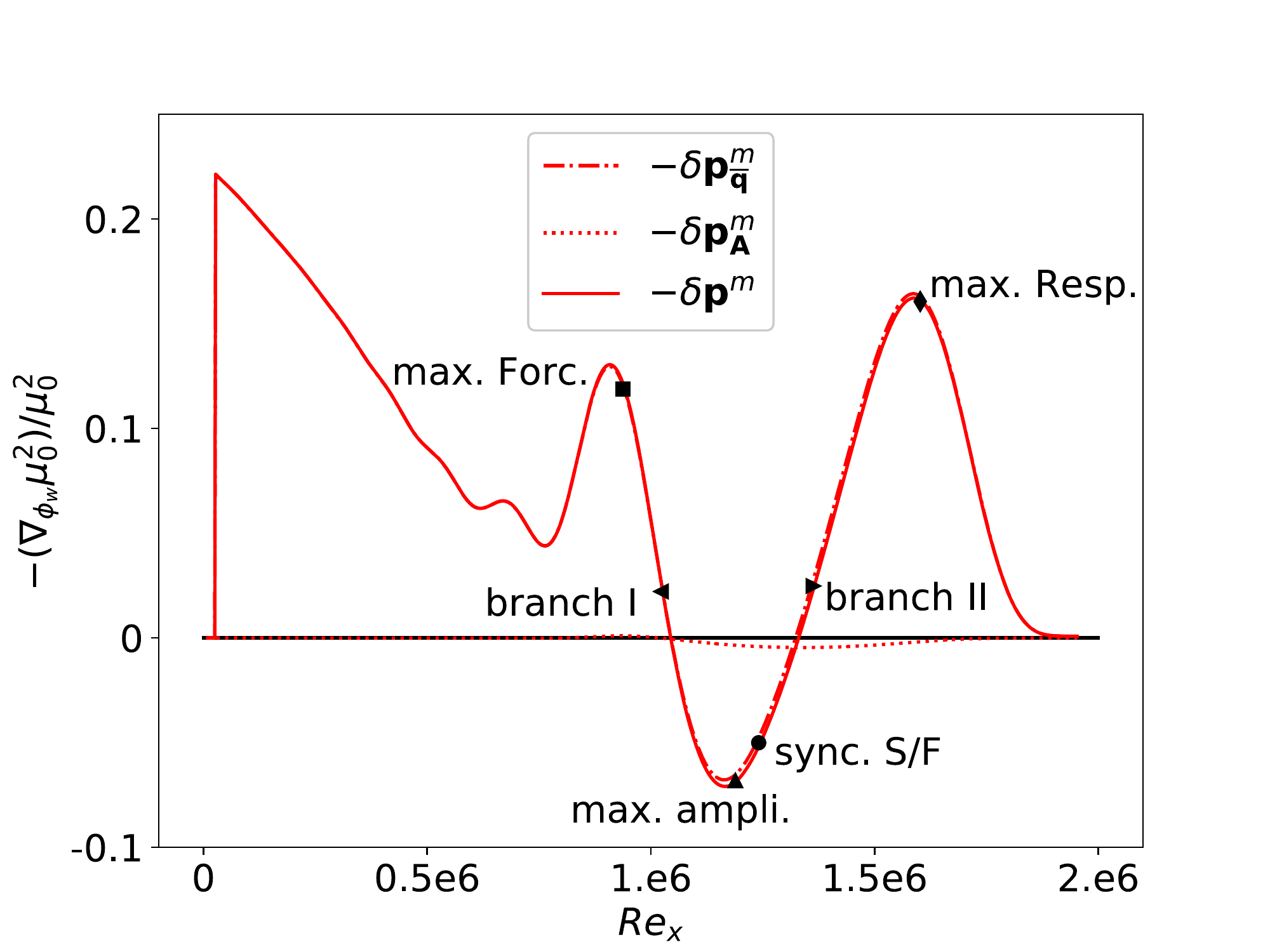}
\caption{Optimal wall heat flux profile to damp the second Mack mode i.e. opposite of the sensitivity of the optimal gain of the second Mack mode to wall heat flux $ -\delta \mathbf{p}^{m} = -\delta \bm{\phi}_w$. Base-flow component $- \delta \mathbf{p}_{\overline{\mathbf{q}}}^m $ and Jacobian component $ -\delta \mathbf{p}_{\mathbf{A}}^m $ are also indicated, the latter being almost zero.}
\label{sens2dfluxbis}
\end{figure}

The sensitivity analysis and all the subsequent steps performed for the steady wall-blowing control are repeated in the case of the steady wall heat flux control. The optimal wall heat flux profile to damp the second Mack mode is plotted in Figure \ref{sens2dfluxbis}(a). As the gradient with respect to wall-blowing, the contribution from the Jacobian variation in the gradient $\nabla_{\bm{\phi}_w} \mu_0^2 $ is negligible.

\begin{figure}
\centering
\subfloat[\label{gain2dflux}]{\includegraphics[width=0.5\textwidth]{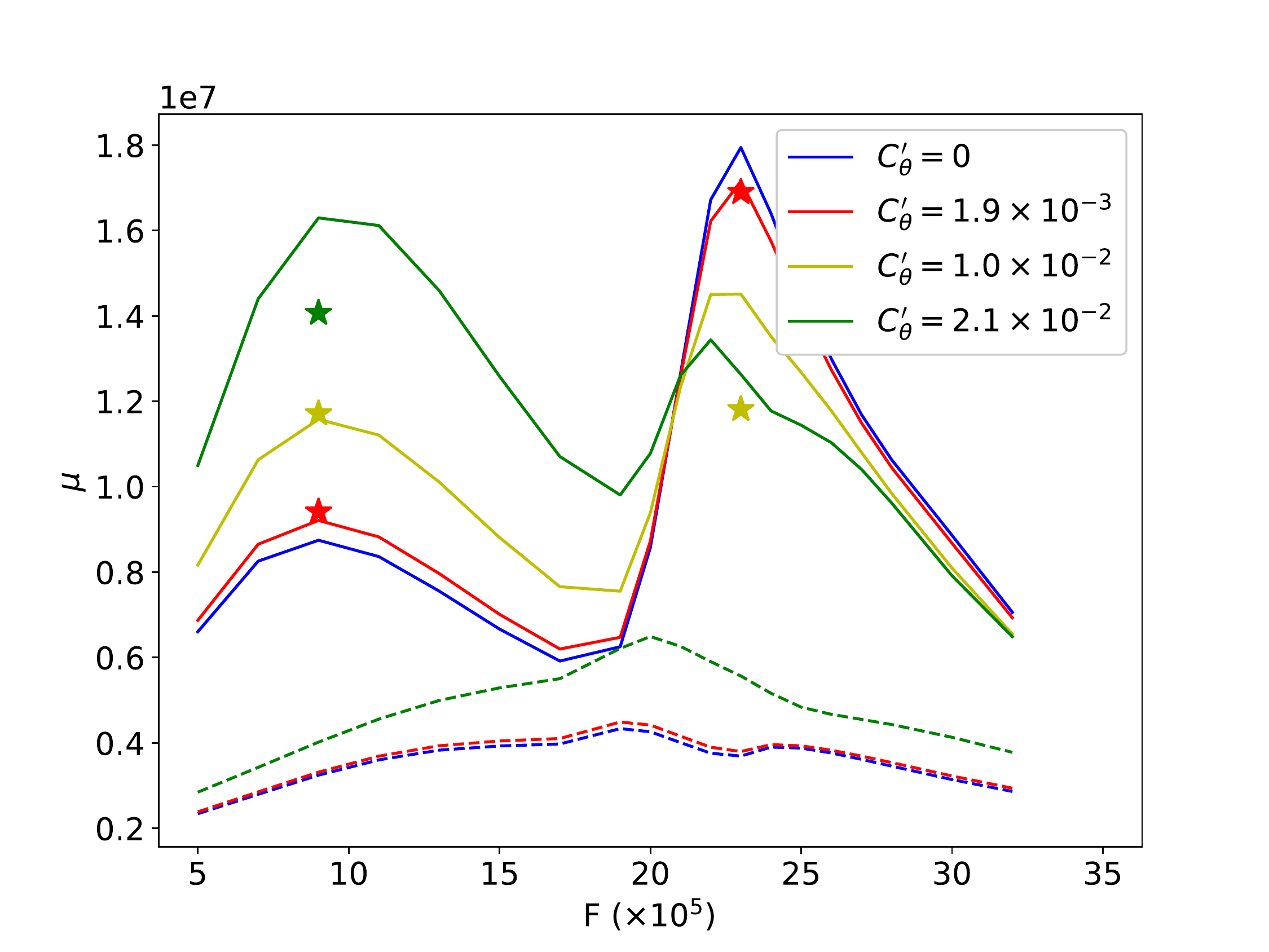}}
{\subfloat[\label{dfdchuflux}]{\includegraphics[width=0.5\linewidth]{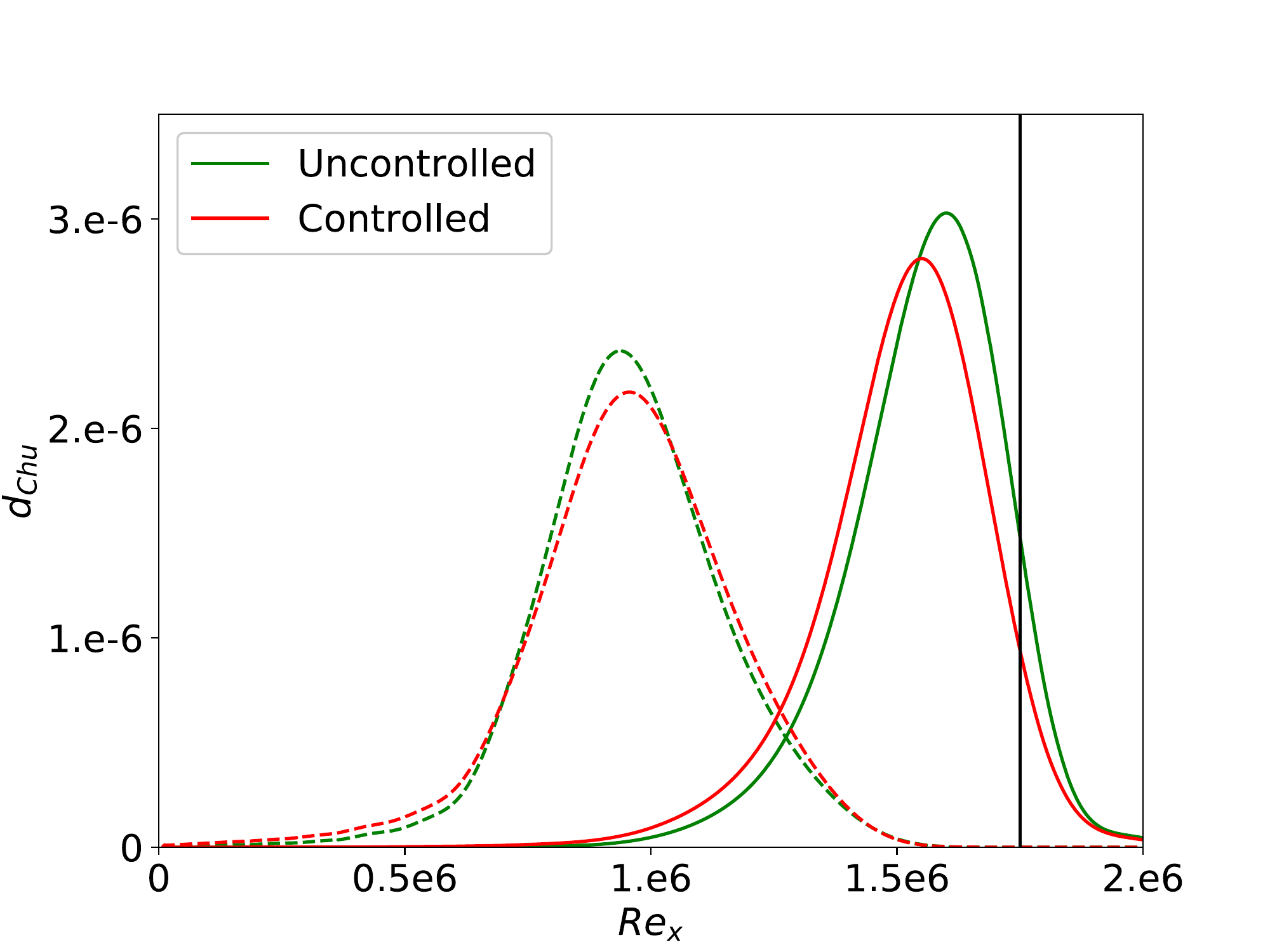}}}
\caption{Resolvent analysis for different base-flows controlled with the optimal wall heat flux profile $-\nabla_{\bm{\phi}_w} \mu_0^2 /\mu_0^2$. (a) Optimal gain $\mu$ for $ \beta=0$ as a function of frequency $F$ for $C_\theta'=0$ (blue),  $C_\theta'=1.9 \times 10^{-3}$ (red), $C_\theta'=1.0 \times 10^{-2}$ (yellow) and $C_\theta'=2.1 \times 10^{-2}$ (green). Solid lines indicate the optimal gain $\mu_0$ while dashed lines indicate the first suboptimal gain $\mu_1$. The stars indicate the linear gain predicted from the gradient. (b) Energy density $d_\text{Chu}$ of the optimal forcing (dashed lines) and response (solid lines) of the second Mack mode without control (green) and for the controlled base-flow at $C_\theta'=1.0 \times 10^{-2}$ (red). Black vertical line indicates the end of the optimisation domain for resolvent analysis.}
\label{fluxgaindf}
\end{figure}

The gradient shows two heating and one cooling zones. First, the largest sensitivity is close to the leading edge. This large sensitivity upstream of the forcing was already highlighted by \citet{fedorov2014numerical} through the analysis of the effect of a local volume energy source term. Then, local maxima of the gradient in the heating zone seems to correspond with the maximal forcing and response zones. The same behaviour is noticed at a different frequency (see appendix \ref{sec:appsize}). Then, the cooling sensitivity region is located in the wavemaker region where forcing and response overlap, more precisely between branch I and branch II of mode S. Previous studies \citep{zhao2018numerical, batista2020local} have found that cooling upstream of the synchronisation point and heating downstream damp the second Mack mode. In the present result, the shift between cooling and heating regions of sensitivity is not located at the synchronisation point (but close to branch II of mode S). This discrepancy may be explained by the fact that, contrary to previous studies where the forcing structure was kept fixed when control was applied, we let the forcing adapt and be optimised as the control is applied (see Figure \ref{dfdchuflux}).

We now prescribe the heat flux profile given by $-\nabla_{\bm{\phi}_w} \mu_0^2$ at the wall, compute a new base-flow and repeat the resolvent analysis. To quantify the wall heat flux control applied to the base-flow, similarly to the blowing momentum coefficient $C_\theta$, we define an energy coefficient $C_\theta'$ based on the ratio of energy injected at the wall over the free-stream energy deficit:
\begin{equation}
C_\theta' = \frac{ \int_{y=0} \lambda |\frac{\partial T}{\partial y}| \, dx}{\int_{x=x_{out}} (\rho_\infty E_\infty U_\infty - \rho E u) \, dy},
\label{cmu2}
\end{equation}
with $\lambda \frac{\partial T}{\partial y}$ the heat flux injected (the uncontrolled case being adiabatic). 
To understand the effect of the control profile shown in Figure \eqref{sens2dfluxbis}, we compare local stability (spatial LST) analysis results applied on the uncontrolled base-flow and on the controlled base-flow with the full stabilising gradient at $C_\theta' = 1.0 \times 10^{-2}$ .

\begin{figure}
\centering
\subfloat[\label{crflux2}]{\includegraphics[width=0.5\textwidth]{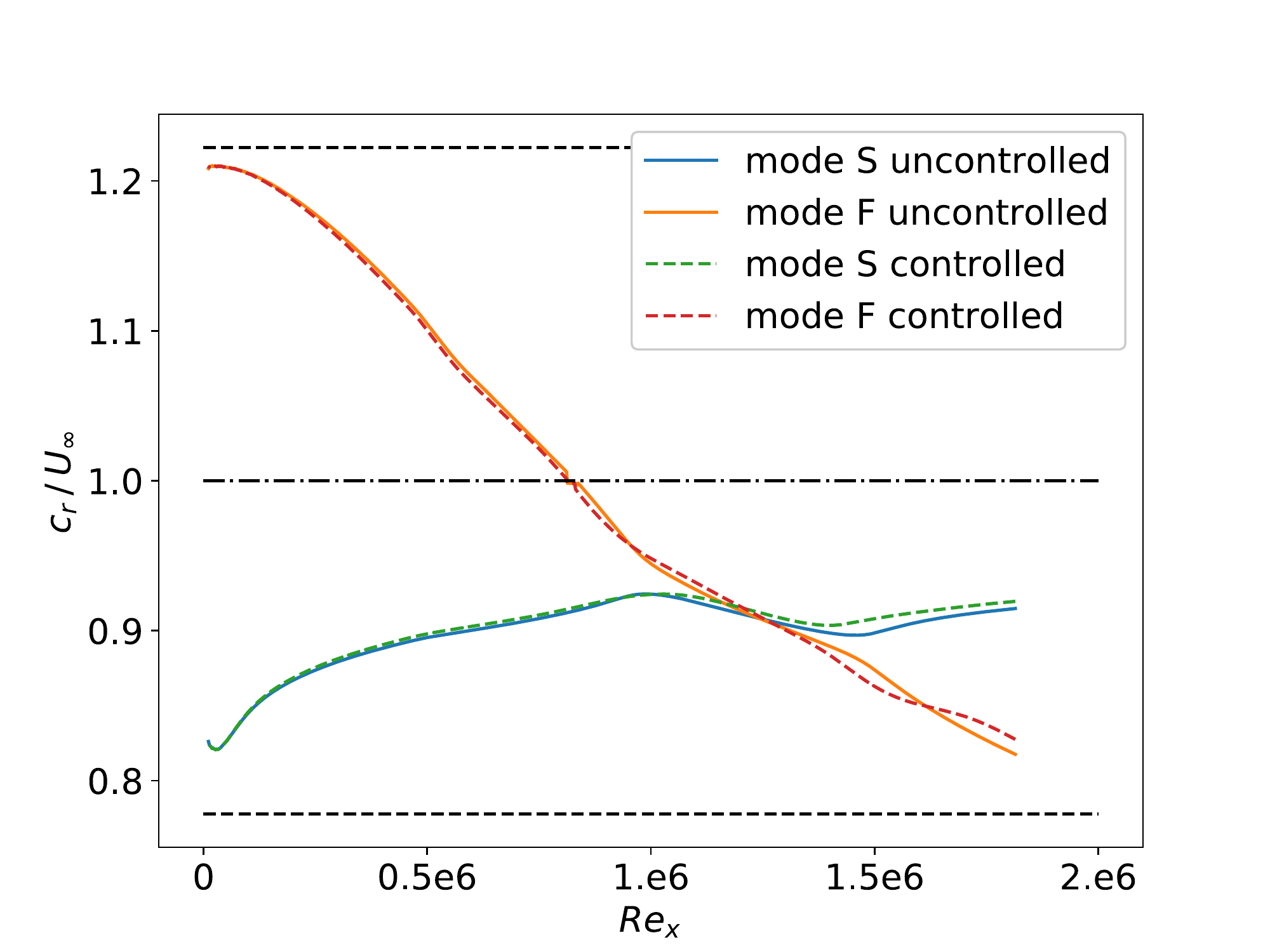}}
{\subfloat[\label{alphaiflux2}]{\includegraphics[width=0.5\linewidth]{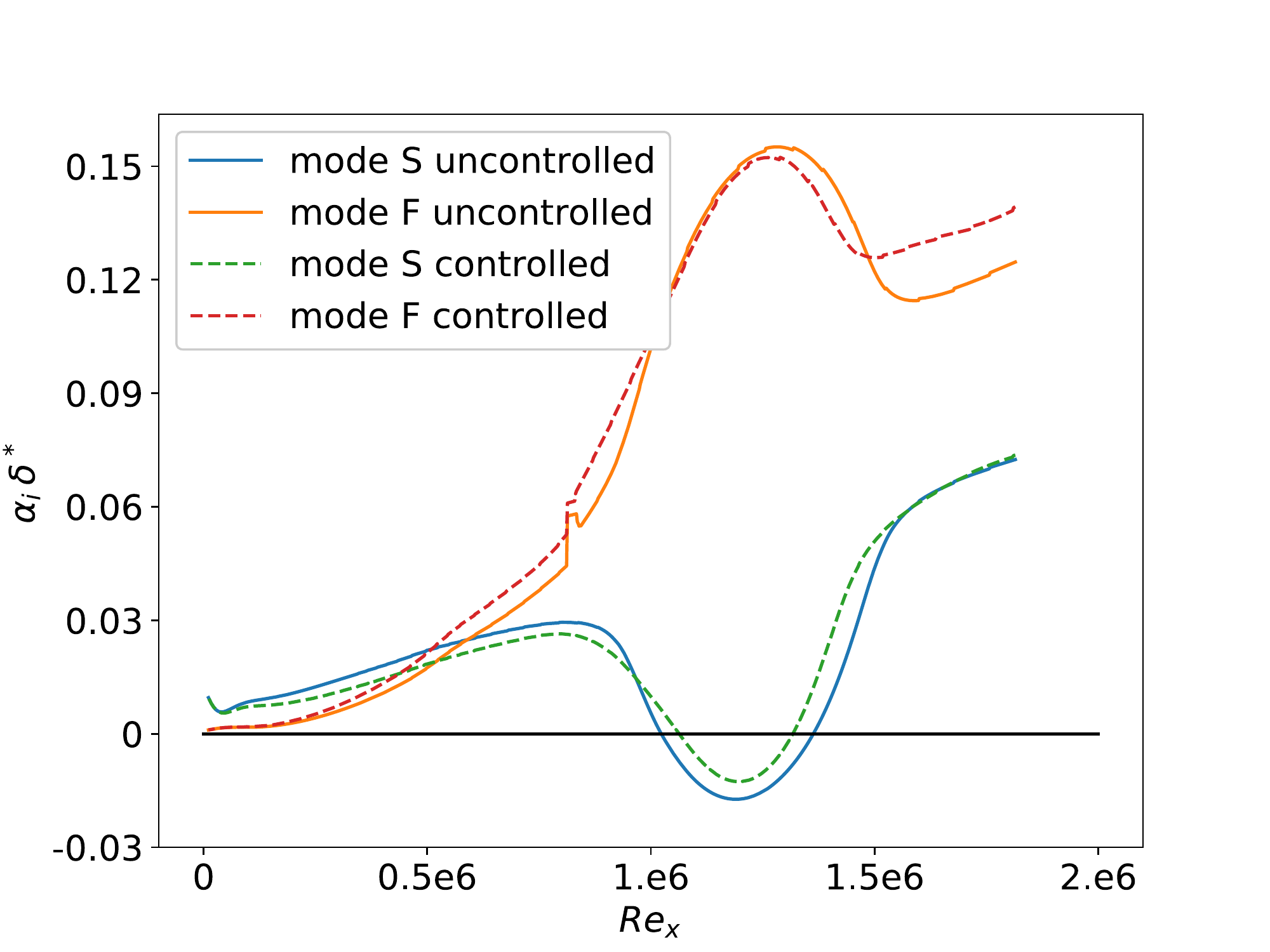}}}
\caption{Local stability analysis of the uncontrolled base-flow (solid lines) and the controlled base-flow (dashed lines) with the optimal wall heat flux profile $-\nabla_{\bm{\phi}_w} \mu_0^2  /\mu_0^2$ for $C_\theta' = 1.0 \times 10^{-2}$. (a) Phase velocity $c_r/U_\infty$. Dashed lines denote respectively from top to bottom the phase velocities $1 + 1/M_\infty$, $1$ and $1-1/M_\infty$. (b) Amplification rate $\alpha_i  \delta^*$.}
\label{LSTflux2}
\end{figure}

Phase velocities and amplification rates of modes S and F are shown in Figure \ref{LSTflux2} with and without the application of the stabilising wall profile. Unlike the blowing/suction case, the phase velocity of both modes F and S of the controlled base-flow now slightly deviate from the original ones in the control region. This results in a shorter synchronisation region between mode F and S leading to a shorter unstable region for the amplification rate of mode S induced by both the modification of branch I and II locations. \citet{zhao2018numerical} noticed the same behaviour for the phase velocity in the case of heating and cooling strips control with a stronger increase of the phase velocity of mode F.

In the following, we compare the predictions obtained by the gradient approach and the exact results obtained by computing the modified base-flow and the resolvent analysis.

We plot in Figure \ref{gain2dflux} the optimal gain for $ \beta=0 $ as a function of frequency $ F $ computed for the controlled base-flow with different values of $C_\theta'$. The maximal optimal gain reduction for different values of $C_\theta'$ is reported in the Table \ref{tab:flux}. At $C_\theta' = 1.9 \times 10^{-4}$, the linear prediction remains accurate while above $C_\theta' = 1.0 \times 10^{-2}$, the nonlinear ($0.809$) and linear ($0.658$) predicted optimal gain ratios of the second Mack mode start to significantly deviate from each others. The linear predictions for the two-dimensional first Mack mode remain accurate for larger $C_\theta'$. The control applied on the base-flow reduces the optimal gain for the frequencies around the second Mack mode but strongly increases it for the lower frequencies corresponding to the two-dimensional first Mack mode up to the point where this mode becomes the dominant one at $C_\theta' = 2.1 \times 10^{-2}$. Furthermore, the first suboptimal gain $\mu_1$ for the different blowing momentum coefficients are also plotted in Figure \ref{gain2dflux}. While the optimal gain $\mu_0$ is effectively damped with an increased wall-temperature control, the suboptimal gain $\mu_1$ increases. As for blowing/suction control, it appears that the optimal response alone will not be sufficient to describe the dynamics of the boundary layer for a large amplitude wall-temperature control.

The Chu's energy densities of the optimal forcing and response with heating/cooling control at $C_\theta' = 1.66 \times 10^{-2}$ are plotted in Figure \ref{dfdchuflux} and compared without control. Both the optimal forcing and response density exhibit a larger support with control than without. Comparing with the blowing/suction case, there is both a downstream shift of the optimal forcing and an upstream shift of the optimal response. This can be understood from spatial LST results, where Branch I is seen to be shifted downstream and Branch II upstream.

\begin{table}
    \begin{center}
\def~{\hphantom{0}}
        \begin{tabular}{c|c|c|c}
$C_\theta'$ & Gain $\mu_0$ & Gain ratio & Linear gain ratio  \\ 
     \hline
     $1.9 \times 10^{-4}$ & $1.789 \times 10^7$  & $0.997$ & $0.994$ \\
     $1.9 \times 10^{-3}$ & $1.716 \times 10^7$  & $0.956$ & $0.942$ \\
     $1.0 \times 10^{-2}$ & $1.451 \times 10^7$  & $0.809$ & $0.658$ 
    \end{tabular}
    \caption{Optimal gain evolution of the second Mack mode with respect to the optimal wall heat flux profile $-\nabla_{\bm{\phi}_w} \mu_0^2  /\mu_0^2$ at various momentum coefficient $C_\theta'$ intensity. Gain ratio is computed by $\mu_0(C_\theta' \neq 0)/\mu_0(C_\theta' = 0)$. The expected linear gain ratio is computed by $\sqrt{\mu_0^2- \epsilon \|\nabla_{\mathbf{p}} \mu_0^2 \|^2}/\mu_0$ with $\epsilon $ a function of $ C_\theta'$.}
    \label{tab:flux}
    \end{center}
\end{table}

\subsection{Design of a wall actuator targeting all instabilities} \label{sec:sens3Dapplied}

\begin{figure}
\centering
\subfloat[\label{machbl}]{\includegraphics[width=0.32\textwidth]{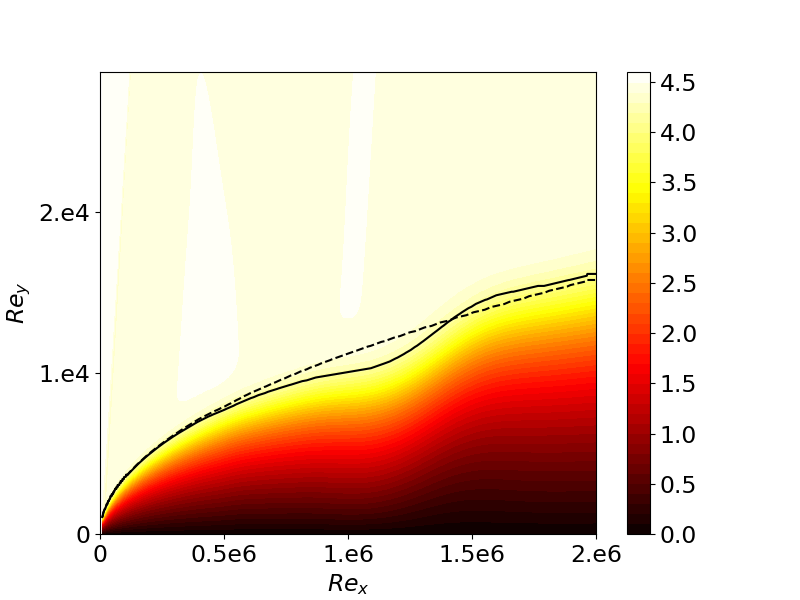}}
{\subfloat[\label{machf2}]{\includegraphics[width=0.32\linewidth]{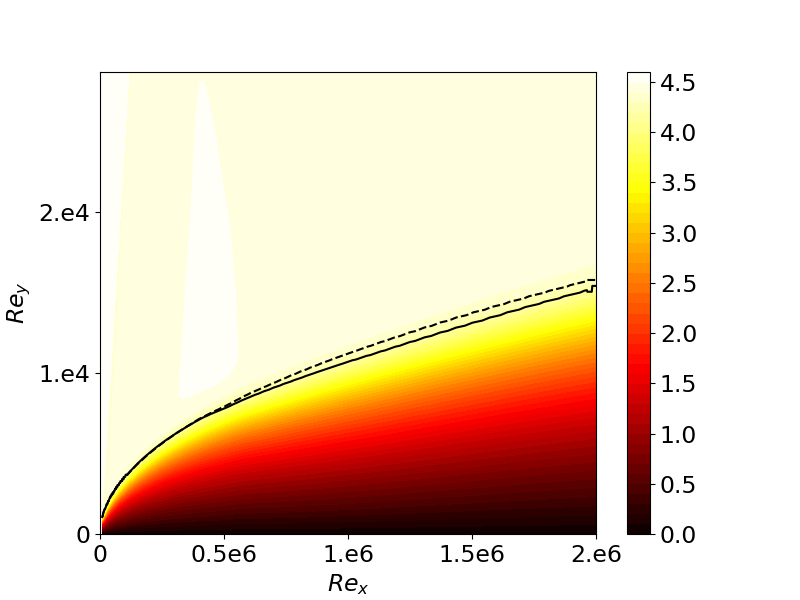}}}
{\subfloat[\label{machf1}]{\includegraphics[width=0.32\linewidth]{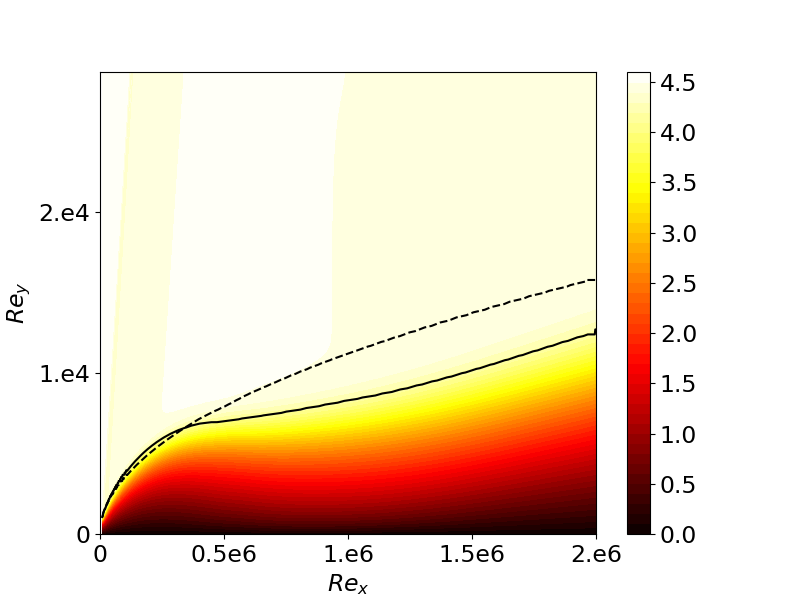}}}
{\subfloat[\label{Tbl}]{\includegraphics[width=0.32\linewidth]{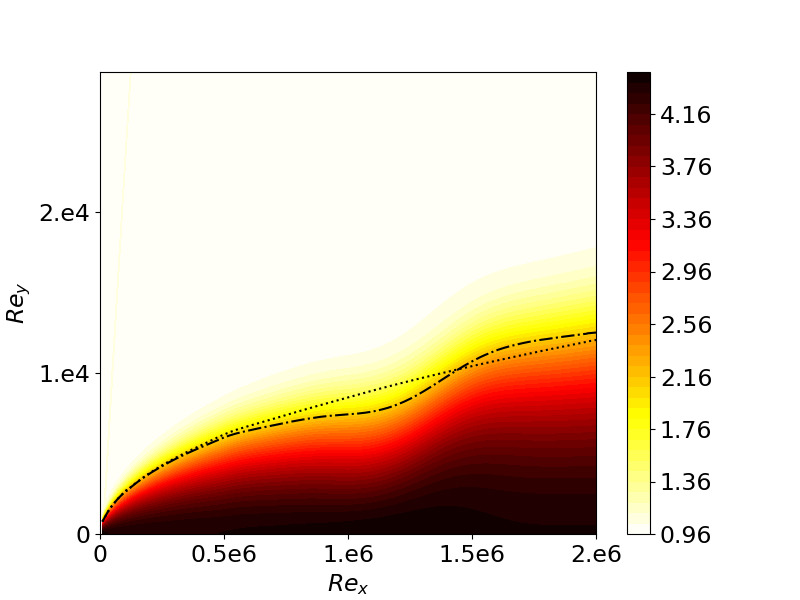}}}
{\subfloat[\label{Tf2}]{\includegraphics[width=0.32\linewidth]{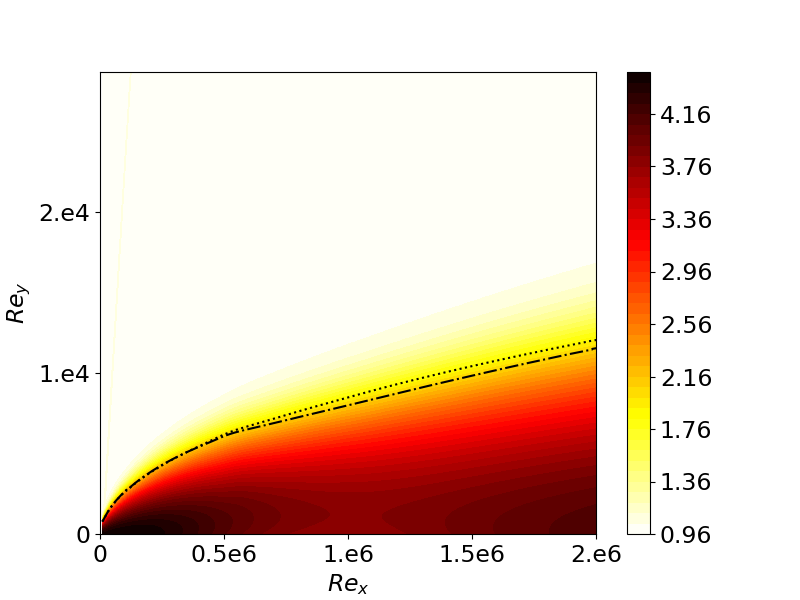}}}
{\subfloat[\label{Tf1}]{\includegraphics[width=0.32\linewidth]{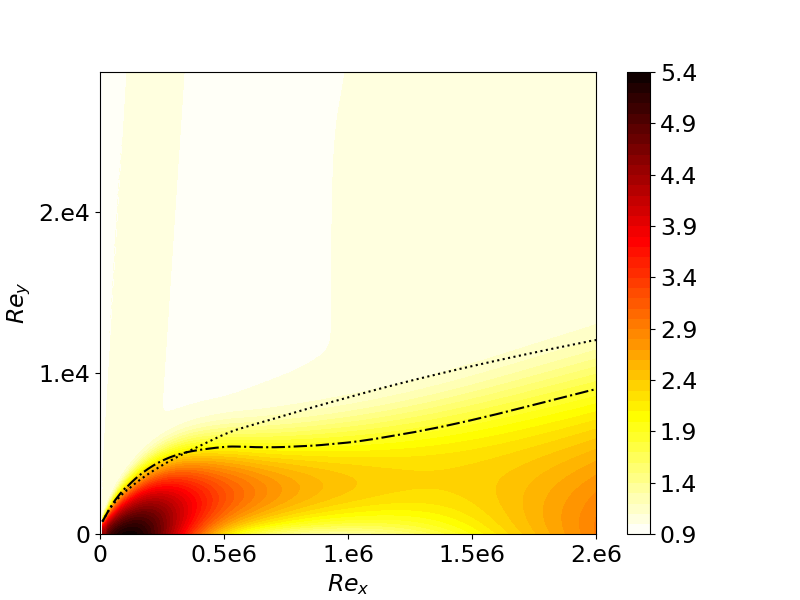}}}
\caption{Mach number (a-c) and temperature (d-f) of the controlled base-flows. (a,d) Blowing control given by the wall-velocity profile $-\nabla_{\mathbf{v}_w} \mu_0^2/\mu_0^2$ for the second Mack mode at $C_\theta=3.2 \times 10^{-4}$. (b,e) Heat flux control given by the wall heat flux profile $-\nabla_{\bm{\phi}_w} \mu_0^2/\mu_0^2$ for the first Mack mode at $C_\theta'=6.4 \times 10^{-3}$. (c,f) Heat flux control given by the wall heat flux profile $-\nabla_{\bm{\phi}_w} \mu_0^2/\mu_0^2$ for the first Mack mode at $C_\theta'=4.0 \times 10^{-2}$. In (a-c), solid line indicates the boundary layer thickness with control and dashed line indicates the boundary layer thickness of the uncontrolled base-flow (see Figure \ref{mach}). In (d-f), dash-dotted line indicates the total energy deficit thickness $\delta_E = \int_{y=0}^{+\infty} \left(1- \frac{\rho E u}{\rho_\infty E_\infty U_\infty}\right)\,dy$ and dotted line indicates $\delta_E$ of the uncontrolled base-flow.}
\label{bsfc}
\end{figure}

\begin{figure}
\centering
\subfloat[\label{gainch}]{\includegraphics[width=0.32\linewidth]{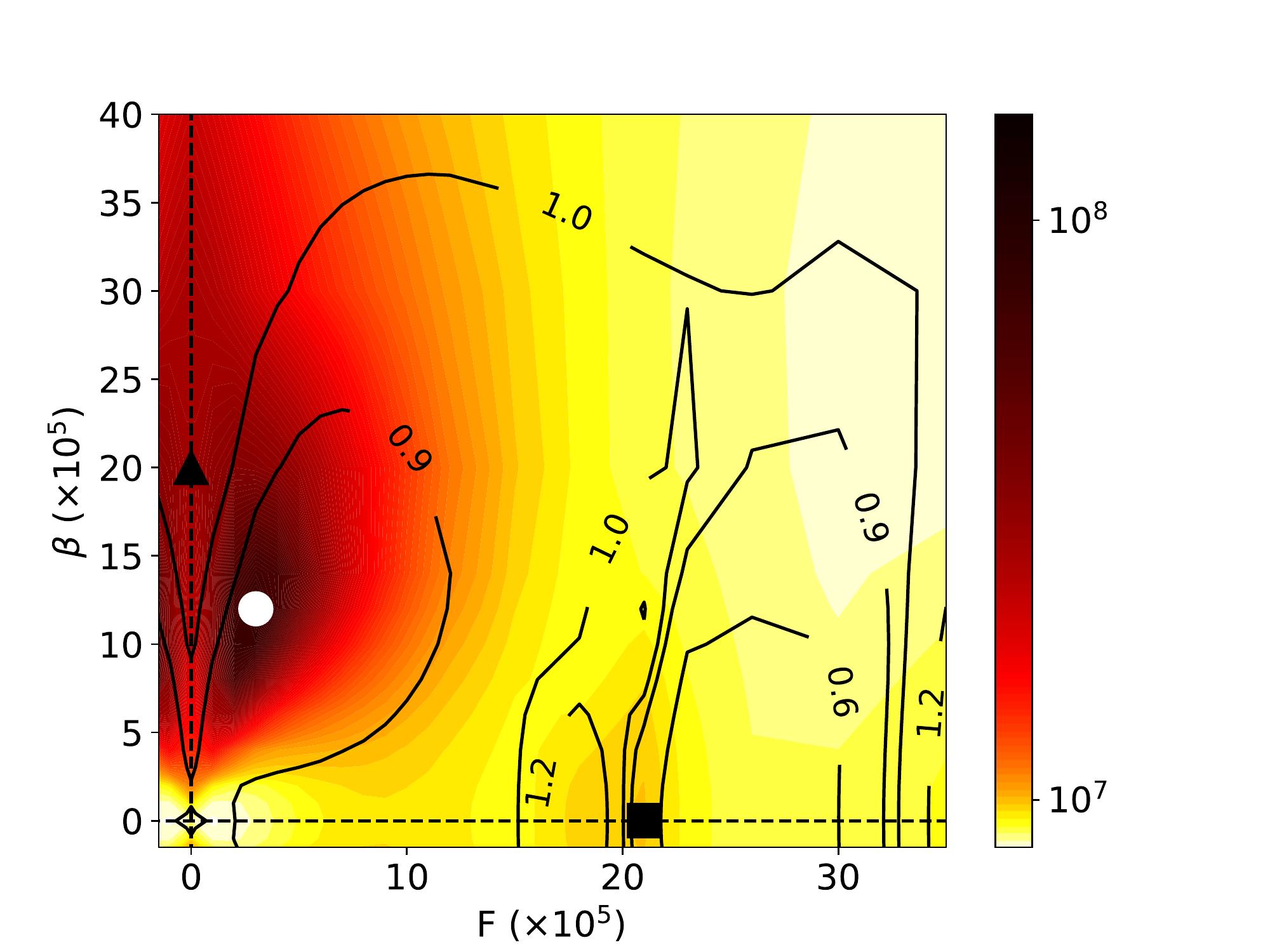}}
{\subfloat[\label{gaincf}]{\includegraphics[width=0.32\textwidth]{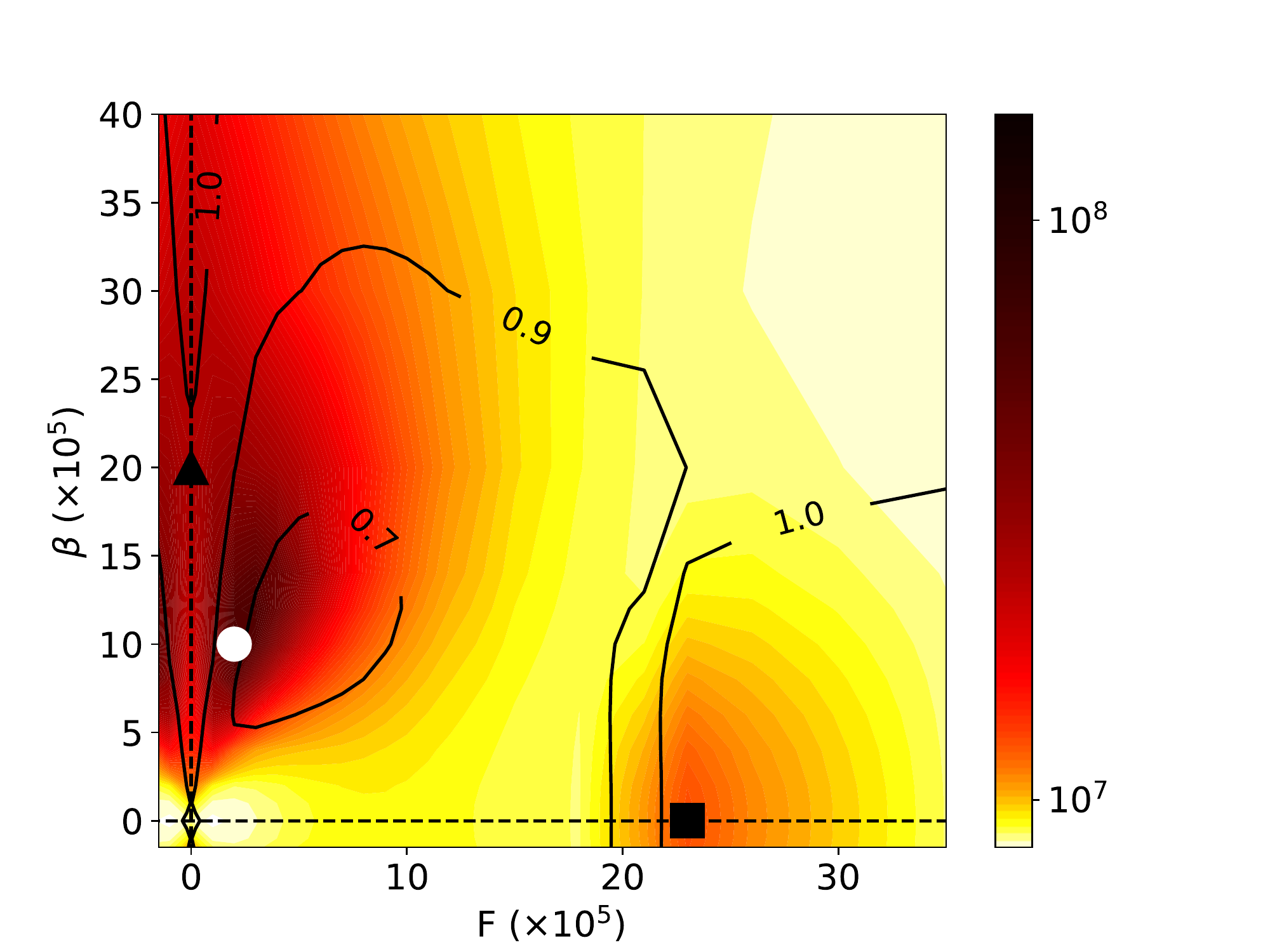}}}
{\subfloat[\label{gaincf2}]{\includegraphics[width=0.32\linewidth]{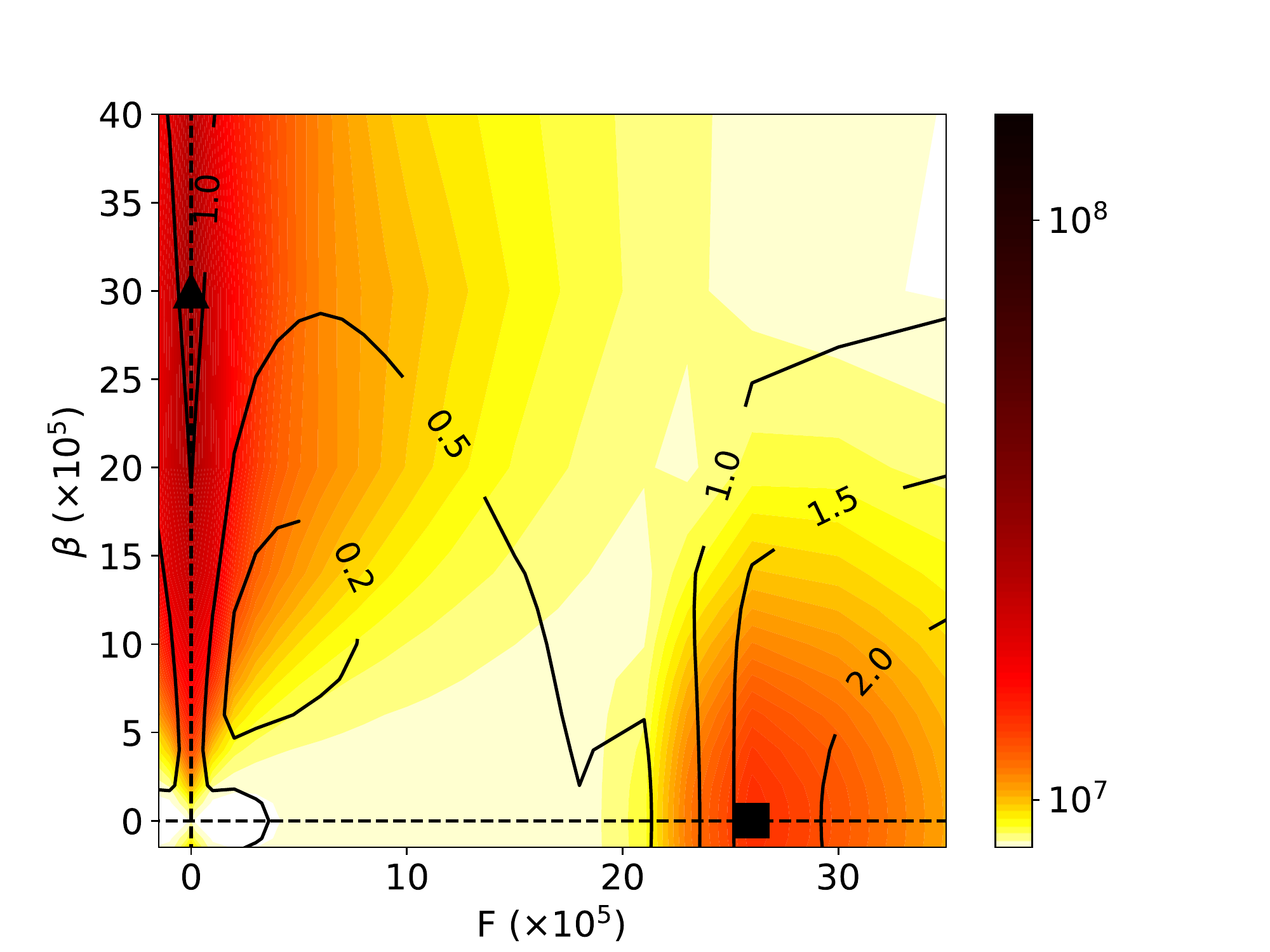}}}
\caption{Optimal gain $\mu_0$ of controlled boundary layers. White circle denotes the first Mack mode, black triangle denotes the streaks and black square denotes the second Mack mode. Contour lines denote the ratio of the optimal gain with control over the optimal gain without control $\mu_0(C_\theta' \neq 0)/\mu_0(C_\theta' = 0)$ given by Figure \ref{gain3d}. (a) Blowing / suction control with wall-velocity profile $-\nabla_{\mathbf{v}_w} \mu_0^2/\mu_0^2$ for the second Mack mode at $C_\theta=3.2 \times 10^{-4}$. (b) Heat flux control with wall heat flux profile $-\nabla_{\bm{\phi}_w} \mu_0^2/\mu_0^2$ for the first Mack mode at $C_\theta'=6.4 \times 10^{-3}$. (c) Heat flux control given with wall heat flux profile $-\nabla_{\bm{\phi}_w} \mu_0^2/\mu_0^2$ for the first Mack mode at $C_\theta'=4.0 \times 10^{-2}$.}
\label{gain2cs2}
\end{figure}

Once the sensitivity regions of the three instabilities to wall control have been identified (\S \ref{sec:sens3D}), the initial steps for the design of an optimal wall-control actuator can be performed. We first consider the full stabilising profiles given by the gradients $-\nabla_{\mathbf{p}} \mu_0^2/\mu_0^2$ (Figure \ref{sens3d3}) of largest magnitude for  blowing/suction and cooling/heating:
\begin{itemize}
    \item The wall-normal blowing/suction profile $-\nabla_{\mathbf{v}_w} \mu_0^2/\mu_0^2$ of the second Mack mode.
    \item The wall heat flux profile $-\nabla_{\bm{\phi}_w} \mu_0^2/\mu_0^2$ of the first Mack mode.
\end{itemize}
We apply these profiles at the wall, compute a new base-flow and repeat the resolvent analysis over the whole range of spanwise wavenumbers and frequencies, to assess the overall performance of each control strategy. We perform these computations at the finite control amplitude $C_\theta = 3.2 \times 10^{-4}$ for blowing/suction control, at $C_\theta' = 6.4 \times 10^{-3}$ and $C_\theta' = 4.0 \times 10^{-2}$ for heat flux control. Mach number and temperature of the controlled base-flows are plotted in Figure \ref{bsfc}, resolvent gain maps in Figure \ref{gain2cs2}. Table \ref{tab:gain3D} summarises the optimal gain reduction for all the actuators considered. 

\begin{table}
    \begin{center}
\def~{\hphantom{0}}
    \begin{tabular}{c|c|c|c|c|c}
     Type of wall control & Intensity & 1st mode  & 2nd mode  & Streaks & Global peak \\
      &  & gain ratio &  gain ratio &  gain ratio & ratio\\
     \hline
    $-\nabla_{\mathbf{v}_w} \mu_0^2$ for 2nd Mack mode & $C_\theta=3.2\times 10^{-4}$  & $0.80$ & $\mathbf{0.53}$ & $1.07$ & $0.80$ \\
    $-\nabla_{\bm{\phi}_w} \mu_0^2$ for 1st Mack mode & $C_\theta'=6.4 \times 10^{-3}$  & $\mathbf{0.69}$ & $1.06$ & $0.99$ & $0.69$ \\
    $-\nabla_{\bm{\phi}_w} \mu_0^2$ for 1st Mack mode & $C_\theta'=4.0 \times 10^{-2}$  & $\mathbf{-}$ & $1.27$ & $1.06$ & $0.42$ \\
    Heating strip  & $C_\theta'=0.9 \times 10^{-2}$  & $0.83$ & $\mathbf{0.72}$ & $0.98$ & $0.83$ \\
    Cooling strip & $C_\theta'=0.9 \times 10^{-2}$  & $\mathbf{0.75}$ & $1.03$ & $0.93$ & $0.75$ \\
    Heating $\&$ Cooling strips & $C_\theta'=6.4 \times 10^{-3}$  & $\mathbf{0.83}$ & $0.85$ & $0.97$ & $0.83$ \\
    Heating $\&$ Cooling strips & $C_\theta'=0.9 \times 10^{-2}$  & $0.81$ & $\mathbf{0.73}$ & $0.92$ & $0.81$
    \end{tabular}
    \caption{Optimal gains evolution with respect to different wall control profiles. Gain ratio is computed by $\mu_0(C_\theta \neq 0)/\mu_0(C_\theta = 0)$. Note that 1st (resp. 2nd and Streaks) gain ratio refers to the maximum gain over $ (F,\beta)$ in the region of the 1st Mack (resp. 2nd Mack and Streaks) mode. The last column, global peak ratio, represents the global maximum over the whole domain $ (F,\beta)$: $\max \mu_0(C_\theta \neq 0)/\max \mu_0(C_\theta = 0)$.}
    \label{tab:gain3D}
    \end{center}
\end{table}

For blowing/suction control, we observe in Figure \ref{gainch} that the first Mack mode has also been reduced (optimal gain ratio of $0.80$) because of the suction upstream. For two-dimensional disturbances, as seen in \S \ref{sec:sens2Dblow}, the second Mack mode is damped around the frequency where the gradient has been computed but increased at lower and higher frequencies resulting in a shift of the second Mack mode to lower frequencies. Therefore, the overall optimal gain peak ratio of the second Mack mode is decreased to $0.54$, however, the blowing/suction mechanism is difficult to apply to realistic configurations.

Considering heat flux control, in Figure \ref{gaincf}, the frequencies around the first Mack mode have been reduced leading to an optimal gain ratio of $0.69$ this mode. Nonetheless, as expected from the heat flux gradients in Figure \ref{sens3d2flux}, the second Mack mode is promoted by wall cooling, frequencies higher than $F=2.2 \times 10^{-4}$ being amplified. Increasing the intensity of the control (Figure \ref{gaincf2}) leads to the same results exhibiting a larger optimal gain for the second Mack mode while the first Mack mode is damped such that the local peak of this instability vanishes making the streaks be the most dominant mechanism (highest optimal gain).

In order to design a wall control actuator which damps all instabilities, two options appear from the computations above (combining blowing and heating is not considered in this work). The first would be a series of suction actuators in the upstream part of the flat plate which would damp both Mack modes but this option is not selected because of its difficult real application. The second is a series of constant heating and cooling strips with an appropriate location so that they would damp both Mack modes. By looking at the heat flux gradients in Figure \ref{sens3d2flux}, the gradients for both Mack modes go towards the same direction in two locations: heating from the leading edge until $Re_x=0.21 \times 10^6$ and cooling between $Re_x=1.045 \times 10^6$ and $Re_x=1.33 \times 10^6$. Therefore, we apply a constant heating strip at the first location and a constant cooling strip at the second location, the relative intensity of the heating / cooling strips being computed from the integrals (within each strip location) of the sum of the heat flux gradients of both Mack modes. This yields to a similar amplitude for heating and cooling, the cooling being stronger by 1.175 than the heating.

\begin{figure}
\centering
\subfloat[\label{Tchh}]{\includegraphics[width=0.32\linewidth]{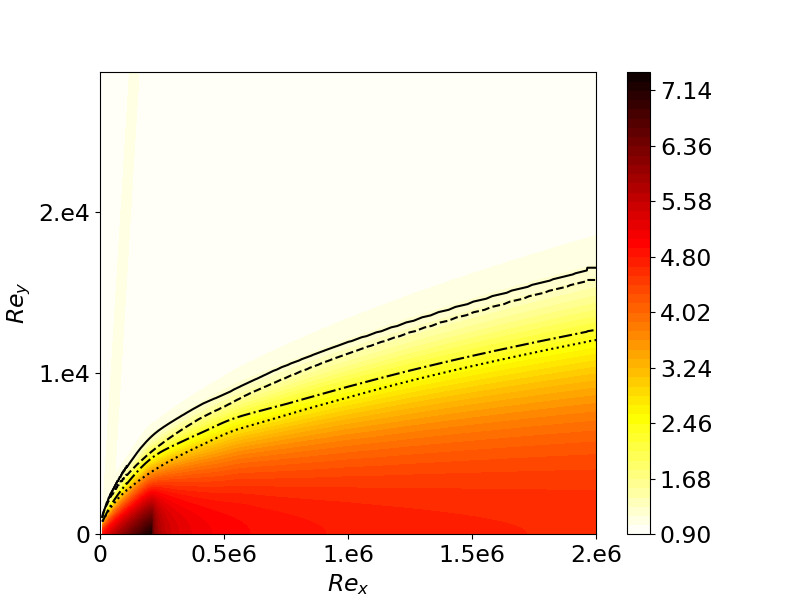}}
{\subfloat[\label{Tccc}]{\includegraphics[width=0.32\linewidth]{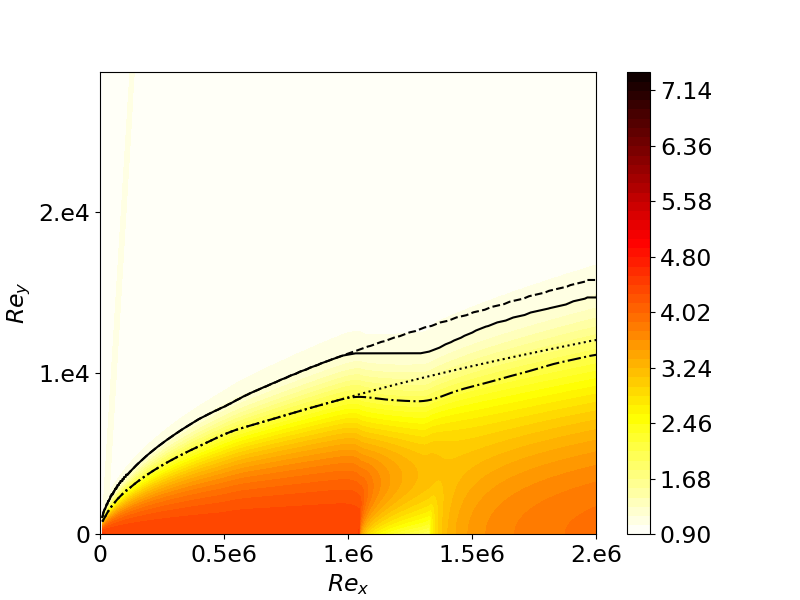}}}
{\subfloat[\label{Tchc}]{\includegraphics[width=0.32\linewidth]{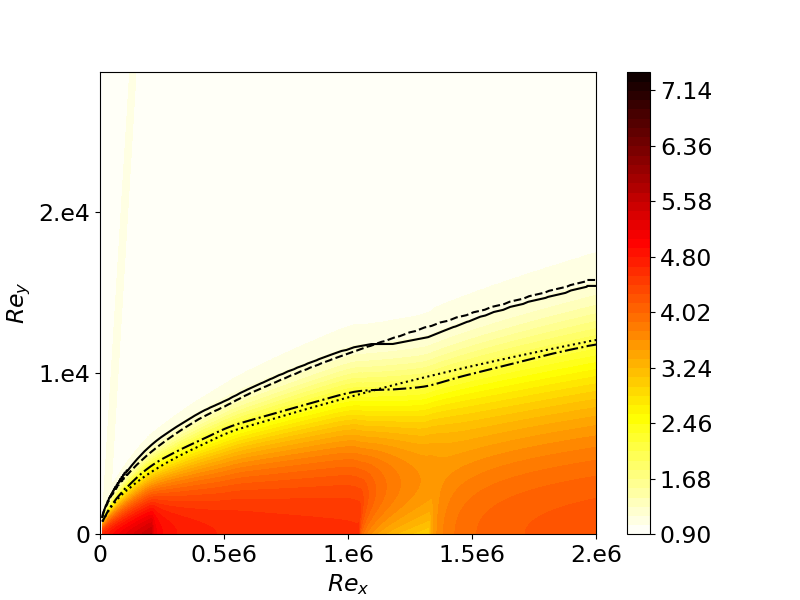}}}
\caption{Temperature of the controlled base-flows. Solid line (resp. dashed) indicates the boundary layer thickness of the uncontrolled (resp. controlled) base-flow. Dotted line (resp. dash-dotted) indicates the total energy deficit thickness $\delta_E$ of the uncontrolled (resp. controlled) base-flow. (a) Heating strip upstream with $C_\theta'=0.9 \times 10^{-2}$. (b) Cooling strip downstream with $C_\theta'=0.9 \times 10^{-2}$. (c) Heating strip upstream and cooling strip downstream with $C_\theta'=0.9 \times 10^{-2}$.}
\label{tcs}
\end{figure}

\begin{figure}
\centering
\subfloat[\label{gainchh}]{\includegraphics[width=0.32\linewidth]{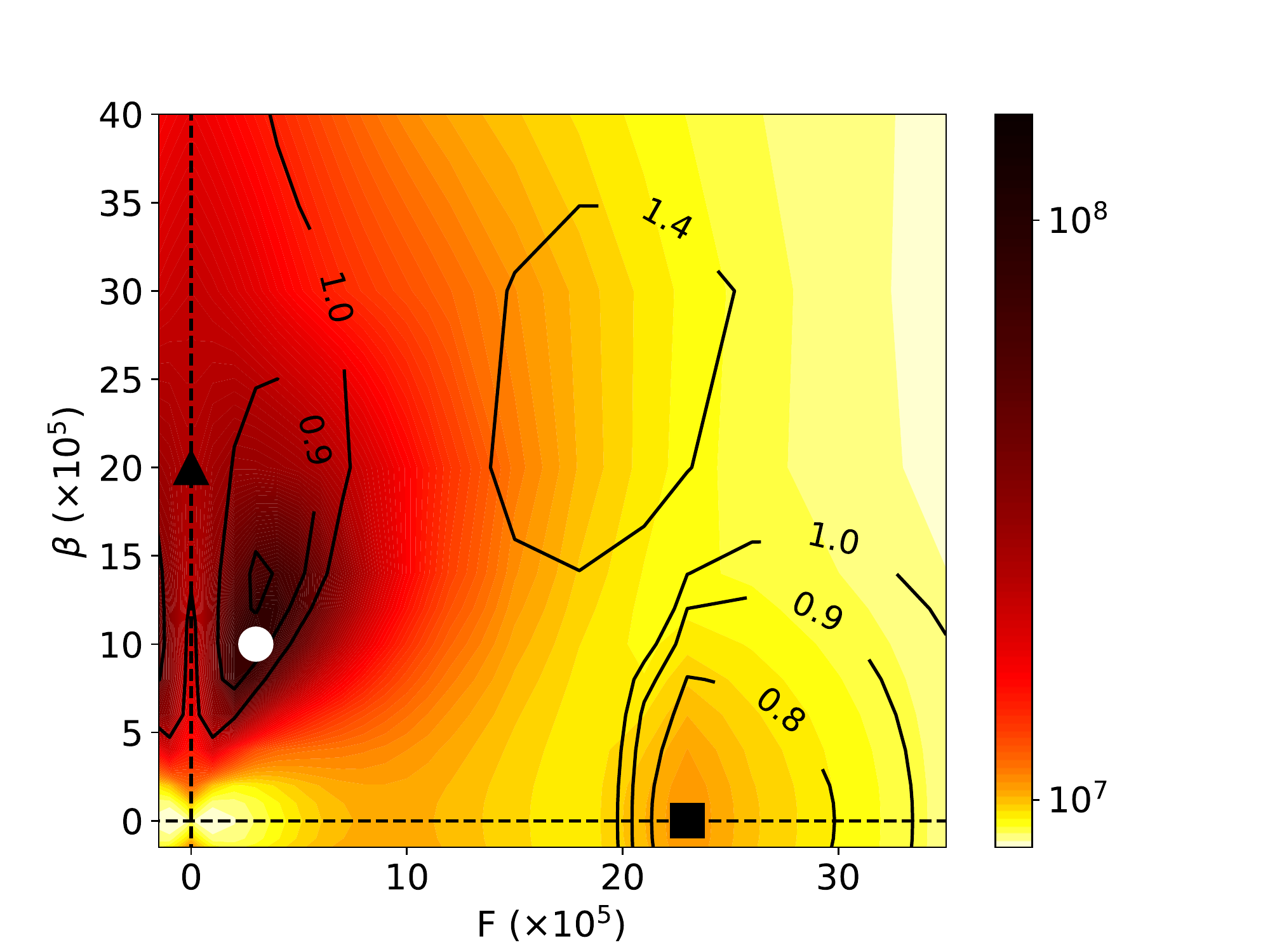}}
{\subfloat[\label{gainccc}]{\includegraphics[width=0.32\textwidth]{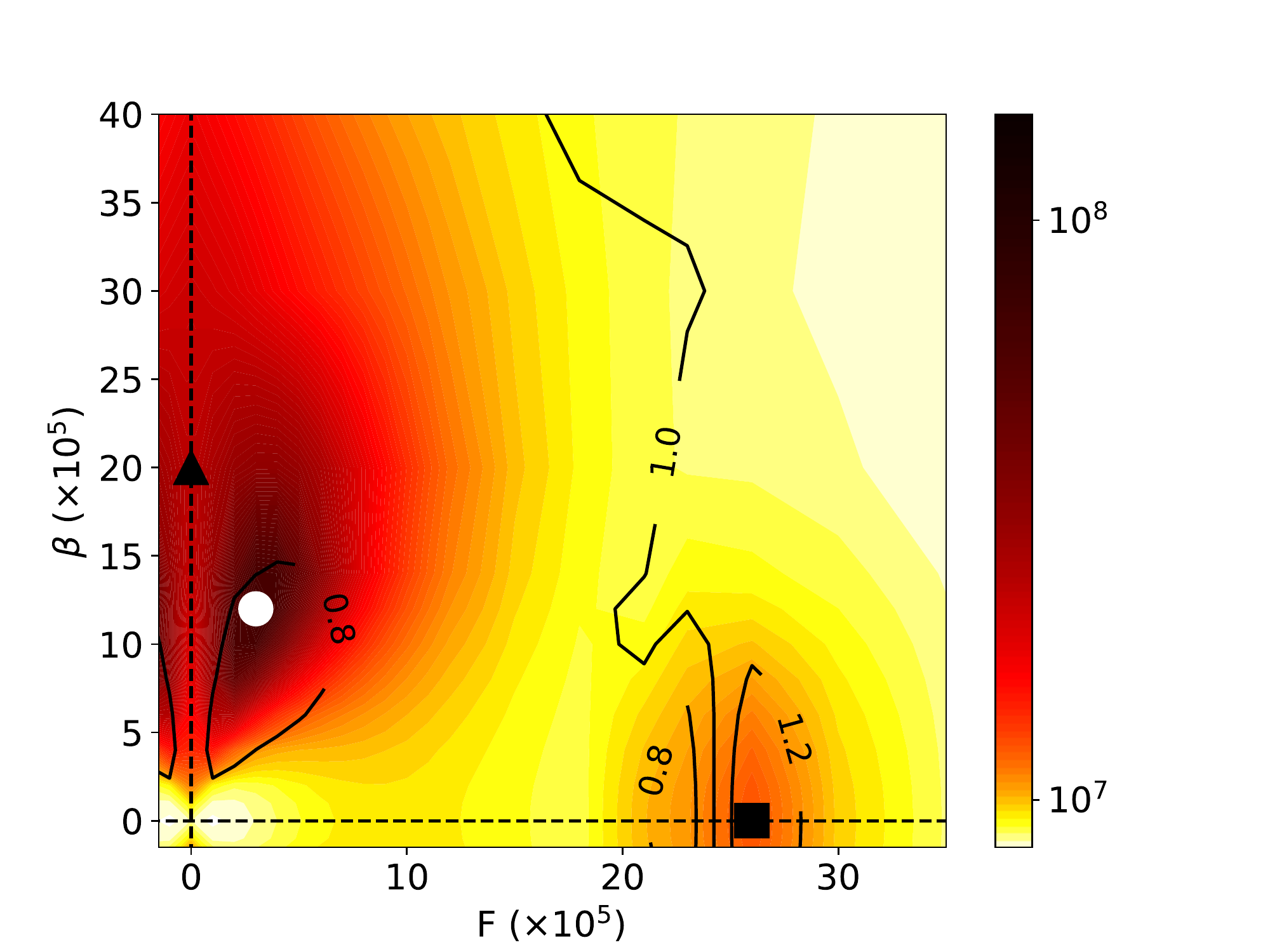}}}
{\subfloat[\label{gainchc}]{\includegraphics[width=0.32\linewidth]{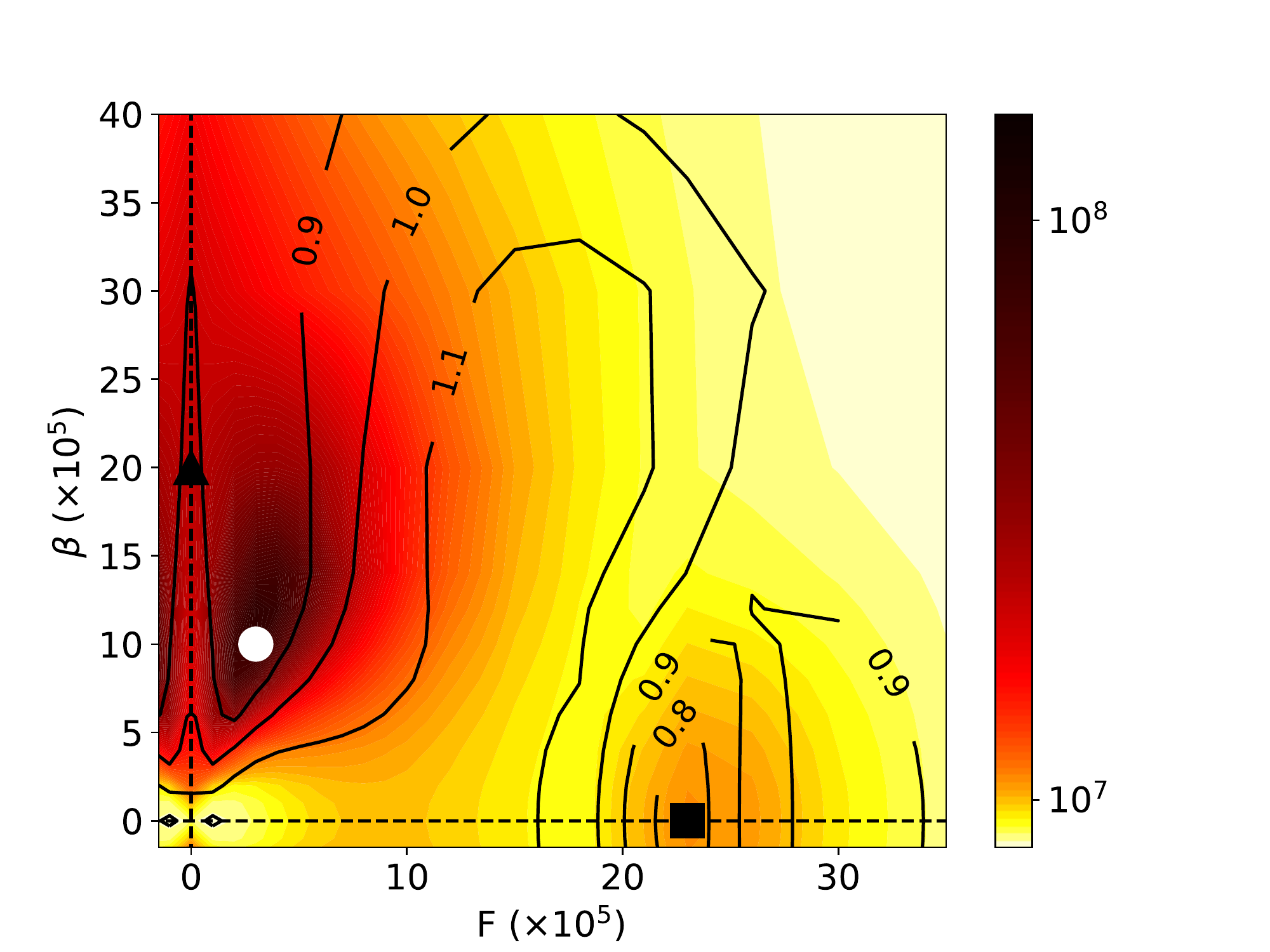}}}
\caption{Optimal gain $\mu_0$ of controlled boundary layers. White circle denotes the first Mack mode, black triangle denotes the streak mode and black square denotes the second Mack mode. Contour lines denote the ratio of the optimal gain with control over the optimal gain without control $\mu_0(C_\theta' \neq 0)/\mu_0(C_\theta' = 0)$ given by Figure \ref{gain3d}. (a)  Heating strip upstream with $C_\theta'=0.9 \times 10^{-2}$. (b) Cooling strip downstream with $C_\theta'=0.9 \times 10^{-2}$. (c) Heating strip upstream and cooling strip downstream with $C_\theta'=0.9 \times 10^{-2}$.}
\label{gain2cs3}
\end{figure}

We first study the effect of both strips independently and then together, at the same level of $C_\theta'$. The temperature fields of the base-flows are plotted in Figure \ref{tcs} and the resolvent gain maps in Figure \ref{gain2cs3}. First, for a single heating strip (Figure \ref{gainchh}), two specific $F-\beta$ regions are damped: the first Mack mode / Streaks region and the second Mack mode region. There is also a region in the $F-\beta$ plane, where the instabilities are strengthened by the control, but the instabilities were weak initially in that region, so this actuator indeed manages to mitigate all instabilities. Then, for a single cooling strip (Figure \ref{gainccc}), instabilities for all wavenumbers and frequencies below $F=2.4 \times 10^{-4}$ are damped. Similarly to the cases where the full heat flux gradient profile was prescribed at the wall, the second Mack mode is shifted to higher frequencies and amplified. Therefore, this actuator is very efficient to damp the first Mack mode but it is not robust to mitigate all the instabilities as a second Mack mode appearing at a larger frequency might be promoted by the cooling strip. Eventually, the control through a heating strip upstream and a cooling strip downstream (Figure \ref{gainchc}) leads to a similar resolvent map and reduces by the same amount the optimal gains of both Mack modes than with a single heating strip. Moreover, the streaks are more damped with the additional cooling strip. Therefore, for the same spent energy, depending on the context, two strips may represent a better actuator as each strip requires to produce a smaller heat flux than a single heating strip but the mechanism would involve both a heating and a cooling strip.

To conclude, a relevant initial design for a wall actuator would be a thin steady heating strip located close to the leading edge (and possibly a second cooling strip downstream) which would modify the base-flow so that the boundary layer is less receptive to streaks, first and second Mack modes. However, only linear computations were performed in this work. To properly design an optimal finite amplitude actuator, one must perform a full nonlinear evaluation (base-flow/stability/sensitivity analyses).

\section{Conclusion}

Optimal steady wall-blowing and wall-heating actuator locations for a Mach $4.5$ boundary layer over an adiabatic flat plate have been found. Firstly, resolvent analyses have been performed around a base-flow highlighting the three main instabilities: the streaks, first and second Mack modes. An adjoint-based optimisation technique then allowed to identify the optimal steady wall-actuators to mitigate/strengthen the various instabilities. 

Wall blowing/suction control induces streamwise momentum modifications in the corresponding sensitive region of the flow i.e. between the critical layer and the boundary layer thickness for both Mack modes. Wall heating/cooling control acts in the region of the flow sensitive to temperature variations for the first Mack mode i.e. within the boundary layer close to the wall but fails to induce temperature modifications above the critical layer where it is more optimal for the second Mack mode. For both Mack modes, the variations of optimal gain are driven by the streamwise momentum modifications always counterbalanced by the entropy modifications, except for the wall heat flux control of the second Mack mode where it is the entropy which promotes the optimal gain modification.

For steady wall-blowing control, several conclusions are drawn. First, the second Mack mode is the most sensitive to blowing/suction control. Suction control is optimal to damp the mode if located upstream of the synchronisation point and conversely for blowing control, in agreement with previous results found for roughness control. Secondly, the second Mack mode is optimally damped by a local suction device located in the region of branch I of the local mode S while the optimal region for the first Mack mode is around the maximum amplification rate of the local first mode. A steady blowing can damp the second Mack mode only if it is applied in the region of branch II of mode S. The application of the optimal wall suction and blowing control, computed for the second Mack mode, in the non-linear regime (finite amplitude control) yields a reduction of the optimal gain of the second Mack mode but shifts its peak to lower or higher frequencies, triggers a small decrease of the first Mack mode gains and lets the streaks unaffected. Therefore, only a local suction actuator would efficiently damp all instabilities.

The stability and sensitivity analyses have been repeated to find an optimal steady wall heat flux actuator. The first Mack mode appears as the most sensitive instability to heating/cooling control. In agreement with previous findings \citep{mack1993effect}, the first and second Mack modes have an opposite sensitivity with respect to wall temperature changes, however, wall heating located close to the leading edge damps both Mack modes. In the downstream region, a large cooling region damps the first Mack mode while only a local cooling strip located in the unstable region of mode S damps the second Mack mode but this is suboptimal in comparison with the leading edge region. The application of the optimal wall heating and cooling control, computed for the first Mack mode, at a non-linear regime yields a strong reduction of the first Mack mode gains but strongly amplifies the second Mack mode. The application of a single local steady heating strip close to the leading edge (and possibly a cooling strip downstream in the unstable region of mode S) manages to damp all the instabilities and might be considered as an actuator to delay transition towards turbulence for a various range of frequencies and spanwise wavenumbers.

\section*{Acknowledgements}
This work is funded by the French Agency for Innovation and Defence (AID) as part of the UK-FR PhD program. G. R. acknowledges funding from the Air Force Office of Scientific Research (AFOSR)/European Office of Aerospace Research and Development (EOARD) (Award FA8655-21-1-7009). We are grateful to X. Chanteux for the share of his LST code.

\section*{Declaration of Interests}
The authors report no conflict of interest.

\appendix{}

\section{Local stability analysis} \label{sec:lst}

In this study, local stability analysis is applied only to characterise the gradients given by the sensitivity of the optimal gains but is never required to perform the sensitivity analysis itself.

Linear local stability theory (LST) is valid only on parallel flows in the streamwise direction. Under this assumption, an ansatz of the fluctuations is made of the form
\begin{equation}
\mathbf{q}' = \mathbf{\hat{q}}(y) e^{i(\alpha x - \omega t+ \beta z)}.
\label{loc}
\end{equation}
For the spatial LST \citep{schmid2002stability}, the frequency $\omega$ is set real ($\omega = \omega_r$) and the streamwise wavenumber $\alpha$ is complex ($\alpha = \alpha_r + i \alpha_i$). Linearising the Navier-Stokes equations and injecting \eqref{loc} leads to a dispersion relation between $\alpha$, $\omega$, $Re_x$ and $M$. Fixing the three latter parameters, the dispersion relation gives the complex value of $\alpha$. Depending on the sign of $\alpha_i$, the disturbances exponentially decay ($\alpha_i>0$) or grow ($\alpha_i<0$) in the streamwise direction. The phase velocity $c$ is defined as
\begin{equation}
c = c_r + i c_i = \frac{\omega}{\alpha} = \frac{\omega_r \alpha_r}{|\alpha|^2} - i \frac{\omega_r \alpha_i}{|\alpha|^2}.
\label{loc2}
\end{equation}

The linear local stability (LST) numerical method relies on a Chebyshev collocation method for the wall-normal direction and the dispersion relation is solved through LAPACK library. An isothermal boundary condition is prescribed for the linearised disturbances ($T'=0$ at wall). The code has been developed by \citet{chanteux2022construction} following \citet{saint2020prevision} framework and validated on a supersonic boundary layer by \citet{nibourel2023reactive}.

\section{Details of the sensitivity to base-flow modifications} \label{sec:bfvar}

The derivation of the sensitivity of the optimal gain to base-flow modifications is detailed:

\begin{eqnarray}
\langle \frac{\partial \mathcal{L}}{\partial \mathbf{\overline{q}}}, \delta \mathbf{\overline{q}} \rangle &=& \langle \bm{\lambda}_1, \mu_i \frac{\partial (\mathbf{A} \mathbf{\check{q}}_i)}{\partial \mathbf{q}}  \delta \mathbf{\overline{q}} \rangle + \langle \bm{\lambda}_2 ,  \mu_i \frac{\partial (\mathbf{Q}_q \mathbf{\check{q}}_i)}{\partial \mathbf{q}}  \delta \mathbf{\overline{q}} + \frac{\partial (\mathbf{A^*} \mathbf{\check{a}})}{\partial \mathbf{q}}  \delta \mathbf{\overline{q}}\rangle - \langle \bm{\lambda}_3 , \mu_i^2 \frac{\partial (\mathbf{Q}_f \mathbf{\check{f}}_i)}{\partial \mathbf{q}}  \delta \mathbf{\overline{q}} \rangle \nonumber \\
&=& \langle \mathbf{\check{a}} , \mu_i \frac{\partial (\mathbf{A} \mathbf{\check{q}}_i)}{\partial \mathbf{q}} \delta \mathbf{\overline{q}} \rangle + \langle \mu_i \mathbf{\check{q}}_i ,  \mu_i \frac{\partial (\mathbf{Q}_q \mathbf{\check{q}}_i)}{\partial \mathbf{q}} \delta \mathbf{\overline{q}} + \frac{\partial (\mathbf{A^*} \mathbf{\check{a}})}{\partial \mathbf{q}} \delta \mathbf{\overline{q}}\rangle - \langle \mathbf{\check{f}}_i , \mu_i^2 \frac{\partial (\mathbf{Q}_f \mathbf{\check{f}}_i)}{\partial \mathbf{q}} \delta \mathbf{\overline{q}} \rangle \nonumber \\
&=& \mu_i \langle \left( \frac{\partial (\mathbf{A} \mathbf{\check{q}}_i)}{\partial \mathbf{q}}  \right)^* \mathbf{\check{a}} , \delta \mathbf{\overline{q}} \rangle + \mu_i \langle \left( \frac{\partial (\mathbf{A}^* \mathbf{\check{a}})}{\partial \mathbf{q}}  \right)^* \mathbf{\check{q}}_i,  \delta \mathbf{\overline{q}}\rangle  \nonumber \\ 
& + & \mu_i^2 \langle \left( \frac{\partial (\mathbf{Q}_q \mathbf{\check{q}}_i)}{\partial \mathbf{q}} \right)^* \mathbf{\check{q}}_i, \delta \mathbf{\overline{q}}\rangle  - \mu_i^2 \langle \left( \frac{\partial (\mathbf{Q}_f \mathbf{\check{f}}_i)}{\partial \mathbf{q}} \right)^* \mathbf{\check{f}}_i ,  \delta \mathbf{\overline{q}} \rangle.
\label{sens40}
\end{eqnarray}
By writing the operators in component notations, one can find that $\left( \frac{\partial (\mathbf{A}^* \mathbf{\check{a}})}{\partial \mathbf{q}}  \right)^* \mathbf{\check{q}}_i = \mathbf{\check{a}}^* \left( \frac{\partial (\mathbf{A} \mathbf{\check{q}}_i)}{\partial \mathbf{q}}  \right) $ and therefore the sum of both complex conjugates arises twice the real part:
\begin{eqnarray}
\langle \frac{\partial \mathcal{L}}{\partial \mathbf{\overline{q}}}, \delta \mathbf{\overline{q}} \rangle &=& \langle 2\mu_i \operatorname{Re}\left(\left( \frac{\partial (\mathbf{A} \mathbf{\check{q}}_i)}{\partial \mathbf{q}} \right)^* \mathbf{\check{a}} \right) + \mu_i^2  \left( \frac{\partial (\mathbf{Q}_q \mathbf{\check{q}}_i)}{\partial \mathbf{q}} \right)^* \mathbf{\check{q}}_i  \nonumber \\
&-& \mu_i^2  \left( \frac{\partial (\mathbf{Q}_f \mathbf{\check{f}}_i)}{\partial \mathbf{q}} \right)^* \mathbf{\check{f}}_i ,  \delta \mathbf{\overline{q}} \rangle,
\label{sens4}
\end{eqnarray}
and getting from Eq. \eqref{sensapp} that $\mu_i \mathbf{Q}_q \mathbf{\check{q}}_i = \mathbf{\mathcal{R}}^{-1*} \mathbf{\check{a}}$, one has:
\begin{eqnarray}
\langle \frac{\partial \mathcal{L}}{\partial \mathbf{\overline{q}}}, \delta \mathbf{\overline{q}} \rangle &=& \langle 2\mu_i^2 \operatorname{Re}\left( \left( \frac{\partial (\mathbf{A} \mathbf{\check{q}}_i)}{\partial \mathbf{q}} \right)^* \mathbf{\mathcal{R}}^{*} \mathbf{Q}_q \mathbf{\check{q}}_i \right) + \mu_i^2  \left( \frac{\partial (\mathbf{Q}_q \mathbf{\check{q}}_i)}{\partial \mathbf{q}} \right)^* \mathbf{\check{q}}_i   \nonumber \\ 
&-& \mu_i^2  \left( \frac{\partial (\mathbf{Q}_f \mathbf{\check{f}}_i)}{\partial \mathbf{q}} \right)^* \mathbf{\check{f}}_i,  \delta \mathbf{\overline{q}} \rangle, \nonumber \\
\Rightarrow \nabla_{\mathbf{\overline{q}}} \mu_i ^2 &=&  2\mu_i^2 \operatorname{Re}\left( \mathbf{H}'^* \mathbf{\mathcal{R}}^{*} \mathbf{Q}_q \mathbf{\check{q}}_i \right) + \mu_i^2  \left( \frac{\partial (\mathbf{Q}_q \mathbf{\check{q}}_i)}{\partial \mathbf{q}} \right)^* \mathbf{\check{q}}_i - \mu_i^2  \left( \frac{\partial (\mathbf{Q}_f \mathbf{\check{f}}_i)}{\partial \mathbf{q}} \right)^* \mathbf{\check{f}}_i,
\label{sens4b}
\end{eqnarray}
where $\operatorname{Re}(\cdot)$ is the real part.
Expressing the sparse Hessian operator $\mathbf{H}$ as:
\begin{equation}
H_{ijk} = \left. \frac{\partial^2 R_i}{\partial q_j \partial q_k} \right|_{\mathbf{\overline{q}}},
\label{hesa}
\end{equation}
we get that $\mathbf{H}(\mathbf{\check{q}}, \delta \mathbf{\overline{q}}) = \frac{\partial \left( \mathbf{A} \mathbf{\check{q}} \right)}{\partial \mathbf{q}}  \delta \mathbf{\overline{q}}$. We can then define the matrix $\mathbf{H}'$ as $\mathbf{H}' \delta \mathbf{\overline{q}}= \mathbf{H}(\mathbf{\check{q}}, \delta \mathbf{\overline{q}})$ and then the equation \eqref{sens4b} gives the sensitivity of the optimal gain to base-flow modifications Eq. \eqref{sensapp3}.

\section{Sensitivity to divergence-free base-flow modifications} \label{sec:appdivfree}

From the sensitivity to base-flow modifications $ \nabla_{\mathbf{\overline{q}}} \mu_i^2$ given by Eq. \eqref{sensapp3}, a restriction to divergence-free base-flow variations can be performed. Denoting $\nabla_{\bm{\overline{\rho u}}}^{df} \mu_i^2$ and $\nabla_{\bm{\overline{\rho v}}}^{df} \mu_i^2$ the sensitivities to divergence-free modifications of the base-flow momentum, any divergence-free base-flow modifications $\delta^{df} \overline{\bm{\rho u}}$ and $\delta^{df} \overline{\bm{\rho v}}$ yields:
\begin{equation}
\delta \mu_i^2 = \langle \nabla_{\bm{\overline{\rho u}}}^{df} \mu_i^2, \delta^{df} \overline{\bm{\rho u}} \rangle + \langle \nabla_{\bm{\overline{\rho v}}}^{df} \mu_i^2, \delta^{df} \overline{\bm{\rho v}} \rangle = \langle \nabla_{\bm{\overline{\rho u}}} \mu_i^2, \delta^{df} \overline{\bm{\rho u}} \rangle + \langle \nabla_{\bm{\overline{\rho v}}} \mu_i^2, \delta^{df} \overline{\bm{\rho v}} \rangle.
\label{div0}
\end{equation}
As the sensitivity and the momentum base-flow modifications are both divergence-free, they can be written as functions of scalars:
\begin{eqnarray}
\nabla_{\bm{\overline{\rho u}}}^{df} \mu_i^2 = \frac{\partial \psi}{\partial y}, \;\;\;\;  \nabla_{\bm{\overline{\rho v}}}^{df} \mu_i^2 = -\frac{\partial \psi}{\partial x}, \;\;\;\;  \delta^{df} \overline{\bm{\rho u}} = \frac{\partial \phi}{\partial y}, \;\;\;\;  \delta^{df} \overline{\bm{\rho v}} = - \frac{\partial \phi}{\partial x}.
\label{div1}
\end{eqnarray}
Injecting Eq. \eqref{div1} into Eq. \eqref{div0} and assuming that the equation should be valid for any scalar $\phi$, an integration by parts leads to the Poisson equation:
\begin{equation}
-\Delta \psi = \frac{\partial \nabla_{\bm{\overline{\rho v}}} \mu_i^2}{\partial x} - \frac{\partial \nabla_{\bm{\overline{\rho u}}} \mu_i^2}{\partial y},
\label{div2}
\end{equation}
with the boundary conditions $\nabla \psi \cdot \mathbf{n} = n_y \nabla_{\bm{\overline{\rho u}}} \mu_i^2 -  n_x \nabla_{\bm{\overline{\rho v}}} \mu_i^2$ and $\mathbf{n} = (n_x,n_y)$ the boundary normal.

\section{Wall boundary condition for blowing and heat flux} \label{sec:bcwall}

The implementation of the boundary conditions in the code BROADCAST is described in further details in \citet{poulain2023broadcast}. They are enforced through the addition of ghost cells to prescribe characteristic-based conditions for permeable boundaries or mixed Dirichlet-Neumann conditions for solid boundaries. The wall boundary conditions with non-zero heat-flux and wall-normal velocity is derived below as they play a major role in the computation of the sensitivity.

The boundary layer assumption is performed involving a zero pressure gradient in the wall-normal direction: $\partial p/\partial n=0$ with $n$ the wall-normal direction. From the derivative in the wall-normal direction of the perfect gas law (Eq. \eqref{NS2}), computed at the wall, one gets:
\begin{equation}
\left. \frac{\partial p}{\partial n} \right|_w = \rho_w r \left. \frac{\partial T}{\partial n}\right|_w + r T_w \left.\frac{\partial \rho}{\partial n}\right|_w,
\label{dergas}
\end{equation}
with the subscript $\cdot_w$ indicating the values at the wall. The temperature $T_w$ is replaced by $p_w/(\rho_w r)$ from the perfect gas law. The pressure at the wall $p_w$ is known from the value in the first cell as $\partial p/\partial n=0$ and the temperature gradient writes as a heat flux $\phi=\lambda \partial T/\partial n$. Therefore, assuming a first order extrapolation for the density, Eq. \eqref{dergas} becomes a function of a single unknown, the wall density $\rho_w$:
\begin{equation}
-\frac{r}{2 \lambda p_w} \phi_w \rho_w^2 + \rho_w - \rho_1 = 0,
\label{dergas2}
\end{equation}
with $\rho_1$ the density at the first cell center. For an adiabatic wall condition $\phi_w=0$, we get $\rho_w=\rho_1$, otherwise we obtain $\rho_w = (1-\sqrt{1-2r\phi_w \rho_1/(\lambda p_w)})/(r\phi_w /(\lambda p_w))$. However, the latter expression for $\rho_w$ is not differentiable around $\phi_w=0$. Then, in order to compute the Jacobian and Hessian of an adiabatic flow, which respectively involve the first and second derivative of the wall boundary conditions, the equation Eq. \eqref{dergas2} is solved by a Newton method to compute $\rho_w$, this iterative method being then linearised by Algorithmic Differentiation to build the derivative operators. Eventually, all the conservative variables are prescribed at the wall: $\rho_w$, $(\rho u)_w = 0$ (no-slip), $(\rho v)_w = \rho_w \times v_w$ (with a non-zero $v_w$ for a wall-normal blowing/suction condition) and $(\rho E)_w = p_w/(\gamma -1) + 0.5 \rho_w v_w^2$.

\section{Validation on a low Mach number boundary layer} \label{sec:valid}

The discrete linear sensitivity of the optimal gain to steady blowing is validated on an adiabatic flat plate at low Mach number regime against \citet{brandt2011effect} where the sensitivity analysis had been derived in a continuous and incompressible framework.

The freestream Mach number is $M=0.1$, the Reynolds number is $Re=6 \times 10^5$. The computational domain extends from $Re_{x,\text{in}} = 4300$ to $Re_{x,\text{out}} = 7.5 \times 10^5$, the optimisation for the resolvent mode is restrained up to $Re_x \leq 6 \times 10^5$ inside $\mathbf{Q}_q$ and $\mathbf{P}$. The height of the domain is about $ 9\,\delta$ with $\delta$ the displacement thickness at $Re_x = 6 \times 10^5$. The following boundary conditions are applied: Blasius solution prescribed at the inlet (non-reflecting subsonic), adiabatic no-slip wall at the bottom, non-reflecting condition at the top and non-reflecting subsonic condition at the outlet with the free-stream pressure taken as reference (zero pressure gradient assumed). A Cartesian mesh similar to the hypersonic boundary layer case is chosen with $(N_x, N_y) = (1000, 150)$. 

\begin{figure}
\centering
\includegraphics[width=0.5\textwidth]{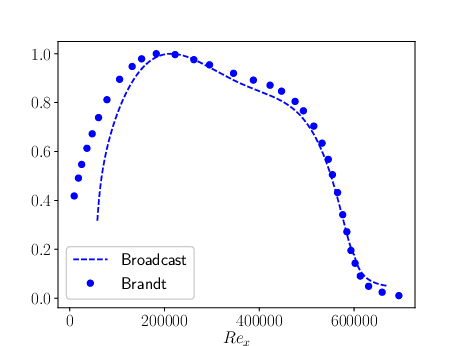}
\caption{Normalised sensitivity of the optimal gain of the streaks ($\beta \delta = 0.94$) to wall blowing $\nabla_{\mathbf{v}_w} \mu^2$. Comparison with \citet{brandt2011effect}.}
\label{sensInc}
\end{figure}

In \citet{brandt2011effect}, the largest optimal gain at zero frequency is obtained for the streaks at $\beta \delta = 0.94$. The sensitivity of this 3D mode to steady wall blowing is plotted in Figure \ref{sensInc}. They show good agreement while we compare two different frameworks: discrete linearisation in a low Mach compressible framework in our work while \citet{brandt2011effect} used a continuous linearisation in the incompressible framework. However, slight discrepancies are noticed in the leading edge part which can be explained by the fact that the leading edge was included inside the computational domain used in \citet{brandt2011effect} while we start at $Re_{x,\text{in}} = 4300$.

\section{Gradient dependence with frequency} \label{sec:appsize}

\begin{figure}
\centering
\subfloat[\label{gain2dshort}]{\includegraphics[width=0.5\textwidth]{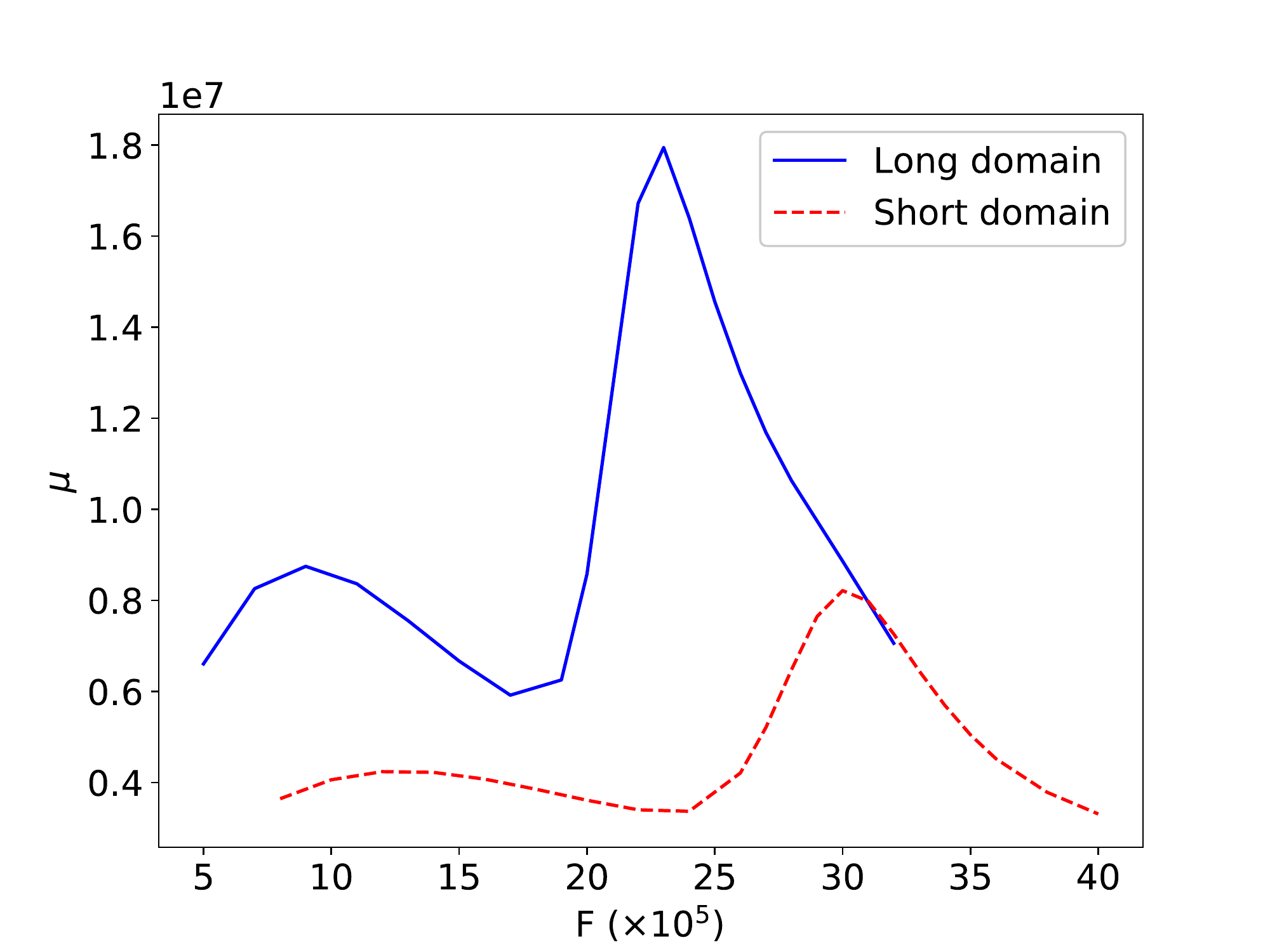}}
{\subfloat[\label{dfdchushort}]{\includegraphics[width=0.5\linewidth]{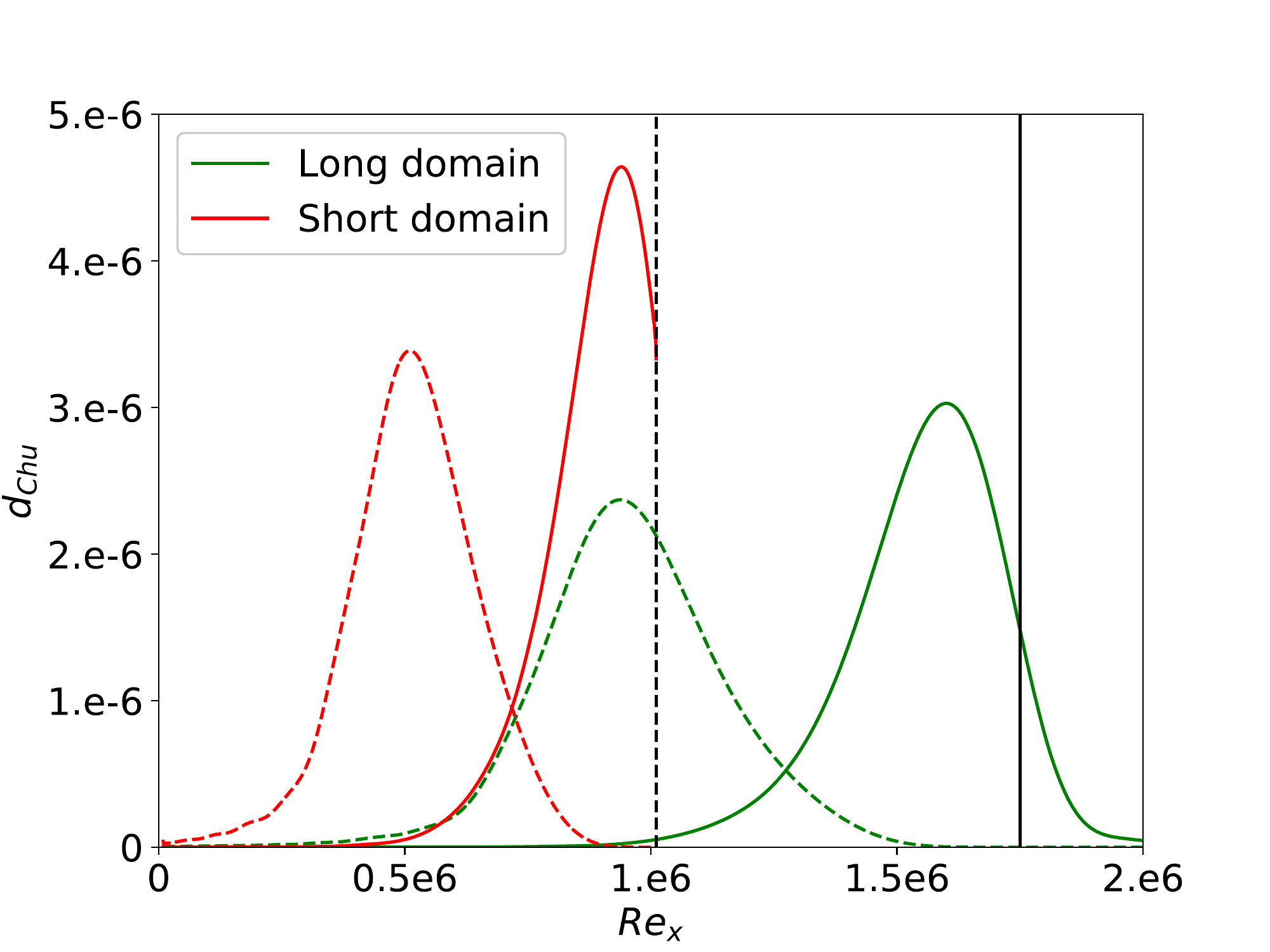}}}
\caption{Resolvent analysis on short and long domains. (a) Two-dimensional optimal gain $\mu_0$ with respect to the frequency $F$ for short (red) and long (blue) domains. (b) Energy density $d_\text{Chu}$ of the optimal forcing (dashed lines) and response (solid lines) of the second Mack mode for the short (red) and long (green) domains. Black vertical lines indicate the end of the optimisation domain for resolvent analysis for the short domain (dashed line) and for the long domain (solid line).}
\label{resolventshort}
\end{figure}

\begin{figure}
\centering
\subfloat[\label{sensshortvw}]{\includegraphics[width=0.5\textwidth]{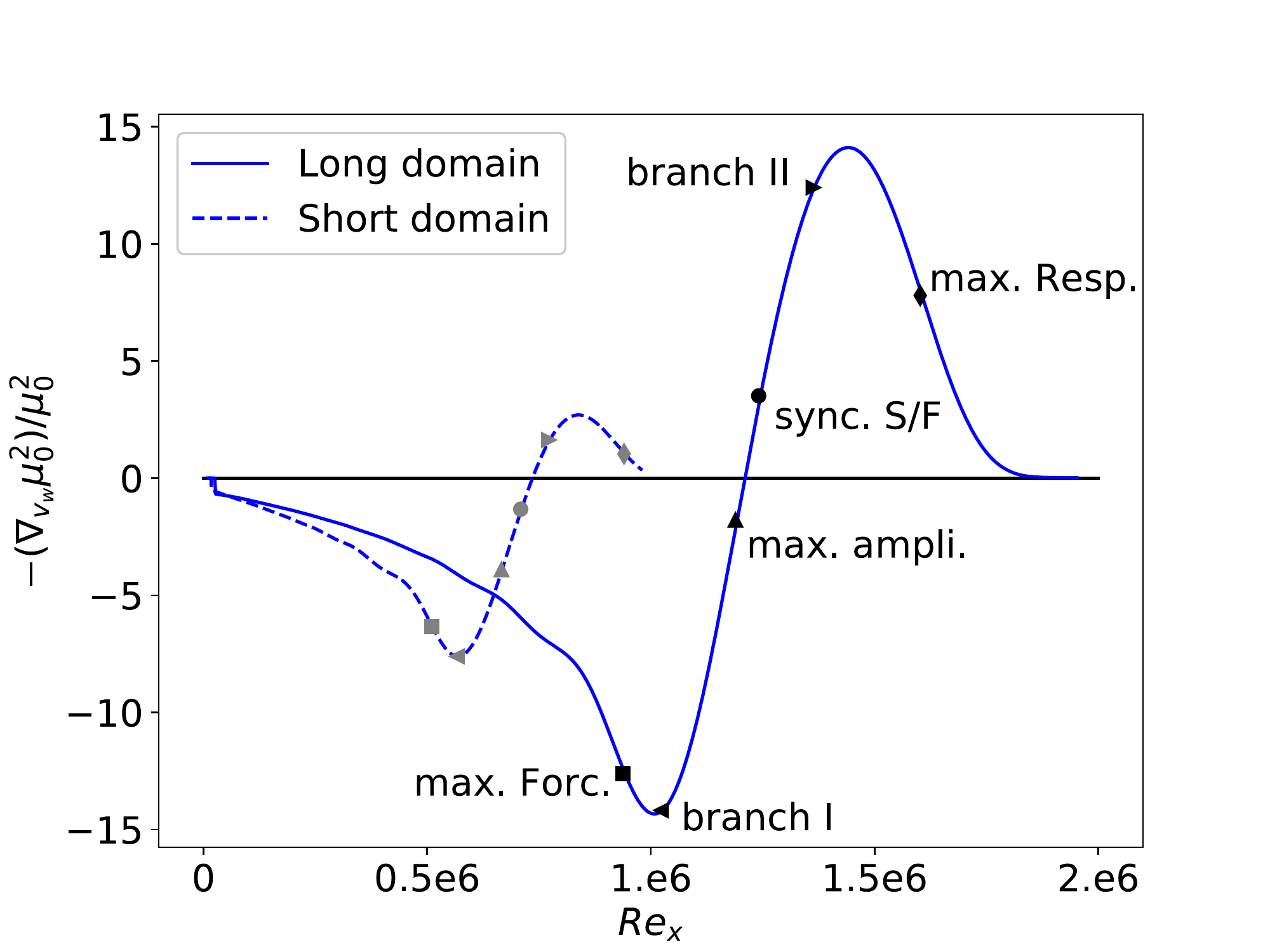}}
{\subfloat[\label{sensshorttw}]{\includegraphics[width=0.5\linewidth]{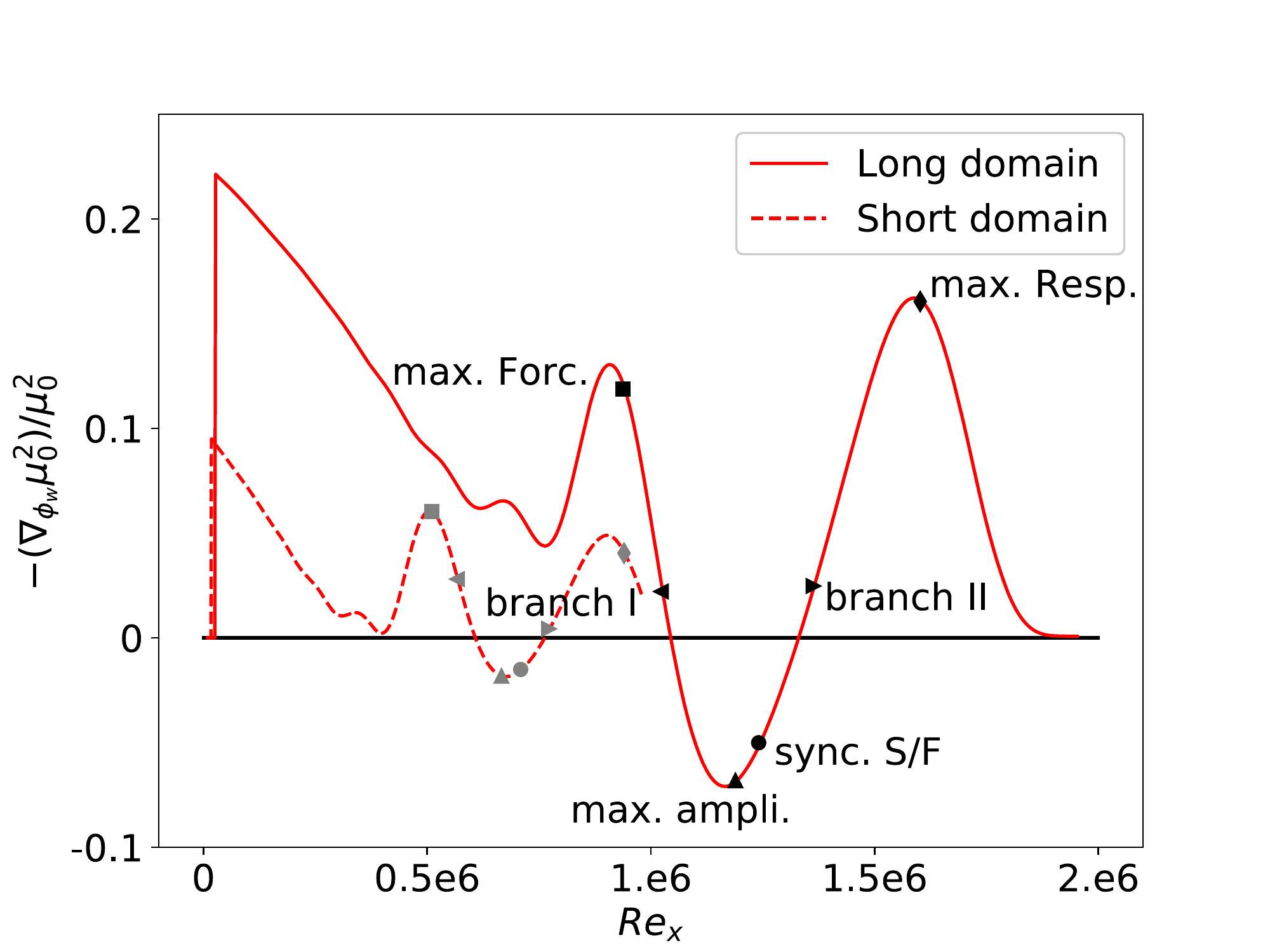}}}
\caption{Opposite of the sensitivity of the optimal gain for short (dashed lines) and long (solid lines) domains. (a) $-\nabla_{\mathbf{v}_w} \mu_0^2/\mu_0^2$ for short and long domains. (b) $-\nabla_{\bm{\phi}_w} \mu_0^2/\mu_0^2$ for short and long domains.}
\label{sensshort}
\end{figure}

To check how the gradients evolve with the instability frequency, 2D stability and sensitivity are repeated on a shorter domain which ends at $Re_{x,\text{out}} = 1 \times 10^6$ (half of the previous one). From Figure \ref{gain2dshort}, the second Mack mode is maximal for larger frequencies ($F=3 \times 10^{-4}$). The gradients to steady wall control (Figure \ref{sensshort}) preserve a very similar trend in terms of variations along $Re_x$ however their relative amplitude vary. 
Eventually, the gradient zones highlighted by local or global stability analyses remain similar on both domain length.

\section{Gradient to wall temperature and to wall heat-flux} \label{sec:appheatflux}

\begin{figure}
\centering
\includegraphics[width=0.5\textwidth]{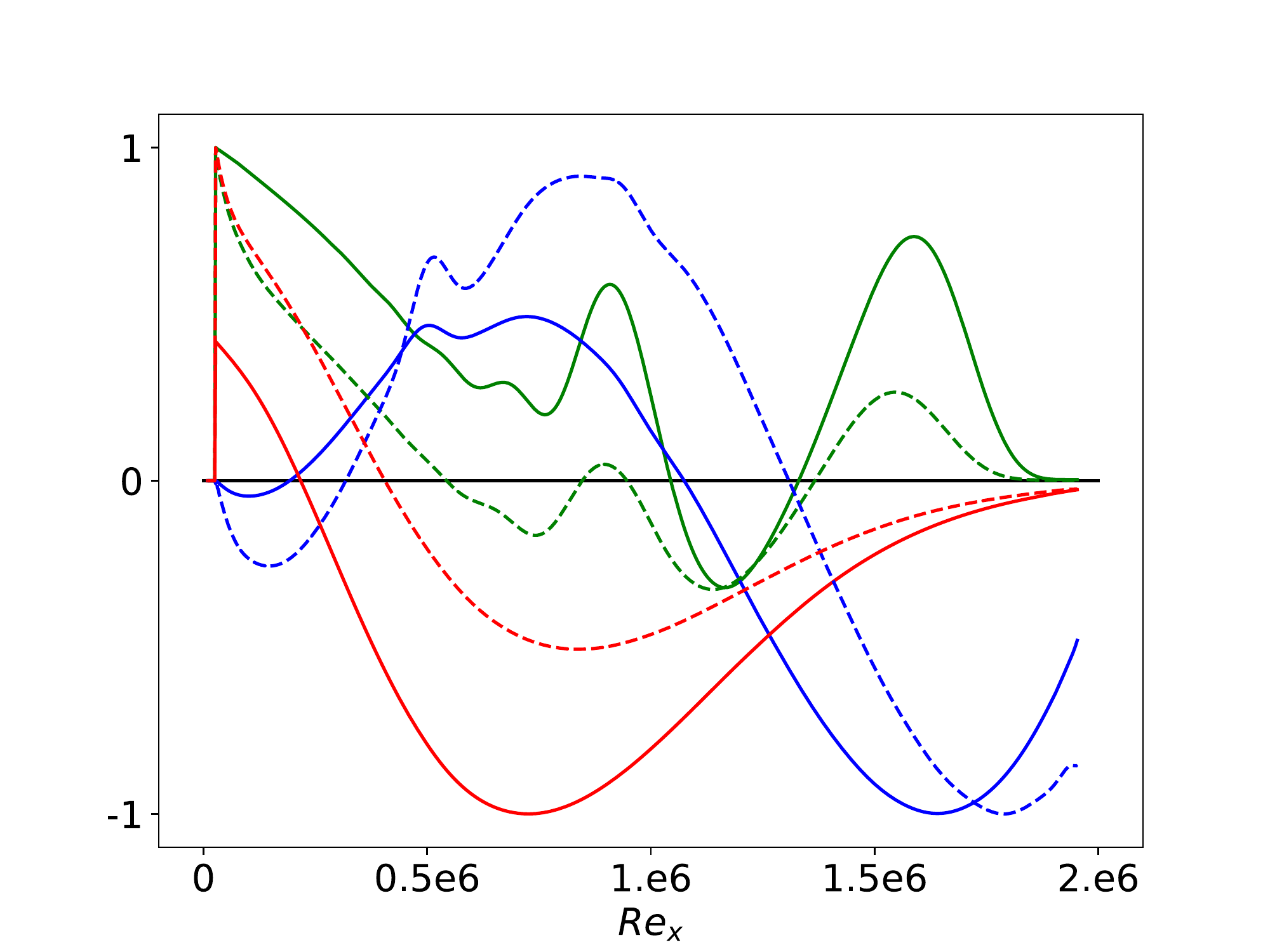}
\caption{Comparison of the opposite of the  sensitivity of the optimal gain for the streaks (blue), the first (red) and second (green) Mack modes to steady wall heat flux $-\nabla_{\bm{\phi}_w} \mu_0^2$ (solid lines) with to wall heating $-\nabla_{\mathbf{T}_w} \mu_0^2$ (dashed lines) normalised by their maximum.}
\label{sensfluxtemp}
\end{figure}

The gradient of the optimal gain to steady wall-temperature $-\nabla_{\mathbf{T}_w} \mu_0^2$ and to steady wall heat flux $-\nabla_{\bm{\phi}_w} \mu_0^2$ are compared in Figure \ref{sensfluxtemp}. The trend for the three instabilities are similar however, it might be noticed that the gradient to wall heat flux for the first Mack mode is relatively much lower at the leading edge than to wall temperature. It may be partially explained by the largest difference at leading edge between the prescribed constant wall temperature ($\mathbf{T}_w = 4.395 T_\infty$) for the isothermal case ($\nabla_{\mathbf{T}_w} \mu_0^2$ case) and the adiabatic wall temperature ($\nabla_{\bm{\phi}_w} \mu_0^2$ case). Therefore, both gradients provide the same knowledge on the sensitivity location. The analysis of the gradient to wall heat flux has been performed in this work as the input data during an experiment would be a heat flux and not a temperature.

\bibliographystyle{jfm}
\bibliography{biblioAP2}

\begin{thebibliography}{65}
\expandafter\ifx\csname natexlab\endcsname\relax\def\natexlab#1{#1}\fi
\def\au#1{#1} \def\ed#1{#1} \def\yr#1{#1}\def\at#1{#1}\def\jt#1{\textit{#1}}
  \def\bt#1{#1}\def\bvol#1{\textbf{#1}} \def\vol#1{#1} \def\pg#1{#1}
  \def\publ#1{#1}\def\arxiv#1{#1}\def\org#1{#1}\def\st#1{\textit{#1}}

\bibitem[Airiau {\em et~al.\/}(2003)Airiau, Bottaro, Walther \&
  Legendre]{airiau2003methodology}
{\sc \au{Airiau, Christophe}, \au{Bottaro, Alessandro}, \au{Walther, Steeve} \&
  \au{Legendre, Dominique}} \yr{2003}  \at{A methodology for optimal laminar
  flow control: Application to the damping of tollmien--schlichting waves in a
  boundary layer}.  \jt{Physics of Fluids}  \bvol{15}~(5),  \pg{1131--1145}.

\bibitem[Amestoy {\em et~al.\/}(2001)Amestoy, Duff, L'Excellent \&
  Koster]{amestoy2001fully}
{\sc \au{Amestoy, Patrick}, \au{Duff, Iain}, \au{L'Excellent, Jean-Yves} \&
  \au{Koster, Jacko}} \yr{2001}  \at{A fully asynchronous multifrontal solver
  using distributed dynamic scheduling}.  \jt{SIAM Journal on Matrix Analysis
  and Applications}  \bvol{23}~(1),  \pg{15--41}.

\bibitem[Balakumar \& Hall(1999)]{balakumar1999optimum}
{\sc \au{Balakumar, Ponnampalam} \& \au{Hall, Philip}} \yr{1999}  \at{Optimum
  suction distribution for transition control}.  \jt{Theoretical and
  computational fluid dynamics}  \bvol{13}~(1),  \pg{1--19}.

\bibitem[Balay {\em et~al.\/}(2019)Balay, Abhyankar, Adams, Brown, Brune,
  Buschelman, Dalcin, Dener, Eijkhout, Gropp {\em et~al.\/}]{balay2019petsc}
{\sc \au{Balay, Satish}, \au{Abhyankar, Shrirang}, \au{Adams, Mark}, \au{Brown,
  Jed}, \au{Brune, Peter}, \au{Buschelman, Kris}, \au{Dalcin, Lisandro},
  \au{Dener, Alp}, \au{Eijkhout, Victor}, \au{Gropp, W} \& \au{others}}
  \yr{2019}  \at{Petsc users manual}.  \jt{ANL-95/11} .

\bibitem[Batista \& Kuehl(2020)]{batista2020local}
{\sc \au{Batista, Armani} \& \au{Kuehl, Joseph}} \yr{2020}  \at{Local wall
  temperature effects on the second-mode instability}.  \jt{Journal of
  Spacecraft and Rockets}  \bvol{57}~(3),  \pg{580--595}.

\bibitem[Boujo(2021)]{boujo2021second}
{\sc \au{Boujo, Edouard}} \yr{2021}  \at{Second-order adjoint-based sensitivity
  for hydrodynamic stability and control}.  \jt{Journal of Fluid Mechanics}
  \bvol{920}.

\bibitem[Brandt {\em et~al.\/}(2011)Brandt, Sipp, Pralits \&
  Marquet]{brandt2011effect}
{\sc \au{Brandt, Luca}, \au{Sipp, Denis}, \au{Pralits, Jan} \& \au{Marquet,
  Olivier}} \yr{2011}  \at{Effect of base-flow variation in noise amplifiers:
  the flat-plate boundary layer}.  \jt{Journal of Fluid Mechanics}  \bvol{687},
   \pg{503--528}.

\bibitem[Bugeat {\em et~al.\/}(2019)Bugeat, Chassaing, Robinet \&
  Sagaut]{bugeat20193d}
{\sc \au{Bugeat, Benjamin}, \au{Chassaing, Jean-Camille}, \au{Robinet,
  Jean-Christophe} \& \au{Sagaut, Pierre}} \yr{2019}  \at{3\textsc{D} global
  optimal forcing and response of the supersonic boundary layer}.  \jt{Journal
  of Computational Physics}  \bvol{398},  \pg{108888}.

\bibitem[Chanteux {\em et~al.\/}(2022)Chanteux, B{\'e}gou, Deniau \&
  Vermeersch]{chanteux2022construction}
{\sc \au{Chanteux, Xavier}, \au{B{\'e}gou, Guillaume}, \au{Deniau, Hugues} \&
  \au{Vermeersch, Olivier}} \yr{2022} Construction and application of
  transition prediction databased method for 2nd mode on sharp cone.  \bt{In
  {\em AIAA AVIATION 2022 Forum\/}},  \pg{p. 3470}.

\bibitem[Chu(1965)]{chu1965energy}
{\sc \au{Chu, Boa-Teh}} \yr{1965}  \at{On the energy transfer to small
  disturbances in fluid flow (part i)}.  \jt{Acta Mechanica}  \bvol{1}~(3),
  \pg{215--234}.

\bibitem[Cinnella \& Content(2016)]{cinnella2016high}
{\sc \au{Cinnella, Paola} \& \au{Content, C{\'e}dric}} \yr{2016}
  \at{High-order implicit residual smoothing time scheme for direct and large
  eddy simulations of compressible flows}.  \jt{Journal of Computational
  Physics}  \bvol{326},  \pg{1--29}.

\bibitem[Crivellini \& Bassi(2011)]{crivellini2011implicit}
{\sc \au{Crivellini, Andrea} \& \au{Bassi, Francesco}} \yr{2011}  \at{An
  implicit matrix-free discontinuous galerkin solver for viscous and turbulent
  aerodynamic simulations}.  \jt{Computers \& fluids}  \bvol{50}~(1),
  \pg{81--93}.

\bibitem[Dalcin {\em et~al.\/}(2011)Dalcin, Paz, Kler \&
  Cosimo]{dalcin2011parallel}
{\sc \au{Dalcin, Lisandro}, \au{Paz, Rodrigo}, \au{Kler, Pablo} \& \au{Cosimo,
  Alejandro}} \yr{2011}  \at{Parallel distributed computing using
  \textsc{P}ython}.  \jt{Advances in Water Resources}  \bvol{34}~(9),
  \pg{1124--1139}.

\bibitem[Fedorov \& Khokhlov(2002)]{fedorov2002receptivity}
{\sc \au{Fedorov, Alexander} \& \au{Khokhlov, Andrew}} \yr{2002}
  \at{Receptivity of hypersonic boundary layer to wall disturbances}.
  \jt{Theoretical and Computational Fluid Dynamics}  \bvol{15}~(4),
  \pg{231--254}.

\bibitem[Fedorov {\em et~al.\/}(2014)Fedorov, Ryzhov, Soudakov \&
  Utyuzhnikov]{fedorov2014numerical}
{\sc \au{Fedorov, Alexander}, \au{Ryzhov, Alexander}, \au{Soudakov, Vitaly} \&
  \au{Utyuzhnikov, Sergey}} \yr{2014}  \at{Numerical simulation of the effect
  of local volume energy supply on high-speed boundary layer stability}.
  \jt{Computers \& Fluids}  \bvol{100},  \pg{130--137}.

\bibitem[Fedorov \& Tumin(2011)]{fedorov2011high}
{\sc \au{Fedorov, Alexander} \& \au{Tumin, Anatoli}} \yr{2011}  \at{High-speed
  boundary-layer instability: old terminology and a new framework}.  \jt{AIAA
  \textsc{J}ournal}  \bvol{49}~(8),  \pg{1647--1657}.

\bibitem[Fong {\em et~al.\/}(2014)Fong, Wang \& Zhong]{fong2014numerical}
{\sc \au{Fong, Kahei~Danny}, \au{Wang, Xiaowen} \& \au{Zhong, Xiaolin}}
  \yr{2014}  \at{Numerical simulation of roughness effect on the stability of a
  hypersonic boundary layer}.  \jt{Computers \& Fluids}  \bvol{96},
  \pg{350--367}.

\bibitem[George \& Sujith(2011)]{george2011chu}
{\sc \au{George, K~Joseph} \& \au{Sujith, RI}} \yr{2011}  \at{On \textsc{C}hu's
  disturbance energy}.  \jt{Journal of Sound and Vibration}  \bvol{330}~(22),
  \pg{5280--5291}.

\bibitem[Guo {\em et~al.\/}(2021)Guo, Gao, Jiang \& Lee]{guo2021sensitivity}
{\sc \au{Guo, Peixu}, \au{Gao, Zhenxun}, \au{Jiang, Chongwen} \& \au{Lee,
  Chun-Hian}} \yr{2021}  \at{Sensitivity analysis on supersonic-boundary-layer
  stability subject to perturbation of flow parameters}.  \jt{Physics of
  Fluids}  \bvol{33}~(8),  \pg{084111}.

\bibitem[Hanifi {\em et~al.\/}(1996)Hanifi, Schmid \&
  Henningson]{hanifi1996transient}
{\sc \au{Hanifi, Ardeshir}, \au{Schmid, Peter} \& \au{Henningson, Dan}}
  \yr{1996}  \at{Transient growth in compressible boundary layer flow}.
  \jt{Physics of Fluids}  \bvol{8}~(3),  \pg{826--837}.

\bibitem[Hascoet \& Pascual(2013)]{hascoet2013tapenade}
{\sc \au{Hascoet, Laurent} \& \au{Pascual, Val{\'e}rie}} \yr{2013}  \at{The
  tapenade automatic differentiation tool: principles, model, and
  specification}.  \jt{ACM Transactions on Mathematical Software (TOMS)}
  \bvol{39}~(3),  \pg{1--43}.

\bibitem[Hern{\'a}ndez {\em et~al.\/}(2007)Hern{\'a}ndez, Rom{\'a}n, Tom{\'a}s
  \& Vidal]{hernandez2007krylov}
{\sc \au{Hern{\'a}ndez, Vicente}, \au{Rom{\'a}n, Jose}, \au{Tom{\'a}s,
  Andr{\'e}s} \& \au{Vidal, Vicente}} \yr{2007}  \at{Krylov-\textsc{S}chur
  methods in slepc}.  \jt{Universitat Politecnica de Valencia, Tech. Rep.
  STR-7} .

\bibitem[Huang \& Wu(2016)]{huang2016effect}
{\sc \au{Huang, ZhangFeng} \& \au{Wu, Xuesong}} \yr{2016} The effect of local
  steady suction on the stability and transition of boundary layer on a flat
  plate.  \bt{In {\em 8th AIAA Flow Control Conference\/}},  \pg{p. 3471}.

\bibitem[Huerre \& Monkewitz(1990)]{huerre1990local}
{\sc \au{Huerre, Patrick} \& \au{Monkewitz, Peter~A}} \yr{1990}  \at{Local and
  global instabilities in spatially developing flows}.  \jt{Annual
  \textsc{R}eview of \textsc{F}luid \textsc{M}echanics}  \bvol{22}~(1),
  \pg{473--537}.

\bibitem[Joslin(1998)]{joslin1998overview}
{\sc \au{Joslin, Ronald}} \yr{1998}  \bt{Overview of laminar flow control}.
  {\em Tech. Rep.\/} TP-1998-20705.  \org{NASA}.

\bibitem[Kazakov {\em et~al.\/}(1995)Kazakov, Kogan \&
  Kuparev]{kazakov1995optimization}
{\sc \au{Kazakov, Alexander}, \au{Kogan, Mikhail} \& \au{Kuparev, Vassili}}
  \yr{1995}  \at{Optimization of laminar-turbulent transition delay by means of
  local heating of the surface}.  \jt{Fluid dynamics}  \bvol{30}~(4),
  \pg{563--570}.

\bibitem[Ma \& Zhong(2003)]{ma2003receptivity}
{\sc \au{Ma, Yanbao} \& \au{Zhong, Xiaolin}} \yr{2003}  \at{Receptivity of a
  supersonic boundary layer over a flat plate. part 1. wave structures and
  interactions}.  \jt{Journal of Fluid Mechanics}  \bvol{488},  \pg{31--78}.

\bibitem[Mack(1963)]{mack1963inviscid}
{\sc \au{Mack, Leslie}} \yr{1963}  \at{The inviscid stability of the
  compressible laminar boundary layer}.  \jt{Space Programs Summary}
  \bvol{37},  \pg{23}.

\bibitem[Mack(1993)]{mack1993effect}
{\sc \au{Mack, Leslie}} \yr{1993}  \at{Effect of cooling on boundary-layer
  stability at mach number 3}.  \bt{In {\em Instabilities and turbulence in
  engineering flows\/}},  \pg{pp. 175--188}.  \publ{Springer}.

\bibitem[Malik(1989)]{malik1989prediction}
{\sc \au{Malik, Mujeeb}} \yr{1989}  \at{Prediction and control of transition in
  supersonic and hypersonic boundary layers}.  \jt{AIAA \textsc{J}ournal}
  \bvol{27}~(11),  \pg{1487--1493}.

\bibitem[Marquet {\em et~al.\/}(2008)Marquet, Sipp \&
  Jacquin]{marquet2008sensitivity}
{\sc \au{Marquet, Olivier}, \au{Sipp, Denis} \& \au{Jacquin, Laurent}}
  \yr{2008}  \at{Sensitivity analysis and passive control of cylinder flow}.
  \jt{Journal of Fluid Mechanics}  \bvol{615},  \pg{221--252}.

\bibitem[Mart{\'\i}nez-Cava(2019)]{martinez2019direct}
{\sc \au{Mart{\'\i}nez-Cava, Alejandro}} \yr{2019}  \at{Direct and adjoint
  methods for highly detached flows}. PhD thesis, Espacio.

\bibitem[Martinez-Cava {\em et~al.\/}(2020)Martinez-Cava, Chávez-Modena,
  Valero, de~Vicente \& Ferrer]{martinez-cava_sensitivity_2020}
{\sc \au{Martinez-Cava, Alejandro}, \au{Chávez-Modena, Miguel}, \au{Valero,
  Eusebio}, \au{de~Vicente, Javier} \& \au{Ferrer, Esteban}} \yr{2020}
  \at{Sensitivity gradients of surface geometry modifications based on
  stability analysis of compressible flows}.  \jt{Physical Review Fluids}
  \bvol{5}~(6),  \pg{063902}.

\bibitem[Masad(1995)]{masad1995transition2}
{\sc \au{Masad, Jamal}} \yr{1995}  \at{Transition in flow over heat-transfer
  strips}.  \jt{Physics of Fluids}  \bvol{7}~(9),  \pg{2163--2174}.

\bibitem[Masad \& Abid(1995)]{masad1995transition}
{\sc \au{Masad, Jamal} \& \au{Abid, Ridha}} \yr{1995}  \at{On transition in
  supersonic and hypersonic boundary layers}.  \jt{International Journal of
  Engineering Science}  \bvol{33}~(13),  \pg{1893--1919}.

\bibitem[Mettot(2013)]{mettot2013linear}
{\sc \au{Mettot, Cl{\'e}ment}} \yr{2013}  \at{Linear stability, sensitivity,
  and passive control of turbulent flows using finite differences}. PhD thesis,
  Palaiseau, Ecole polytechnique.

\bibitem[Mettot {\em et~al.\/}(2014)Mettot, Renac \&
  Sipp]{mettot2014computation}
{\sc \au{Mettot, Cl{\'e}ment}, \au{Renac, Florent} \& \au{Sipp, Denis}}
  \yr{2014}  \at{Computation of eigenvalue sensitivity to base flow
  modifications in a discrete framework: Application to open-loop control}.
  \jt{Journal of Computational Physics}  \bvol{269},  \pg{234--258}.

\bibitem[Mir{\'o}~Mir{\'o} \& Pinna(2018)]{miro2018effect}
{\sc \au{Mir{\'o}~Mir{\'o}, Fernando} \& \au{Pinna, Fabio}} \yr{2018}
  \at{Effect of uneven wall blowing on hypersonic boundary-layer stability and
  transition}.  \jt{Physics of Fluids}  \bvol{30}~(8),  \pg{084106}.

\bibitem[Morkovin(1994)]{morkovin1994transition}
{\sc \au{Morkovin, Mark}} \yr{1994}  \at{Transition in open flow systems-a
  reassessment}.  \jt{Bull. Am. Phys. Soc.}  \bvol{39},  \pg{1882}.

\bibitem[Nibourel {\em et~al.\/}(2023)Nibourel, Leclercq, Demourant, Garnier \&
  Sipp]{nibourel2023reactive}
{\sc \au{Nibourel, Pierre}, \au{Leclercq, Colin}, \au{Demourant, Fabrice},
  \au{Garnier, Eric} \& \au{Sipp, Denis}} \yr{2023}  \at{Reactive control of
  second mack mode in a supersonic boundary layer with free-stream
  velocity/density variations}.  \jt{Journal of Fluid Mechanics}  \bvol{954},
  \pg{A20}.

\bibitem[Oz {\em et~al.\/}(2023)Oz, Goebel, Jewell \& Kara]{oz2023local}
{\sc \au{Oz, Furkan}, \au{Goebel, Thomas~E}, \au{Jewell, Joseph~S} \& \au{Kara,
  Kursat}} \yr{2023}  \at{Local wall cooling effects on hypersonic
  boundary-layer stability}.  \jt{Journal of Spacecraft and Rockets}
  \bvol{60}~(2),  \pg{412--426}.

\bibitem[Park \& Zaki(2019)]{park2019sensitivity}
{\sc \au{Park, Junho} \& \au{Zaki, Tamer}} \yr{2019}  \at{Sensitivity of
  high-speed boundary-layer stability to base-flow distortion}.  \jt{Journal of
  Fluid Mechanics}  \bvol{859},  \pg{476--515}.

\bibitem[Poinsot \& Lele(1992)]{poinsot1992boundary}
{\sc \au{Poinsot, Thierry} \& \au{Lele, Sanjiva}} \yr{1992}  \at{Boundary
  conditions for direct simulations of compressible viscous flows}.
  \jt{Journal of Computational Physics}  \bvol{101}~(1),  \pg{104--129}.

\bibitem[Poulain {\em et~al.\/}(2023)Poulain, Content, Sipp, Rigas \&
  Garnier]{poulain2023broadcast}
{\sc \au{Poulain, Arthur}, \au{Content, C{\'e}dric}, \au{Sipp, Denis},
  \au{Rigas, Georgios} \& \au{Garnier, Eric}} \yr{2023}  \at{Broadcast: A
  high-order compressible \textsc{CFD} toolbox for stability and sensitivity
  using algorithmic differentiation}.  \jt{Computer Physics Communications}
  \bvol{283},  \pg{108557}.

\bibitem[Pralits {\em et~al.\/}(2000)Pralits, Airiau, Hanifi \&
  Henningson]{pralits2000sensitivity}
{\sc \au{Pralits, Jan}, \au{Airiau, Christophe}, \au{Hanifi, Ardeshir} \&
  \au{Henningson, Dan}} \yr{2000}  \at{Sensitivity analysis using adjoint
  parabolized stability equations for compressible flows}.  \jt{Flow,
  turbulence and combustion}  \bvol{65}~(3),  \pg{321--346}.

\bibitem[Pralits {\em et~al.\/}(2002)Pralits, Hanifi \&
  Henningson]{pralits2002adjoint}
{\sc \au{Pralits, Jan}, \au{Hanifi, Ardeshir} \& \au{Henningson, Dan}}
  \yr{2002}  \at{Adjoint-based optimization of steady suction for disturbance
  control in incompressible flows}.  \jt{Journal of Fluid Mechanics}
  \bvol{467},  \pg{129--161}.

\bibitem[Reed \& Nayfeh(1986)]{reed1986numerical}
{\sc \au{Reed, Helen} \& \au{Nayfeh, Ali}} \yr{1986}
  \at{Numerical-perturbation technique for stability of flat-plate boundary
  layers with suction}.  \jt{AIAA \textsc{J}ournal}  \bvol{24}~(2),
  \pg{208--214}.

\bibitem[Reynolds \& Saric(1986)]{reynolds1986experiments}
{\sc \au{Reynolds, Gregory} \& \au{Saric, William}} \yr{1986}  \at{Experiments
  on the stability of the flat-plate boundary layer with suction}.  \jt{AIAA
  \textsc{J}ournal}  \bvol{24}~(2),  \pg{202--207}.

\bibitem[Roman {\em et~al.\/}(2015)Roman, Campos, Romero \&
  Tom{\'a}s]{roman2015slepc}
{\sc \au{Roman, Jose}, \au{Campos, Carmen}, \au{Romero, Eloy} \& \au{Tom{\'a}s,
  Andr{\'e}s}} \yr{2015}  \at{Slepc users manual}.  \jt{D. Sistemes
  Inform{\`a}tics i Computaci{\'o} Universitat Polit{\`e}cnica de Val{\`e}ncia,
  Valencia, Spain, Report No. DSIC-II/24/02} .

\bibitem[Saint-James(2020)]{saint2020prevision}
{\sc \au{Saint-James, Julien}} \yr{2020}  \at{Pr{\'e}vision de la transition
  laminaire-turbulent dans le code elsa. extension de la m{\'e}thode des
  paraboles aux parois chauff{\'e}es}. PhD thesis, Institut Sup{\'e}rieur de
  l'A{\'e}ronautique et de l'Espace (ISAE).

\bibitem[Schmid {\em et~al.\/}(2002)Schmid, Henningson \&
  Jankowski]{schmid2002stability}
{\sc \au{Schmid, Peter~J}, \au{Henningson, Dan~S} \& \au{Jankowski, DF}}
  \yr{2002}  \at{Stability and transition in shear flows. applied mathematical
  sciences, vol. 142}.  \jt{Appl. Mech. Rev.}  \bvol{55}~(3),  \pg{B57--B59}.

\bibitem[Sciacovelli {\em et~al.\/}(2021)Sciacovelli, Passiatore, Cinnella \&
  Pascazio]{sciacovelli2021assessment}
{\sc \au{Sciacovelli, Luca}, \au{Passiatore, Donatella}, \au{Cinnella, Paola}
  \& \au{Pascazio, Giuseppe}} \yr{2021}  \at{Assessment of a high-order
  shock-capturing central-difference scheme for hypersonic turbulent flow
  simulations}.  \jt{Computers \& Fluids}  \bvol{230},  \pg{105134}.

\bibitem[Shen {\em et~al.\/}(2009)Shen, Zha \& Chen]{shen2009high}
{\sc \au{Shen, Yiqing}, \au{Zha, Gecheng} \& \au{Chen, Xiangying}} \yr{2009}
  \at{High order conservative differencing for viscous terms and the
  application to vortex-induced vibration flows}.  \jt{Journal of Computational
  Physics}  \bvol{228}~(22),  \pg{8283--8300}.

\bibitem[Sidorenko {\em et~al.\/}(2015)Sidorenko, Gromyko, Bountin, Polivanov
  \& Maslov]{sidorenko2015effect}
{\sc \au{Sidorenko, Andrey}, \au{Gromyko, Yu}, \au{Bountin, Dmitry},
  \au{Polivanov, Pavel} \& \au{Maslov, Anatoly}} \yr{2015}  \at{Effect of the
  local wall cooling/heating on the hypersonic boundary layer stability and
  transition}.  \jt{Progress in Flight Physics--Volume 7}  \bvol{7},
  \pg{549--568}.

\bibitem[Sipp \& Marquet(2013{\natexlab{{\em a\/}}})]{sipp2013characterization}
{\sc \au{Sipp, Denis} \& \au{Marquet, Olivier}} \yr{2013{\natexlab{{\em a\/}}}}
   \at{Characterization of noise amplifiers with global singular modes: the
  case of the leading-edge flat-plate boundary layer}.  \jt{Theoretical and
  Computational Fluid Dynamics}  \bvol{27}~(5),  \pg{617--635}.

\bibitem[Sipp \& Marquet(2013{\natexlab{{\em b\/}}})]{sippmarquet}
{\sc \au{Sipp, D.} \& \au{Marquet, O.}} \yr{2013{\natexlab{{\em b\/}}}}
  \at{Characterization of noise amplifiers with global singular modes: The case
  of the leading-edge flat-plate boundary layer.}  \jt{Theoretical and
  Computational Fluid Dynamics}  \bvol{27}.

\bibitem[Soudakov {\em et~al.\/}(2015)Soudakov, Fedorov \&
  Egorov]{soudakov2015stability}
{\sc \au{Soudakov, Vitaly}, \au{Fedorov, Alexander} \& \au{Egorov, Ivan}}
  \yr{2015}  \at{Stability of high-speed boundary layer on a sharp cone with
  localized wall heating or cooling}.  \jt{Progress in Flight Physics--Volume
  7}  \bvol{7},  \pg{569--584}.

\bibitem[Stuckert {\em et~al.\/}(1995)Stuckert, Lin \&
  Herbert]{stuckert1995nonparallel}
{\sc \au{Stuckert, Greg}, \au{Lin, Nay} \& \au{Herbert, Thorwald}} \yr{1995}
  Nonparallel effects in hypersonic boundary layer stability.  \bt{In {\em 33rd
  Aerospace Sciences Meeting and Exhibit\/}},  \pg{p. 776}.

\bibitem[Sutherland(1893)]{sutherland1893lii}
{\sc \au{Sutherland, William}} \yr{1893}  \at{Lii. the viscosity of gases and
  molecular force}.  \jt{The London, Edinburgh, and Dublin Philosophical
  Magazine and Journal of Science}  \bvol{36}~(223),  \pg{507--531}.

\bibitem[Walther {\em et~al.\/}(2001)Walther, Airiau \&
  Bottaro]{walther2001optimal}
{\sc \au{Walther, Steeve}, \au{Airiau, Christophe} \& \au{Bottaro, Alessandro}}
  \yr{2001}  \at{Optimal control of tollmien--schlichting waves in a developing
  boundary layer}.  \jt{Physics of Fluids}  \bvol{13}~(7),  \pg{2087--2096}.

\bibitem[Wang \& Zhong(2009)]{wang2009effect}
{\sc \au{Wang, Xiaowen} \& \au{Zhong, Xiaolin}} \yr{2009}  \at{Effect of wall
  perturbations on the receptivity of a hypersonic boundary layer}.
  \jt{Physics of fluids}  \bvol{21}~(4),  \pg{044101}.

\bibitem[Wang {\em et~al.\/}(2011)Wang, Zhong \& Ma]{wang2011response}
{\sc \au{Wang, Xiaowen}, \au{Zhong, Xiaolin} \& \au{Ma, Yanbao}} \yr{2011}
  \at{Response of a hypersonic boundary layer to wall blowing-suction}.
  \jt{AIAA \textsc{J}ournal}  \bvol{49}~(7),  \pg{1336--1353}.

\bibitem[Wang {\em et~al.\/}(2019)Wang, Ferrer, Mart{\'\i}nez-Cava, Zheng \&
  Valero]{wang2019enhanced}
{\sc \au{Wang, Yinzhu}, \au{Ferrer, Esteban}, \au{Mart{\'\i}nez-Cava,
  Alejandro}, \au{Zheng, Yao} \& \au{Valero, Eusebio}} \yr{2019}  \at{Enhanced
  stability of flows through contraction channels: Combining shape optimization
  and linear stability analysis}.  \jt{Physics of Fluids}  \bvol{31}~(7),
  \pg{074109}.

\bibitem[Zhao {\em et~al.\/}(2018)Zhao, Wen, Tian, Long \&
  Yuan]{zhao2018numerical}
{\sc \au{Zhao, Rui}, \au{Wen, Chihyung}, \au{Tian, Xudong}, \au{Long, Tiehan}
  \& \au{Yuan, Wu}} \yr{2018}  \at{Numerical simulation of local wall heating
  and cooling effect on the stability of a hypersonic boundary layer}.
  \jt{International Journal of Heat and Mass Transfer}  \bvol{121},
  \pg{986--998}.

\bibitem[Zuccher {\em et~al.\/}(2004)Zuccher, Luchini \&
  Bottaro]{zuccher2004algebraic}
{\sc \au{Zuccher, Simone}, \au{Luchini, Paolo} \& \au{Bottaro, Alessandro}}
  \yr{2004}  \at{Algebraic growth in a blasius boundary layer: optimal and
  robust control by mean suction in the nonlinear regime}.  \jt{Journal of
  Fluid Mechanics}  \bvol{513},  \pg{135--160}.

\end{thebibliography}

\end{document}